\documentclass[superscriptaddress, amsmath, amssymb, aps, twocolumn, longbibliography]{revtex4-1}
\usepackage{graphicx}
\usepackage{dcolumn}
\usepackage{amsthm}
\usepackage{bm}

\pdfoutput=1
\usepackage{graphicx,graphics,epsfig,subfigure,times,bm,bbm,amssymb,amsmath,amsthm,mathrsfs,MnSymbol}
\usepackage{gensymb}
\usepackage{amsfonts}
\usepackage[matrix,frame,arrow]{xypic}
\usepackage[pdfstartview=FitH]{hyperref}
\hypersetup{
	colorlinks=true,       
	linkcolor=red,          
	citecolor=magenta,        
	filecolor=magenta,      
	urlcolor=cyan,           
	runcolor=cyan}
\usepackage{epstopdf}
\usepackage[pdftex]{color}
\usepackage{braket}
\usepackage{enumerate}
\usepackage[normalem]{ulem}
\usepackage[usenames,dvipsnames]{xcolor}
\usepackage{multirow}
\usepackage{mathtools}

\definecolor{orange}{rgb}{1,0.5,0}

\newcommand{\ignore}[1]{}
\usepackage{geometry}

\ignore{
	\documentclass[eprintnumbers,amsmath,amssymb,onecolumn,a4paper,caption ]{article}
	\usepackage{amsfonts}
	\usepackage{amssymb}
	\usepackage{mathrsfs}
	\usepackage{mathbbold}
	\usepackage{bbm}
	\usepackage{mathrsfs}
	\usepackage{dcolumn}
	\usepackage{bm}
	\usepackage{times,epsfig,amssymb,amsmath}
	\usepackage{float}
	\usepackage{subfigure}
	\usepackage{geometry}\geometry{left=2.5cm,right=2.5cm,top=3cm,bottom=3cm}

	\usepackage{color}

}

\usepackage{changes}
\definechangesauthor[name={Yun-Hao Shi},color=blue]{syh}

\begin{document}

\title{Probing spin hydrodynamics on a superconducting quantum simulator}

\author{Yun-Hao Shi}
\thanks{These authors contributed equally to this work.}
\affiliation{Institute of Physics, Chinese Academy of Sciences, Beijing 100190, China}
\affiliation{School of Physical Sciences, University of Chinese Academy of Sciences, Beijing 100049, China}
\affiliation{Beijing Academy of Quantum Information Sciences, Beijing 100193, China}

\author{Zheng-Hang Sun}
\thanks{These authors contributed equally to this work.}
\affiliation{Institute of Physics, Chinese Academy of Sciences, Beijing 100190, China}
\affiliation{School of Physical Sciences, University of Chinese Academy of Sciences, Beijing 100049, China}

\author{Yong-Yi Wang}
\thanks{These authors contributed equally to this work.}
\affiliation{Institute of Physics, Chinese Academy of Sciences, Beijing 100190, China}
\affiliation{School of Physical Sciences, University of Chinese Academy of Sciences, Beijing 100049, China}

\author{Zheng-An Wang}
\affiliation{Beijing Academy of Quantum Information Sciences, Beijing 100193, China}
\affiliation{Hefei National Laboratory, Hefei 230088, China}

\author{Yu-Ran Zhang}
\affiliation{School of Physics and Optoelectronics, South China University of Technology, Guangzhou 510640, China}

\author{Wei-Guo Ma}
\affiliation{Institute of Physics, Chinese Academy of Sciences, Beijing 100190, China}
\affiliation{School of Physical Sciences, University of Chinese Academy of Sciences, Beijing 100049, China}

\author{Hao-Tian Liu}
\affiliation{Institute of Physics, Chinese Academy of Sciences, Beijing 100190, China}
\affiliation{School of Physical Sciences, University of Chinese Academy of Sciences, Beijing 100049, China}

\author{Kui Zhao}
\affiliation{Beijing Academy of Quantum Information Sciences, Beijing 100193, China}

\author{Jia-Cheng Song}
\affiliation{Institute of Physics, Chinese Academy of Sciences, Beijing 100190, China}
\affiliation{School of Physical Sciences, University of Chinese Academy of Sciences, Beijing 100049, China}

\author{Gui-Han Liang}
\affiliation{Institute of Physics, Chinese Academy of Sciences, Beijing 100190, China}
\affiliation{School of Physical Sciences, University of Chinese Academy of Sciences, Beijing 100049, China}

\author{Zheng-Yang Mei}
\affiliation{Institute of Physics, Chinese Academy of Sciences, Beijing 100190, China}
\affiliation{School of Physical Sciences, University of Chinese Academy of Sciences, Beijing 100049, China}

\author{Jia-Chi Zhang}
\affiliation{Institute of Physics, Chinese Academy of Sciences, Beijing 100190, China}
\affiliation{School of Physical Sciences, University of Chinese Academy of Sciences, Beijing 100049, China}

\author{Hao Li}
\affiliation{Beijing Academy of Quantum Information Sciences, Beijing 100193, China}

\author{Chi-Tong Chen}
\affiliation{Institute of Physics, Chinese Academy of Sciences, Beijing 100190, China}
\affiliation{School of Physical Sciences, University of Chinese Academy of Sciences, Beijing 100049, China}

\author{Xiaohui Song}
\affiliation{Institute of Physics, Chinese Academy of Sciences, Beijing 100190, China}

\author{Jieci Wang}
\affiliation{Department of Physics and Key Laboratory of Low Dimensional Quantum Structures and Quantum Control of Ministry of Education, Hunan Normal University, Changsha, Hunan 410081, China}

\author{Guangming Xue}
\affiliation{Beijing Academy of Quantum Information Sciences, Beijing 100193, China}

\author{Haifeng Yu}
\affiliation{Beijing Academy of Quantum Information Sciences, Beijing 100193, China}

\author{Kaixuan Huang}
\email{huangkx@baqis.ac.cn}
\affiliation{Beijing Academy of Quantum Information Sciences, Beijing 100193, China}

\author{Zhongcheng Xiang}
\email{zcxiang@iphy.ac.cn}
\affiliation{Institute of Physics, Chinese Academy of Sciences, Beijing 100190, China}
\affiliation{School of Physical Sciences, University of Chinese Academy of Sciences, Beijing 100049, China}
\affiliation{Hefei National Laboratory, Hefei 230088, China}

\author{Kai Xu}
\email{kaixu@iphy.ac.cn}
\affiliation{Institute of Physics, Chinese Academy of Sciences, Beijing 100190, China}
\affiliation{School of Physical Sciences, University of Chinese Academy of Sciences, Beijing 100049, China}
\affiliation{Beijing Academy of Quantum Information Sciences, Beijing 100193, China}
\affiliation{Hefei National Laboratory, Hefei 230088, China}
\affiliation{Songshan Lake Materials Laboratory, Dongguan, Guangdong 523808, China}
\affiliation{CAS Center for Excellence in Topological Quantum Computation, UCAS, Beijing 100190, China}

\author{Dongning Zheng}
\affiliation{Institute of Physics, Chinese Academy of Sciences, Beijing 100190, China}
\affiliation{School of Physical Sciences, University of Chinese Academy of Sciences, Beijing 100049, China}
\affiliation{Hefei National Laboratory, Hefei 230088, China}
\affiliation{Songshan Lake Materials Laboratory, Dongguan, Guangdong 523808, China}
\affiliation{CAS Center for Excellence in Topological Quantum Computation, UCAS, Beijing 100190, China}

\author{Heng Fan}
\email{hfan@iphy.ac.cn}
\affiliation{Institute of Physics, Chinese Academy of Sciences, Beijing 100190, China}
\affiliation{School of Physical Sciences, University of Chinese Academy of Sciences, Beijing 100049, China}
\affiliation{Beijing Academy of Quantum Information Sciences, Beijing 100193, China}
\affiliation{Hefei National Laboratory, Hefei 230088, China}
\affiliation{Songshan Lake Materials Laboratory, Dongguan, Guangdong 523808, China}
\affiliation{CAS Center for Excellence in Topological Quantum Computation, UCAS, Beijing 100190, China}

\begin{abstract}
\noindent Characterizing the nature of hydrodynamical transport properties in quantum dynamics provides valuable insights into the fundamental understanding of exotic non-equilibrium phases of matter. Experimentally simulating infinite-temperature transport on large-scale complex quantum systems is of considerable interest. Here, using a controllable and coherent superconducting quantum simulator, we experimentally realize the analog quantum circuit, which can efficiently prepare the Haar-random states, and probe spin transport at infinite temperature. We observe diffusive spin transport during the unitary evolution of the ladder-type quantum simulator with ergodic dynamics. Moreover, we explore the transport properties of the systems subjected to strong disorder or a tilted potential, revealing signatures of anomalous subdiffusion in accompany with the breakdown of thermalization. Our work demonstrates a scalable method of probing infinite-temperature spin transport on analog quantum simulators, which paves the way to study other intriguing out-of-equilibrium phenomena from the perspective of transport.
\end{abstract}
\pacs{Valid PACS appear here}
\maketitle



\noindent
\textbf{\large{Introduction}}

\noindent Transport properties of quantum many-body systems driven out of equilibrium 
are of significant interest in several active areas of modern physics, including the ergodicity of quantum systems~\cite{Nandkishore:2015vg,PhysRevLett.114.160401,PhysRevLett.117.040601,PhysRevX.13.011033} and quantum magnetism~\cite{Scheie:2021vl,PhysRevLett.106.220601,PhysRevLett.127.107201}. Understanding these properties is crucial to unveil the non-equilibrium dynamics of isolated quantum systems~\cite{RevModPhys.93.025003,Eisert:2015ws}. One essential property of transport is the emergence of classical hydrodynamics in microscopic quantum dynamics, 
which shows the power-law tail of autocorrelation functions~\cite{RevModPhys.93.025003}. The rate of the power-law decay, referred as to the transport exponent, characterizes the universal classes of hydrodynamics. In $d$-dimensional quantum systems, in addition to generally expected diffusive transport with the exponent $d/2$ in non-integrable systems~\cite{Peng:2023vn,PhysRevB.90.094417,PhysRevB.104.054415}, more attentions have been attracted by the anomalous superdiffusive~\cite{Scheie:2021vl,Ljubotina:2017ub,Wei:2022vj,Joshi:2022tr,2023arXiv230609333R} or subdiffusive transport~\cite{PhysRevLett.114.160401,PhysRevLett.117.040601,PhysRevLett.125.245303,De-Nardis:2022ud,PhysRevResearch.2.033124}, with the exponent larger or smaller than $d/2$, respectively. 

Over the last few decades, considerable strides have been made in enhancing the scalability, controllability, and coherence of noisy intermediate-scale quantum (NISQ) devices based on superconducting qubits~\cite{Ma:2019um,doi:10.1126/science.ade7651,Xiang2023,Gu:2017ut}. With these advancements, several novel phenomena in non-equilibrium dynamics of quantum many-body systems have been observed, such as quantum thermalization~\cite{PhysRevLett.127.020602,PhysRevLett.128.160502}, ergodicity breaking~\cite{Roushan:2017vp,Guo:2021wr,PhysRevLett.127.240502,Zhang:2023vv}, time crystal~\cite{Zhang:2022wf,Mi:2022tv,doi:10.1126/sciadv.abm7652}, and information scrambling~\cite{Mi:2021ta,Braumuller:2022wj}. More importantly, in this platform, the beyond-classical computation has been demonstrated by sampling the final Haar-random states of randomized sequences of gate operations~\cite{Neill:2018wy,Boixo:2018un,Arute:2019ts,PhysRevLett.127.180501,2023arXiv230411119M}. Recently, a method of measuring autocorrelation functions at infinite temperature based on the Haar-random states has been proposed, which opens up a practical application of pseudo-random quantum circuits for simulating hydrodynamics on NISQ devices~\cite{PhysRevLett.126.230501,2022arXiv220812243K}. 

In this work, using a ladder-type superconducting quantum simulator with up to 24 qubits, we first demonstrate that in addition to the digital pseudo-random circuits~\cite{Neill:2018wy,Boixo:2018un,Arute:2019ts,PhysRevLett.127.180501,2023arXiv230411119M,PhysRevLett.126.230501,2022arXiv220812243K}, a unitary evolution governed by a time-independent Hamiltonian, i.e., an analog quantum circuit, can also generate quantum states randomly chosen from the Haar measure, i.e., the Haar-random states, for measuring the infinite-temperature autocorrelation functions~\cite{Choi:2023tt,Karamlou:2024ua,Yanay:2020tb}. Subsequently, we study the properties of spin transport on the superconducting quantum simulator via the measurement of autocorrelation functions by using the Haar-random states. Notably, we observe a clear signature of the diffusive transport on the qubit ladder,  which is a non-integrable system~\cite{PhysRevLett.128.160502,PhysRevB.90.094417,PhysRevB.104.054415}.

Upon subjecting the qubit ladder to disorder, 
a transition from delocalized phases to the many-body localization (MBL) occurs as the strength of disorder increases~\cite{PhysRevResearch.2.013163}. By measuring the autocorrelation functions, we experimentally probe an anomalous subdiffusive transport with intermediate values of the disorder strength. The observed signs of subdiffusion are consistent with recent numerical results, and can be explained as a consequence of Griffth-like region on the delocalized side of the MBL transition~\cite{PhysRevLett.114.160401,PhysRevLett.117.040601,PhysRevB.93.224205,PhysRevB.93.134206,PhysRevB.96.104205,Luitz:2017uz}. 

Finally, we explore spin transport on the qubit ladder with a linear potential, and it is expected that Stark MBL occurs when the potential gradients are sufficiently large~\cite{PhysRevLett.127.240502,Morong:2021ul,PhysRevLett.127.240502,PhysRevLett.122.040606,Nieuwenburg:2019vb,PhysRevB.104.205122,PhysRevB.102.054206}. With a large gradient, the conservation of the dipole moment emerges~\cite{PhysRevLett.127.240502,PhysRevB.102.054206}, associated with the phenomena known as the Hilbert space fragmentation~\cite{PhysRevB.103.L100202,PhysRevB.101.174204,PhysRevX.10.011047}. Recent theoretical works reveal a subdiffusion in the dipole-moment conserving systems~\cite{PhysRevLett.125.245303,PhysRevResearch.2.033124}. In this experiment, we present evidence of a subdiffusive regime of spin transport in the tilted qubit ladder. 

~\\

\begin{figure*}
	\centering
	\includegraphics[width=0.96\linewidth]{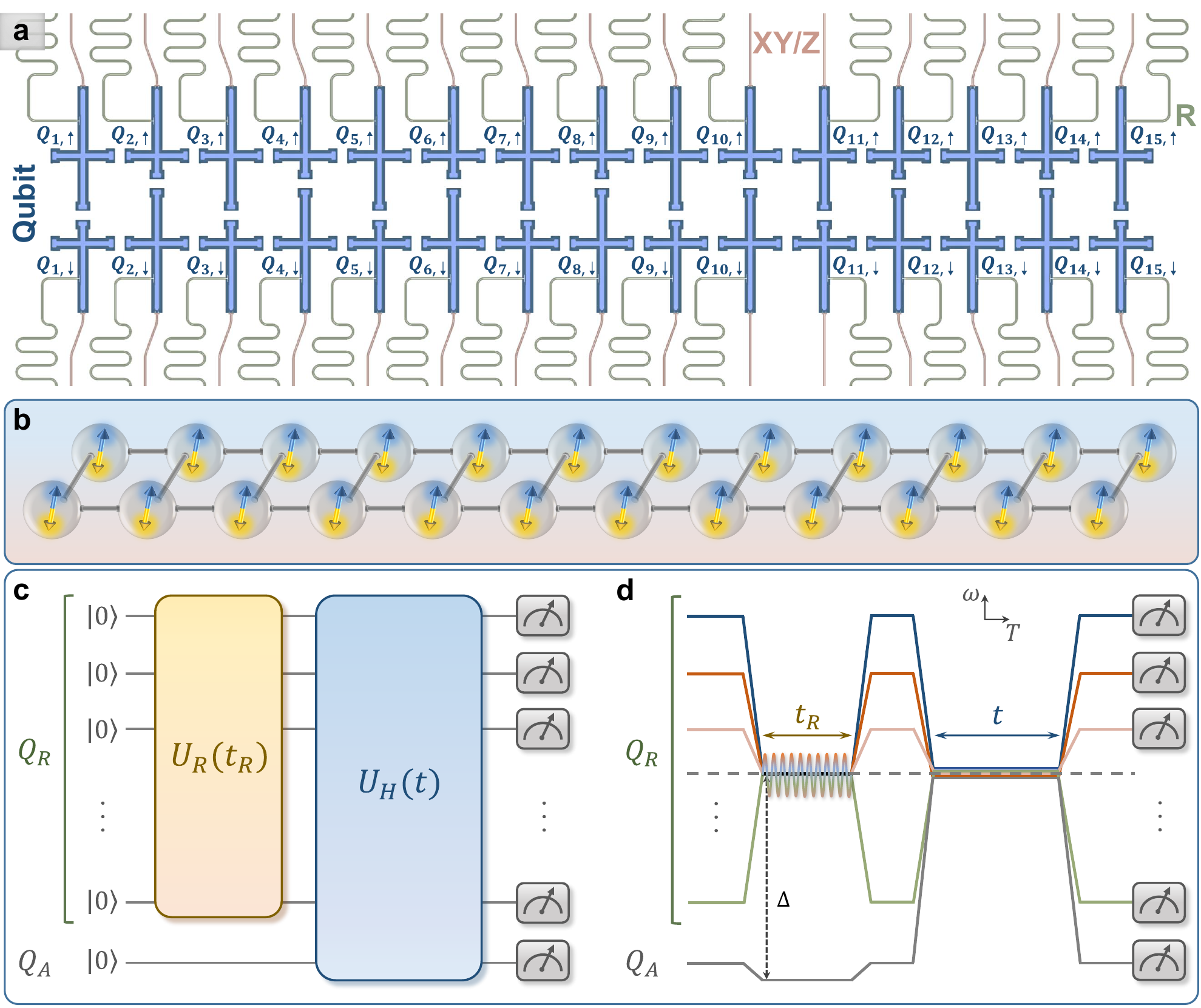}\\
	\caption{\textbf{Superconducting quantum simulator and experimental pulse sequences.} \textbf{a}, The schematic showing the ladder-type superconducting quantum simulator, consisting of 30 qubits (the blue region), labeled $Q_{1,\uparrow}$ to $Q_{15,\uparrow}$, and $Q_{1,\downarrow}$ to $Q_{15,\downarrow}$. Each qubit is coupled to a separate readout resonator (the green region), and has an individual control line (the red region) for both the XY and Z controls. \textbf{b}, Schematic diagram of the simulated 24 spins coupled in a ladder. The blue and yellow double arrows represent the infinite-temperature spin hydrodynamics without preference for spin orientations. \textbf{c}, Schematic diagram of the quantum circuit for measuring the autocorrelation functions at infinite temperature. All qubits are initialized at the state $|0\rangle$. Subsequently, an analog quantum circuit $\hat{U}_{R}(t_{R})$ acts on the set of qubits $Q_{R}$ to generate Haar-random states. This is followed by a time evolution of all qubits, i.e., $\hat{U}_{H}(t) = \exp(-i\hat{H}t)$ with $\hat{H}$ being the Hamiltonian of the system, in which the properties of spin transport are of our interest. \textbf{d}, Experimental pulse sequences corresponding to the quantum circuit in \textbf{c}, displayed in the frequency ($\omega$) versus time ($T$) domain. To realize $\hat{U}_{R}(t_{R})$,  qubits in the set $Q_{R}$ are tuned to the working point (dashed horizontal line) via Z pulses, and simultaneously, the resonant microwave pulses represented as the sinusoidal line are applied to $Q_{R}$ through the XY control lines. Meanwhile, the qubit $Q_{A}$ is detuned from the working point with a large value of the frequency gap $\Delta$. To realize the subsequent evolution $\hat{U}_{H}(t)$ with the Hamiltonian~\eqref{H_int}, all qubits are tuned to the working point.}\label{fig1}
\end{figure*}

\noindent
\textbf{\large{Results}}
\\
\textbf{Experimental setup and protocol}
\\
\noindent Our experiments are performed on a programmable superconducting quantum simulator, consisting of 30 transmon qubits with a geometry of two-legged ladder, see Fig.~\ref{fig1}\textbf{a} and \textbf{b}.  The nearest-neighbor qubits are coupled by a fixed capacitor, and the effective Hamiltonian of capacitive interactions can be written as~\cite{Xiang2023,Gu:2017ut} (also see Supplementary Note 1)
\begin{eqnarray} \nonumber
\hat{H}_{I}/\hbar &=& \sum_{m\in\{\uparrow,\downarrow\}}\sum_{j=1}^{L-1} J^{\parallel}_{j,m} (\hat{\sigma}_{j,m}^{+}\hat{\sigma}_{j+1,m}^{-}+\text{H.c.}) \\ 
&~&+ \sum_{j=1}^{L} J^{\perp}_{j} (\hat{\sigma}_{j,\uparrow}^{+}\hat{\sigma}_{j,\downarrow}^{-}+\text{H.c.}), 
\label{H_int}
\end{eqnarray}
where $\hbar=h/2\pi$, with $h$ being the Planck constant (in the following we set $\hbar=1$), $L$ is the length of the ladder, $\hat{\sigma}_{j,m}^{+}$ ($\hat{\sigma}_{j,m}^{-}$) is the raising (lowering) operator for the qubit $Q_{j,m}$, and $J^{\parallel}_{j,m}$ ($J^{\perp}_{j}$) refers to the rung (intrachain) hopping strength. For this device, the averaged rung and intrachain hopping strength are $\overline{J^{\parallel}}/2\pi \simeq 7.3~\!\mathrm{MHz}$ and $\overline{J^{\perp}}/2\pi \simeq 6.6~\!\mathrm{MHz}$, respectively.
The XY and Z control lines on the device enable us to realize the drive Hamiltonian $\hat{H}_{d} = \sum_{m\in\{\uparrow,\downarrow\}}\sum_{j=1}^{L} \Omega_{j,m}(e^{-i\phi_{j,m}}\hat{\sigma}_{j,m}^{+}+e^{i\phi_{j,m}}\hat{\sigma}_{j,m}^{-})/2$, and the on-site potential Hamiltonian $\hat{H}_{\text{Z}} = \sum_{m\in\{\uparrow,\downarrow\}}\sum_{j=1}^{L}  w_{j,m}\hat{\sigma}_{j,m}^{+} \hat{\sigma}_{j,m}^{-}$, respectively. Here, $\Omega_{j,m}$ and $\phi_{j,m}$ denote the driving amplitude and the phase of the microwave pulse applied on the qubit $Q_{j,m}$, and $w_{j,m}$ is the effective on-site potential. 

To study spin transport and hydrodynamics, we focus on the equal-site autocorrelation function at infinite temperature, which is defined as 
\begin{eqnarray}
C_{\textbf{r},\textbf{r}} = \frac{1}{D} \text{Tr}[\hat{\rho}_{\textbf{r}}(t)\hat{\rho}_{\textbf{r}}],
\label{C_rr}
\end{eqnarray}
where $\hat{\rho}_{\textbf{r}}$ is a local observable at site $\textbf{r}$, $\hat{\rho}_{\textbf{r}}(t) = e^{i\hat{H}t}\hat{\rho}_{\textbf{r}} e^{-i\hat{H}t}$, and $D$ is the Hilbert dimension of the Hamiltonian $\hat{H}$. Here, for the ladder-type superconducting simulator, we choose $\hat{\rho}_{\textbf{r}} = (\hat{\sigma}^{z}_{1,\uparrow} + \hat{\sigma}^{z}_{1,\downarrow})/2$ ($\textbf{r}=1$)~\cite{PhysRevB.104.054415}, and the autocorrelation function can be rewritten as
\begin{eqnarray}
C_{1,1} = \frac{1}{4}(c_{1,\uparrow;1,\uparrow} + c_{1,\uparrow;1,\downarrow} + c_{1,\downarrow;1,\uparrow}+c_{1,\downarrow;1,\downarrow}),
\label{C_11}
\end{eqnarray}
with $c_{\mu;\nu} = \text{Tr}[\hat{\sigma}_{\mu}^{z}(t)\hat{\sigma}_{\nu}^{z}]/D$ (subscripts $\mu$ and $\nu$ denote the qubit index $1,\uparrow$ or $1,\downarrow$). 

The autocorrelation functions (\ref{C_rr}) at infinite temperature can be expanded as the average of $C_{\textbf{r},\textbf{r}}(|\psi_{0}\rangle) = \langle \psi_{0} |\hat{\rho}_{\textbf{r}}(t)\hat{\rho}_{\textbf{r}}  | \psi_{0}\rangle$ over different $| \psi_{0}\rangle$ in $z$-basis. In fact, the dynamical behavior of an individual $C_{\textbf{r},\textbf{r}}(|\psi_{0}\rangle)$ is sensitive to the choice of $| \psi_{0}\rangle$ under some circumstances (see Supplementary Note 7 for the dependence of $C_{\textbf{r},\textbf{r}}(|\psi_{0}\rangle)$ on $|\psi_{0}\rangle$ in the qubit ladder with a linear potential as an example). To experimentally probe the generic properties of spin transport at infinite temperature, one can obtain (\ref{C_rr}) by measuring and averaging $C_{\textbf{r},\textbf{r}}(|\psi_{0}\rangle)$ with different $| \psi_{0}\rangle$~\cite{Joshi:2022tr}. Alternatively, we employ a more efficient method to measure (\ref{C_rr}) without the need of sampling different $| \psi_{0}\rangle$. 
Based on the results in ref.~\cite{PhysRevLett.126.230501} (also see Methods), the autocorrelation function $c_{\mu;\nu}$ can be indirectly measured by using the quantum circuit as shown in Fig.~\ref{fig1}\textbf{c}, i.e., 
\begin{eqnarray}
c_{\mu;\nu} \simeq \langle\psi_{\nu}^{R}(t) | \hat{\sigma}^{z}_{\mu} |\psi_{\nu}^{R}(t)\rangle, 
\label{measure_c}
\end{eqnarray}
where $|\psi_{\nu}^{R}(t)\rangle = \hat{U}_{H}(t) [|0\rangle_{\nu}\otimes |\psi^{R}\rangle] $ with $|\psi^{R}\rangle = \hat{U}_{R}\bigotimes_{i\in Q_{R}} |0\rangle_{i}$, and $\hat{U}_{R}$ being a unitary evolution generating Haar-random states. For example, to experimentally obtain $c_{1,\downarrow;1,\uparrow}$, we choose $Q_{1,\uparrow}$ as $Q_{A}$, and the remainder qubits as the $Q_{R}$. After performing the pulse sequences as shown in Fig.~\ref{fig1}\textbf{d}, we measure the qubit $Q_{1,\downarrow}$ at $z$-basis to obtain the expectation value of the observable $\hat{\sigma}_{1,\downarrow}^{z}$. 


~\\
\noindent  \textbf{Observation of diffusive transport}

\begin{figure}
	\includegraphics[width=0.97\linewidth]{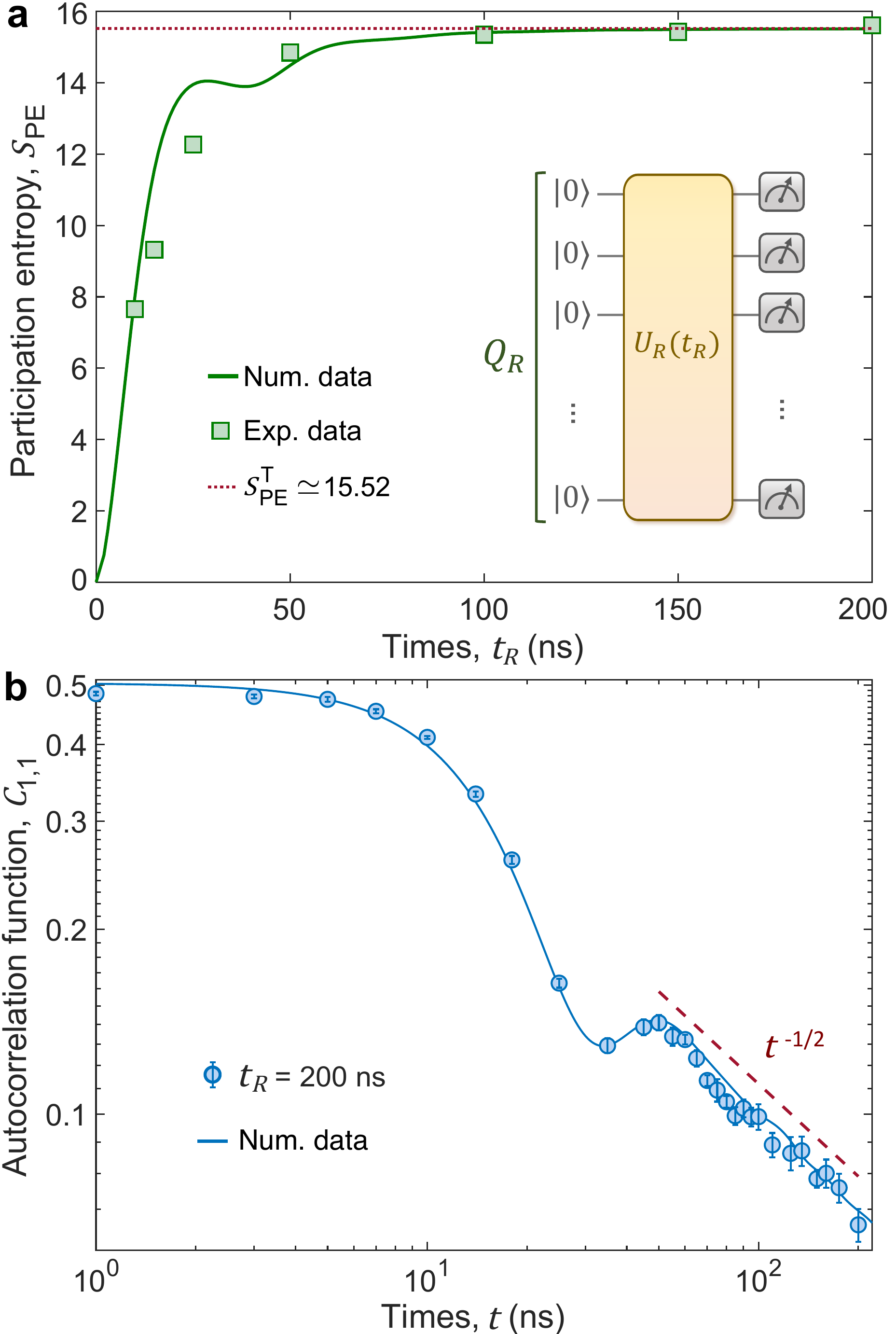}\\
	\caption{\textbf{Observation of diffusive transport.} \textbf{a}, Experimental verification of preparing the  states via the time evolution of participation entropy. Here, we chose $Q_{R} = \{Q_{1,\uparrow},Q_{2,\uparrow},...,Q_{12,\uparrow},Q_{2,\downarrow},Q_{3,\downarrow},...,Q_{12,\downarrow}\}$ with total 23 qubits. The inset of \textbf{a} shows the corresponding quantum circuit. The dotted horizontal line represents the participation entropy for Haar-random states, i.e., $S_{\text{PE}}^{\text{T}}\simeq 15.519$. \textbf{b}, Experimental results of the autocorrelation function $C_{1,1}(t)$ for the qubit ladder with $L=12$, which are measured by performing the quantum circuit shown in Fig.~\ref{fig1}\textbf{c} and \textbf{d}. Here, we consider the state generated from $\hat{U}_{R}(t_{R})$ with $t_{R}= 200~\!\mathrm{ns}$, which is approximate to a Haar-random state. Markers are experimental data. The solid line is the numerical simulation of the correlation function $C_{1,1}$ at infinite temperature.  The dashed line represents a power-law decay $t^{-1/2}$. Error bars represent the standard deviation.}\label{fig2}
\end{figure}

\noindent In this experiment, we first study spin transport on the 24-qubit ladder consisting of $Q_{1,\uparrow},...,Q_{12,\uparrow}$ and $Q_{1,\downarrow},...,Q_{12,\downarrow}$, described by the Hamiltonian (\ref{H_int}). For a non-integrable model, one expects that diffusive transport $C_{1,1}\propto t^{-1/2} $ occurs~\cite{PhysRevB.104.054415}. To measure the autocorrelation function $C_{1,1}$ defined in Eq.~(\ref{C_11}), we should first perform a quantum circuit generating the required Haar-random states $|\psi^{R}\rangle$. Instead of using the digital pseudo-random circuits in Refs.~\cite{Neill:2018wy,Boixo:2018un,Arute:2019ts,PhysRevLett.127.180501,2023arXiv230411119M,PhysRevLett.126.230501,2022arXiv220812243K}, here we experimentally realize the 
time evolution under the Hamiltonian $\hat{H}_{R}=\hat{H}_{I}+\hat{H}_{d}$, where the parameters $\Omega_{j,m}$ and $\phi_{j,m}$ in $\hat{H}_{d}$ have site-dependent values with the average $\overline{\Omega}/2\pi \simeq 10.4~\!\mathrm{MHz}$ ($\overline{\Omega}/\overline{J^{\parallel}} \simeq 1.4$) and $\overline{\phi}=0$ (see Methods and Supplementary Note 3 for more details), i.e., $\hat{U}_{R}(t_{R}) = \exp (-i\hat{H}_{R}t_{R})$, which is more suitable for our analog quantum simulator. 
To benchmark that the final state $|\psi^{R}\rangle = \hat{U}_{R}(t_{R})|0\rangle$ can approximate the Haar-random states, we measure the participation entropy $S_{\text{PE}} = -\sum_{k=1}^{D} p_{k} \ln p_{k}$, with $D$ being the dimension of Hilbert space, $p_{k} = |\langle k |\psi^{R}\rangle|^{2}$, and $\{| k \rangle \}$ being a computational basis. Figure~\ref{fig2}\textbf{a} shows the results of $S_{\text{PE}}$ with different evolution times $t_{R}$. For the 23-qubit system, the probabilities $p_{k}$ are estimated from the single-shot readout with a number of samples $N_{\text{s}} = 3\times 10^{7}$. It is seen that the $S_{\text{PE}}$ tends to the value for Haar-random states, i.e., $S_{\text{PE}}^{\text{T}} = N\ln2 - 1 + \gamma $ with $N=23$ being the number of qubits and $\gamma\simeq 0.577$ as the Euler's constant~\cite{Boixo:2018un}. Moreover, for the final state 
$|\psi^{R}\rangle$ with $t_{R}=200~\!\mathrm{ns}$, the distribution of probabilities $p_{k}$ satisfies the Porter-Thomas distribution (see Supplementary Note 4).

In Fig.~\ref{fig2}\textbf{b}, we show the dynamics of the autocorrelation function $C_{1,1}$ measured via the quantum circuit in Fig.~\ref{fig1}\textbf{c} with $t_{R}=200~\!\mathrm{ns}$. The experimental data satisfies $C_{1,1}\propto t^{-\alpha}$, with a transport exponent $\alpha \simeq 0.5067$, estimated by fitting the data in the time window $t\in [50~\!\mathrm{ns}, 200~\!\mathrm{ns}]$. 
Our experiments clearly show that spin diffusively transports on the qubit ladder $\hat{H}_{I}$ (\ref{H_int}), and demonstrate that the analog quantum circuit $\hat{U}_{R}(t_{R})$ with $t_{R}=200~\!\mathrm{ns}$ can provide sufficient randomness to measure the autocorrelation function defined in Eq.~(\ref{C_rr}) and probe infinite-temperature spin transport. We also discuss the influence of $t_{R}$ in Supplementary Note 4, numerically showing that the results of $C_{1,1}$ do not substantially change for longer $t_{R}>200~\!\mathrm{ns}$. Moreover, in Supplementary Note 4, we show that for a short evolved time $t_{R}\simeq 15~\!\mathrm{ns}$, the values of the observable defined in Eq.~(\ref{measure_c}) are incompatible with the infinite-temperature autocorrelation functions. Given that the chosen initial state for generating the Haar-random state exhibits a high effective temperature associated with the Hamiltonian $\hat{H}_{R}$, the state $|\psi^{R}\rangle$ would asymptotically converge to the Haar-random state with a sufficiently extended $t_{R}$. However, with $t_{R}\simeq 15~\!\mathrm{ns}$, the time scale is too small to get rid of the coherence, and the value of $S_{\text{PE}}$ for the state $|\psi^{R}\rangle$ is much smaller than the $S_{\text{PE}}^{\text{T}}$ (see Fig.~\ref{fig2}\textbf{a}), suggesting that $|\psi^{R}\rangle$ with $t_{R}\simeq 15~\!\mathrm{ns}$ is far away from the Haar-random state, and cannot be employed to measure the infinite-temperature autocorrelation function (\ref{C_rr}). In the following, we fix $t_{R}=200~\!\mathrm{ns}$, and study spin transport in other systems with ergodicity breaking. 




~\\
\noindent \textbf{Subdiffusive transport with ergodicity breaking}

\begin{figure}
	\centering
	\includegraphics[width=0.97\linewidth]{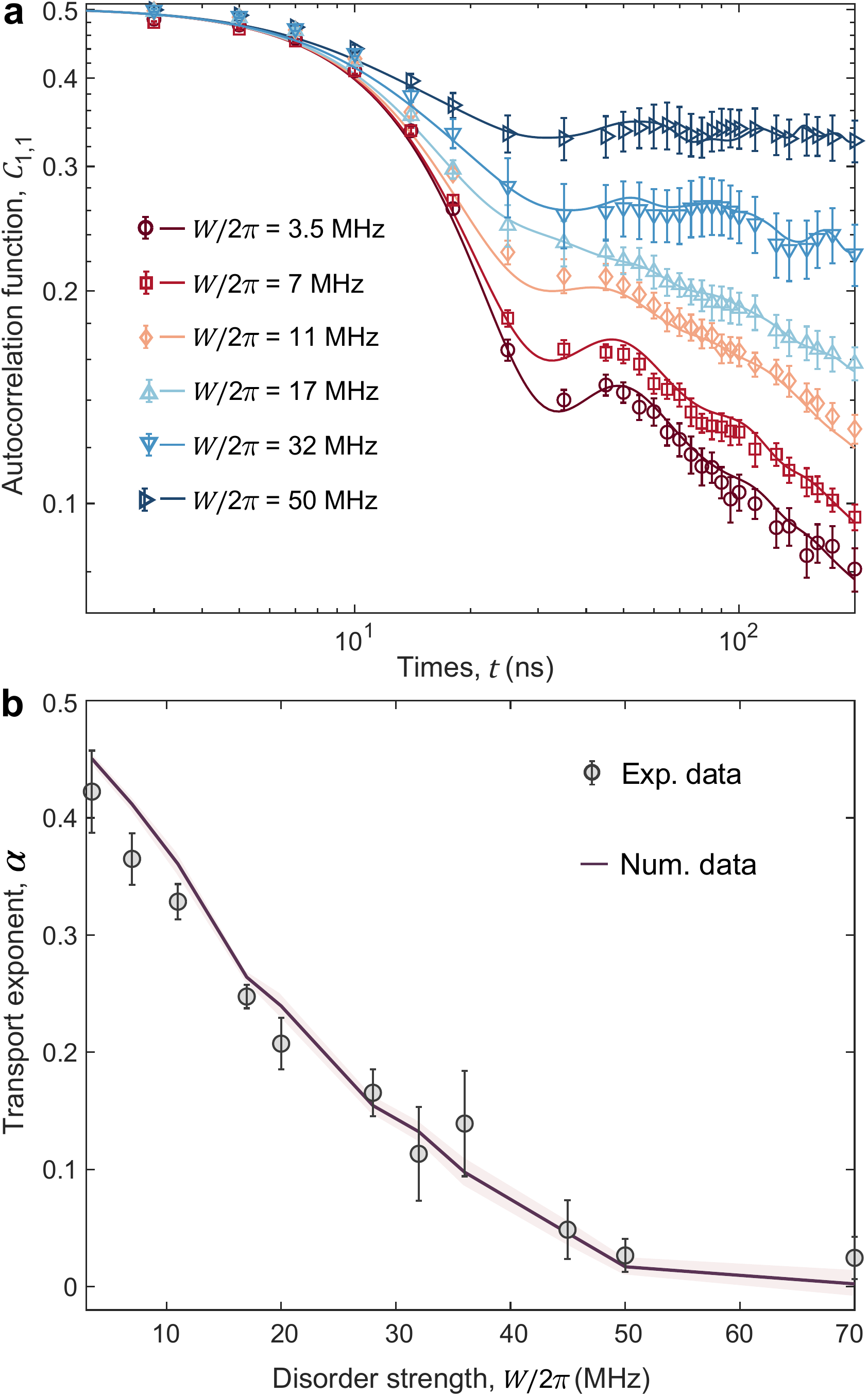}\\
	\caption{\textbf{Subdiffusive transport on the superconducting qubit ladder with disorder.} \textbf{a}, The time evolution of autocorrelation function $C_{1,1}(t)$ for the qubit ladder with $L=12$ and different values of disorder strength $W$, ranging from $W/2\pi=35~\!\mathrm{MHz}$ ($W/\overline{J^{\parallel}}\simeq 0.5$) to $W/2\pi=70~\!\mathrm{MHz}$ ($W/\overline{J^{\parallel}}\simeq 9.6$). Markers (lines) are experimental (numerical) data. \textbf{b}, Transport exponent $\alpha$ as a function of $W$ obtained from fitting the data of $C_{1,1}(t)$. Error bars (experimental data) and shaded regions (numerical data) represent the standard deviation.}\label{fig3}
\end{figure}

\noindent After demonstrating that the quantum circuit shown in Fig.~\ref{fig1}\textbf{c} can be employed to measure the infinite-temperature autocorrelation function $C_{1,1}$, we study spin transport on the superconducting qubit ladder with disorder, whose effective Hamiltonian can be written as $\hat{H}_{D} = \hat{H}_{I} + \sum_{m\in\{\uparrow,\downarrow\}}\sum_{j=1}^{L}  w_{j,m}\hat{\sigma}_{j,m}^{+} \hat{\sigma}_{j,m}^{-}$, with $w_{j,m}$ drawn from a uniform distribution $[-W,W]$, and $W$ being the strength of disorder. For each disorder strength, we consider 10 disorder realizations and plot the dynamics of averaged $C_{1,1}$ with different $W$ are plotted in Fig.~\ref{fig3}\textbf{a}. With the increasing of $W$, and as the system approaches the MBL transition, $C_{1,1}$ decays more slowly. Moreover, the oscillation in the dynamics of $C_{1,1}$ becomes more obvious with larger $W$, which is related to the presence of local integrals of motion in the deep many-body localized phase~\cite{PhysRevLett.111.127201}.

We then fit both the experimental and numerical data with the time window $t\in [50~\!\mathrm{ns},200~\!\mathrm{ns}]$ by adopting the power-law decay $C_{1,1}\propto t^{-\alpha}$. As shown in Fig.~\ref{fig3}\textbf{b}, we observe an anomalous subdiffusive region with the transport exponent $\alpha<1/2$. For the strength of disorder $W/2\pi \gtrsim 50~\!\mathrm{MHz}$, the transport exponent $\alpha\sim 10^{-2}$, indicating the freezing of spin transport and the onset of MBL on the 24-qubit system~\cite{PhysRevLett.114.160401}. Here, we emphasize that the estimated transition point between the subdiffusive regime and MBL is a lower bound since with longer evolved time, the exponent $\alpha$ obtained from the power-law fitting becomes slightly larger (see Supplementary Note 6).

\begin{figure*}[t]
	\centering
	\includegraphics[width=0.97\linewidth]{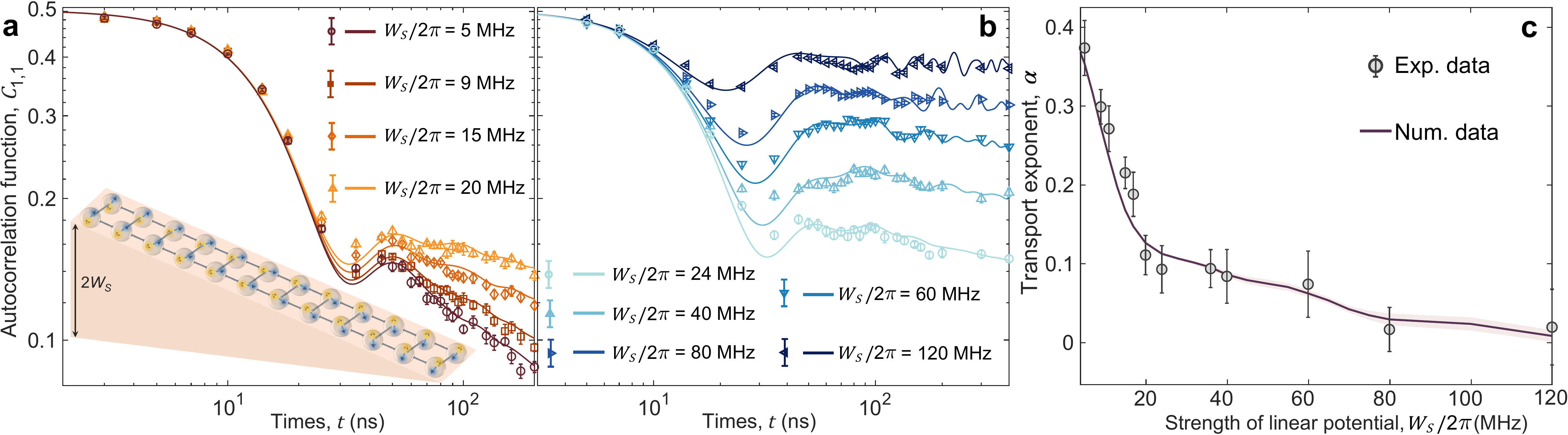}\\
	\caption{\textbf{Subdiffusive transport on the superconducting qubit ladder with linear potential.} \textbf{a}, Time evolution of autocorrelation function $C_{1,1}(t)$ for the tilted qubit ladder with $L=12$ and $W_{S}/2\pi \leq 20~\!\mathrm{MHz}$.  \textbf{b} is similar to \textbf{a} but for the data with $W_{S}/2\pi \geq 24~\!\mathrm{MHz}$. Markers (lines) are experimental (numerical) data. \textbf{c}, Transport exponent $\alpha$ as a function of $W_{S}$. For $W_{S}/2\pi \leq 20~\!\mathrm{MHz}$ and $W_{S}/2\pi \geq 24~\!\mathrm{MHz}$, the exponent $\alpha$ is extracted from fitting the data of $C_{1,1}(t)$ with the time window  $t\in [50~\!\mathrm{ns},200~\!\mathrm{ns}]$ and $t\in [100~\!\mathrm{ns},400~\!\mathrm{ns}]$, respectively. Error bars (experimental data) and shaded regions (numerical data) represent the standard deviation.}\label{fig4}
\end{figure*}

Next, we explore the transport properties on a tilted superconducting qubit ladder, which is subjected to the linear potential $\hat{H}_{L} = \sum_{j=1}^{L} \Delta j \sum_{m\in\{\uparrow,\downarrow\}} \hat{\sigma}_{j,m}^{+} \hat{\sigma}_{j,m}^{-} $, with $\Delta = 2W_{S}/(L-1)$ being the slope of the linear potential (see the tilted ladder in the inset of Fig.~\ref{fig4}\textbf{a}). Thus, the effective Hamiltonian of the tilted superconducting qubit ladder can be written as $\hat{H}_{T}=\hat{H}_{I}+\hat{H}_{L}$. Different from the aforementioned breakdown of ergodicity induced by the disorder, the non-ergodic behaviors induced by the linear potential arise from strong Hilbert-space fragmentation~\cite{PhysRevB.103.L100202,PhysRevB.101.174204,PhysRevX.10.011047}. The ergodicity breaking in the disorder-free system $\hat{H}_{T}$ is known as the Stark MBL~\cite{PhysRevLett.127.240502,Morong:2021ul,PhysRevLett.127.240502,PhysRevLett.122.040606,Nieuwenburg:2019vb,PhysRevB.104.205122,PhysRevB.102.054206}.


We employ the method based on the quantum circuit shown in Fig.~\ref{fig1}\textbf{c} to measure the time evolution of the autocorrelation function $C_{1,1}$ with different slopes of the linear potential. The results are presented in Fig.~\ref{fig4}\textbf{a}~and~\ref{fig4}\textbf{b}. Similar to the system with disorder, the dynamics of $C_{1,1}$ still satisfies $C_{1,1}\propto t^{-\alpha}$ with $\alpha<0.5$, i.e., subdiffusive transport. Figure~\ref{fig4}\textbf{c} displays the transport exponent $\alpha$ with different strength of the linear potential, showing that $\alpha$ asymptotically drops as $W_{S}$ increases. 

Two remarks are in order. First, by employing the same standard for the onset of MBL induced by disorder, i.e., $\alpha\sim 10^{-2}$, the results in Fig.~\ref{fig4}\textbf{c} indicate that the Stark MBL on the tilted 24-qubit ladder occurs when  $W_{S}/2\pi \gtrsim 80~\!\mathrm{MHz}$ ($\Delta/2\pi \gtrsim 14.6~\!\mathrm{MHz}$). Second, in the ergodic side ($W_{S}/2\pi< 80~\!\mathrm{MHz}$ and $W/2\pi<50~\!\mathrm{MHz}$ for the tilted and disordered systems respectively), the transport exponent $\alpha$ exhibits rapid decay with increasing $W_{S}$ up to $W_{S}/2\pi\simeq 20~\!\mathrm{MHz}$ in the tilted system. Subsequently, as $W_{S}$ continues to increase, the decay of $z$ becomes slower. In contrast, for the disordered system, $\alpha$ consistently decreases with increasing disordered strength $W$.  We note that the impact of the emergence of dipole-moment conservation with increasing the slope of linear potential on the spin transport, and its distinction from the transport in disordered systems remain unclear and deserve further theoretical studies.

~\\


\noindent \textbf{\large{Discussion}}

\noindent Based on the novel protocol for simulating the infinite-temperature spin transport using the Haar-random state~\cite{PhysRevLett.126.230501}, we have experimentally probed diffusive transport on a 24-qubit ladder-type programmable superconducting processor. Moreover, when the qubit ladder is subject to sufficiently strong disorder, we observe the signatures of subdiffusive transport, in accompany with the breakdown of ergodicity due to MBL. 

It is worthwhile to emphasize that previous experimental studies of the Stark MBL mainly focus on the dynamics of imbalance~\cite{Morong:2021ul,Scherg:2021tl,PhysRevLett.130.010201}. Different from the disorder-induced MBL with  a power-law decay of imbalance observed in the  subdiffusive Griffith-like region~\cite{PhysRevX.7.041047}, for the Stark MBL, there is no experimental evidence for the power-law decay of imbalance~\cite{Morong:2021ul,Scherg:2021tl,PhysRevLett.130.010201}. Here, by measuring the infinite-temperature autocorrelation function, we provide solid experimental evidence for the subdiffusion in tilted systems, which is induced by the emergence of strong Hilbert-space fragmentation~\cite{PhysRevB.103.L100202,PhysRevB.101.174204,PhysRevX.10.011047}. Theoretically, it has been suggested that for a thermodynamically large system, non-zero tilted potentials, i.e., $\Delta>0$, will lead to a subdiffusive transport with $\alpha\simeq 1/4$~\cite{PhysRevLett.125.245303,PhysRevB.109.115120}. In finite-size systems, both results as shown in Fig.~\ref{fig4} and the cold atom experiments on the tilted Fermi-Hubbard model~\cite{PhysRevX.10.011042} demonstrate a crossover from the diffusive regime to the subdiffusive one. Investigating how this crossover scales with an increasing system size is a further experimental task, which requires for quantum simulators with a larger number of qubits.


Ensembles of Haar-random pure quantum states have several promising applications, including benchmarking quantum devices~\cite{Choi:2023tt,PhysRevA.100.032328} and demonstrating the beyond-classical computation~\cite{Neill:2018wy,Boixo:2018un,Arute:2019ts,PhysRevLett.127.180501,2023arXiv230411119M}. Our work displays a practical application of the randomly distributed quantum state, i.e., probing the infinite-temperature spin transport. In contrast to employing digital random circuits, where the number of imperfect two-qubit gates is proportional to the qubit number~\cite{Boixo:2018un,Arute:2019ts,PhysRevLett.127.180501,2023arXiv230411119M,PhysRevLett.126.230501,2022arXiv220812243K}, the scalable analog circuit adopted in our experiments can also generate multi-qubit Haar-random states useful for simulating hydrodynamics. The protocol employed in our work can be naturally extended to explore the non-trivial transport properties on other analog quantum simulators, including the Rydberg atoms~\cite{Choi:2023tt,RevModPhys.82.2313,Browaeys:2020tl,Henriet2020quantumcomputing}, quantum gas microscopes~\cite{doi:10.1126/science.aaf6725,doi:10.1126/science.aal3837}, and the superconducting circuits with a central resonance bus, which enables long-range interactions~\cite{qc_all_2,PhysRevLett.128.150501,doi:10.1126/science.ade7651}. 

~\\

\noindent \textbf{\large{Methods}}
\\
\noindent \textbf{Derivation of Eq.~(\ref{measure_c})}

\noindent Here, we present the details of the deviation of Eq.~(\ref{measure_c}), which is based on the typicality~\cite{PhysRevLett.126.230501,PhysRevB.104.054415,Jin:2020ue}. 
According to Eq.~(\ref{C_rr}), $c_{\mu;\nu} = \text{Tr}[\hat{\sigma}_{\mu}^{z}(t)\hat{\sigma}_{\nu}^{z}]/D$, with $D=2^{N}$. We define $\hat{N}_{\nu} = (\hat{\sigma}_{\nu}^{z}+1)/2$, and then $c_{\mu;\nu} = \frac{1}{D}\text{Tr}[\hat{\sigma}_{\mu}^{z}(t)\hat{N}_{\nu}]$. By using $\hat{N}_{\nu} = (\hat{N}_{\nu})^{2}$, we have $c_{\mu;\nu} = \frac{1}{D}\text{Tr}[\hat{N}_{\nu}\hat{\sigma}_{\mu}^{z}(t)\hat{N}_{\nu}]$. We note that $\hat{N}_{\nu}$ is an operator which projects the state of the $\nu$-th qubit to the state $|0\rangle$.

According to the typicality~\cite{PhysRevLett.126.230501,PhysRevB.104.054415,Jin:2020ue}, the trace of an operator $\hat{O}$ can be approximated as the expectation value averaged by the pure Haar-random state $|r\rangle$, i.e., 
\begin{eqnarray}
\frac{1}{D}\text{Tr}[\hat{O}] = \langle r| \hat{O}| r \rangle + \mathcal{O}(2^{-N/2}),
\label{typicality}
\end{eqnarray}
with $N$ being the number of qubits. It indicates that the infinite-temperature expectation value $\text{Tr}[\hat{O}]/D$ can be better estimated by the expectation value for the Haar-random state $\langle r| \hat{O}| r \rangle $. Thus, $c_{\mu;\nu} \simeq  \langle r| \hat{N}_{\nu}\hat{\sigma}_{\mu}^{z}(t)\hat{N}_{\nu} | r \rangle =\langle\psi_{\nu}^{R}(t) | \hat{\sigma}^{z}_{\mu} |\psi_{\nu}^{R}(t)\rangle$ for multi-qubit systems. Based on the definition of the projector $\hat{N}_{\nu}$, $\hat{N}_{\nu}|r\rangle$ is a Haar-random state for the whole system except for the $\nu$-th qubit, and in the experiment, only a $(N-1)$-qubit Haar-random state is required.
\\
\\
\noindent \textbf{Numerical simulations}
\\
\noindent
Here, we present the details of the numerical simulations. We calculate the unitary time evolution $|\psi(t+\Delta t)\rangle = e^{-i\hat{H}\Delta t} |\psi(t)\rangle$ by employing the Krylov method~\cite{Luitz:2017uz}. The Krylov subspace is panned by the vectors defined as $\{|\psi(t)\rangle, \hat{H}|\psi(t)\rangle,  \hat{H}^{2}|\psi(t)\rangle, ..., \hat{H}^{(m-1)}|\psi(t)\rangle \}$. Then, the Hamiltonian $\hat{H}$ in the Krylov subspace becomes a $m$-dimensional matrix $\text{H}_{m} = \text{K}_{m}^{\dagger}\text{H}\text{K}_{m}$, where $\text{H}$ denotes the Hamiltonian $\hat{H}$ in the matrix form, and $\text{K}_{m}$ is the matrix whose columns contain the orthonormal basis vectors of the Krylov space. Finally, the unitary time evolution can be approximately simulated in the Krylov subspace as $|\psi(t+\Delta t)\rangle \simeq  \text{K}_{m}^{\dagger}  e^{-i\text{H}_{m} \Delta t} \text{K}_{m} |\psi(t)\rangle$. In our numerical simulations, the dimension of the Krylov subspace $m$ is adaptively adjusted from $m=6$ to $30$, making sure the numerically errors are smaller than $10^{-14}$. 

For the numerical simulation of the $\hat{U}_{R}(t_{R})=e^{-i\hat{H}_{d}t_{R}}$ in Fig.~\ref{fig1}\textbf{c}, based on the experimental data of the XY drive, the parameters in $\hat{H}_{d}$ are $\Omega_{j,m}/2\pi= 10.4\pm 1.6~\!\mathrm{MHz}$, and $\phi_{j,m}\in[-\pi/10,\pi/10]$. 
\\
\\
\noindent \textbf{Details of generating Haar-random states}
\\
\noindent
In this section, we present more details for the generation of faithful Haar-random states. The analog quantum circuit employed to generate Haar-random states is $\hat{U}_{R} = \exp[-i(\hat{H}_{I} + \hat{H}_{d})t]$, where $\hat{H}_{I}$ is given by Eq.~(\ref{H_int}) and  $\hat{H}_{d} = \sum_{m\in\{\uparrow,\downarrow\}}\sum_{j=1}^{L} \Omega_{j,m}(e^{-i\phi_{j,m}}\hat{\sigma}_{j,m}^{+}+e^{i\phi_{j,m}}\hat{\sigma}_{j,m}^{-})/2$ is the drive Hamiltonian.

\begin{figure*}
	\includegraphics[width=0.97\linewidth]{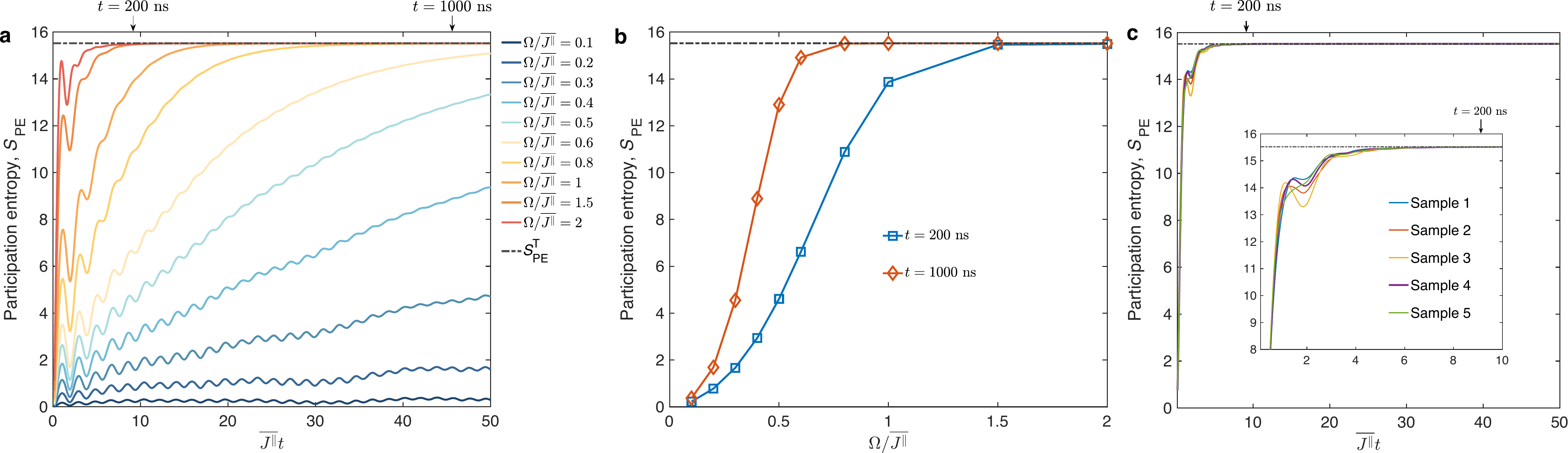}\\
	\caption{\textbf{Impact of driving amplitude and phases of microwave pulse on the generation of Haar-random states.} \textbf{a}, The time evolution of the participation entropy $S_{\text{PE}}$ for different driving amplitude $\Omega$. \textbf{b}, The value of $S_{\text{PE}}$ at two evolved times $t=200~\!\mathrm{ns}$ and $1000~\!\mathrm{ns}$, as a function of $\Omega$. \textbf{c}, The dynamics of $S_{\text{PE}}$ with the phases of the microwave pulse drawn from $[-\pi,\pi]$. Here, we present the numerical data of 5 different samples of the phases. The inset show the dynamics in a shorter time interval. The horizontal dashed line represents the participation entropy for Haar-random states $S_{\text{PE}}^{\text{T}}\simeq 15.519$.  }\label{fig_add1}
\end{figure*}

Here, we first numerically study the influence of the driving amplitude $\Omega_{j,m}$. For convenience, we consider $\phi_{j,m} = 0$ and isotropic driving amplitude, i.e., $\Omega = \Omega_{j,m}$ for all $(j,m)$. We chose $Q_{R} = \{Q_{1,\uparrow},Q_{2,\uparrow},...,Q_{12,\uparrow},Q_{2,\downarrow},Q_{3,\downarrow},...,Q_{12,\downarrow}\}$ with total 23 qubits. The dynamics of participation entropy $S_{\text{PE}}$ for different values of $\Omega$ are plotted in Fig.~\ref{fig_add1}\textbf{a}, and the values of  $S_{\text{PE}}$ with the evolved time $t=200~\!\mathrm{ns}$ and $1000~\!\mathrm{ns}$ are displayed in  Fig.~\ref{fig_add1}\textbf{b}. It is seen that for small $\Omega$, the growth of $S_{\text{PE}}$ is slow and with increasing $\Omega$, it becomes more rapid. In this experiment, we chose $\overline{\Omega}/\overline{J^{\parallel}}\simeq 1.4$ because the participation entropy can achieve $S_{\text{PE}}^{\text{T}}$ with a relatively short evolved time $t\simeq 200~\!\mathrm{ns}$. As $\Omega$ further increases, the time when $S_{\text{PE}}^{\text{T}}$ is reached does not significantly become shorter. Based on above discussions, $\overline{\Omega}/\overline{J^{\parallel}}\simeq 1.4$ is an appropriate choice of the driving amplitude.

Next, we numerically study the influence of the randomness for the phases of driving microwave pulse $\phi_{j,m}$. In this experiment, by using the correction of crosstalk, the randomness of the phases is small, i.e., $\phi_{j,m}\in[-\pi/10,\pi/10]$. Here, we consider the phases with large randomness, i.e., $\phi_{j,m}\in[-\pi,\pi]$. The numerical results for the time evolution of $S_{\text{PE}}$ with 5 samples of $\phi_{j,m}$ are plotted in Fig.~\ref{fig_add1}\textbf{c}. With $\phi_{j,m}\in[-\pi,\pi]$, the participation entropy can still tend to $S_{\text{PE}}^{\text{T}}$ around $200~\!\mathrm{ns}$. Only the short time behaviors are slightly different from each other for the 5 samples (see the inset of Fig.~\ref{fig_add1}\textbf{c}).




\begin{thebibliography}{72}%
	\makeatletter
	\providecommand \@ifxundefined [1]{%
		\@ifx{#1\undefined}
	}%
	\providecommand \@ifnum [1]{%
		\ifnum #1\expandafter \@firstoftwo
		\else \expandafter \@secondoftwo
		\fi
	}%
	\providecommand \@ifx [1]{%
		\ifx #1\expandafter \@firstoftwo
		\else \expandafter \@secondoftwo
		\fi
	}%
	\providecommand \natexlab [1]{#1}%
	\providecommand \enquote  [1]{``#1''}%
	\providecommand \bibnamefont  [1]{#1}%
	\providecommand \bibfnamefont [1]{#1}%
	\providecommand \citenamefont [1]{#1}%
	\providecommand \href@noop [0]{\@secondoftwo}%
	\providecommand \href [0]{\begingroup \@sanitize@url \@href}%
	\providecommand \@href[1]{\@@startlink{#1}\@@href}%
	\providecommand \@@href[1]{\endgroup#1\@@endlink}%
	\providecommand \@sanitize@url [0]{\catcode `\\12\catcode `\$12\catcode
		`\&12\catcode `\#12\catcode `\^12\catcode `\_12\catcode `\%12\relax}%
	\providecommand \@@startlink[1]{}%
	\providecommand \@@endlink[0]{}%
	\providecommand \url  [0]{\begingroup\@sanitize@url \@url }%
	\providecommand \@url [1]{\endgroup\@href {#1}{\urlprefix }}%
	\providecommand \urlprefix  [0]{URL }%
	\providecommand \Eprint [0]{\href }%
	\providecommand \doibase [0]{http://dx.doi.org/}%
	\providecommand \selectlanguage [0]{\@gobble}%
	\providecommand \bibinfo  [0]{\@secondoftwo}%
	\providecommand \bibfield  [0]{\@secondoftwo}%
	\providecommand \translation [1]{[#1]}%
	\providecommand \BibitemOpen [0]{}%
	\providecommand \bibitemStop [0]{}%
	\providecommand \bibitemNoStop [0]{.\EOS\space}%
	\providecommand \EOS [0]{\spacefactor3000\relax}%
	\providecommand \BibitemShut  [1]{\csname bibitem#1\endcsname}%
	\let\auto@bib@innerbib\@empty
	\bibitem [{\citenamefont {Nandkishore}\ and\ \citenamefont
		{Huse}(2015)}]{Nandkishore:2015vg}%
	\BibitemOpen
	\bibfield  {author} {\bibinfo {author} {\bibfnamefont {R.}~\bibnamefont
			{Nandkishore}}\ and\ \bibinfo {author} {\bibfnamefont {D.~A.}\ \bibnamefont
			{Huse}},\ }\bibfield  {title} {\enquote {\bibinfo {title} {{Many-Body
					Localization and Thermalization in Quantum Statistical Mechanics}},}\ }\href
	{\doibase 10.1146/annurev-conmatphys-031214-014726} {\bibfield  {journal}
		{\bibinfo  {journal} {Annual Review of Condensed Matter Physics}\ }\textbf
		{\bibinfo {volume} {6}},\ \bibinfo {pages} {15--38} (\bibinfo {year}
		{2015})}\BibitemShut {NoStop}%
	\bibitem [{\citenamefont {Agarwal}\ \emph {et~al.}(2015)\citenamefont
		{Agarwal}, \citenamefont {Gopalakrishnan}, \citenamefont {Knap},
		\citenamefont {M\"uller},\ and\ \citenamefont
		{Demler}}]{PhysRevLett.114.160401}%
	\BibitemOpen
	\bibfield  {author} {\bibinfo {author} {\bibfnamefont {K.}~\bibnamefont
			{Agarwal}}, \bibinfo {author} {\bibfnamefont {S.}~\bibnamefont
			{Gopalakrishnan}}, \bibinfo {author} {\bibfnamefont {M.}~\bibnamefont
			{Knap}}, \bibinfo {author} {\bibfnamefont {M.}~\bibnamefont {M\"uller}}, \
		and\ \bibinfo {author} {\bibfnamefont {E.}~\bibnamefont {Demler}},\
	}\bibfield  {title} {\enquote {\bibinfo {title} {{Anomalous Diffusion and
					Griffiths Effects Near the Many-Body Localization Transition}},}\ }\href
	{\doibase 10.1103/PhysRevLett.114.160401} {\bibfield  {journal} {\bibinfo
			{journal} {Phys. Rev. Lett.}\ }\textbf {\bibinfo {volume} {114}},\ \bibinfo
		{pages} {160401} (\bibinfo {year} {2015})}\BibitemShut {NoStop}%
	\bibitem [{\citenamefont {\ifmmode \check{Z}\else
			\v{Z}\fi{}nidari\ifmmode~\check{c}\else \v{c}\fi{}}\ \emph
		{et~al.}(2016)\citenamefont {\ifmmode \check{Z}\else
			\v{Z}\fi{}nidari\ifmmode~\check{c}\else \v{c}\fi{}}, \citenamefont
		{Scardicchio},\ and\ \citenamefont {Varma}}]{PhysRevLett.117.040601}%
	\BibitemOpen
	\bibfield  {author} {\bibinfo {author} {\bibfnamefont {M.}~\bibnamefont
			{\ifmmode \check{Z}\else \v{Z}\fi{}nidari\ifmmode~\check{c}\else
				\v{c}\fi{}}}, \bibinfo {author} {\bibfnamefont {A.}~\bibnamefont
			{Scardicchio}}, \ and\ \bibinfo {author} {\bibfnamefont {V.~K.}\ \bibnamefont
			{Varma}},\ }\bibfield  {title} {\enquote {\bibinfo {title} {{Diffusive and
					Subdiffusive Spin Transport in the Ergodic Phase of a Many-Body Localizable
					System}},}\ }\href {\doibase 10.1103/PhysRevLett.117.040601} {\bibfield
		{journal} {\bibinfo  {journal} {Phys. Rev. Lett.}\ }\textbf {\bibinfo
			{volume} {117}},\ \bibinfo {pages} {040601} (\bibinfo {year}
		{2016})}\BibitemShut {NoStop}%
	\bibitem [{\citenamefont {Ljubotina}\ \emph {et~al.}(2023)\citenamefont
		{Ljubotina}, \citenamefont {Desaules}, \citenamefont {Serbyn},\ and\
		\citenamefont {Papi\ifmmode~\acute{c}\else \'{c}\fi{}}}]{PhysRevX.13.011033}%
	\BibitemOpen
	\bibfield  {author} {\bibinfo {author} {\bibfnamefont {M.}~\bibnamefont
			{Ljubotina}}, \bibinfo {author} {\bibfnamefont {J.-Y.}\ \bibnamefont
			{Desaules}}, \bibinfo {author} {\bibfnamefont {M.}~\bibnamefont {Serbyn}}, \
		and\ \bibinfo {author} {\bibfnamefont {Z.}~\bibnamefont
			{Papi\ifmmode~\acute{c}\else \'{c}\fi{}}},\ }\bibfield  {title} {\enquote
		{\bibinfo {title} {{Superdiffusive Energy Transport in Kinetically
					Constrained Models}},}\ }\href {\doibase 10.1103/PhysRevX.13.011033}
	{\bibfield  {journal} {\bibinfo  {journal} {Phys. Rev. X}\ }\textbf {\bibinfo
			{volume} {13}},\ \bibinfo {pages} {011033} (\bibinfo {year}
		{2023})}\BibitemShut {NoStop}%
	\bibitem [{\citenamefont {Scheie}\ \emph {et~al.}(2021)\citenamefont {Scheie},
		\citenamefont {Sherman}, \citenamefont {Dupont}, \citenamefont {Nagler},
		\citenamefont {Stone}, \citenamefont {Granroth}, \citenamefont {Moore},\ and\
		\citenamefont {Tennant}}]{Scheie:2021vl}%
	\BibitemOpen
	\bibfield  {author} {\bibinfo {author} {\bibfnamefont {A.}~\bibnamefont
			{Scheie}}, \bibinfo {author} {\bibfnamefont {N.~E.}\ \bibnamefont {Sherman}},
		\bibinfo {author} {\bibfnamefont {M.}~\bibnamefont {Dupont}}, \bibinfo
		{author} {\bibfnamefont {S.~E.}\ \bibnamefont {Nagler}}, \bibinfo {author}
		{\bibfnamefont {M.~B.}\ \bibnamefont {Stone}}, \bibinfo {author}
		{\bibfnamefont {G.~E.}\ \bibnamefont {Granroth}}, \bibinfo {author}
		{\bibfnamefont {J.~E.}\ \bibnamefont {Moore}}, \ and\ \bibinfo {author}
		{\bibfnamefont {D.~A.}\ \bibnamefont {Tennant}},\ }\bibfield  {title}
	{\enquote {\bibinfo {title} {{Detection of Kardar--Parisi--Zhang
					hydrodynamics in a quantum Heisenberg spin-1/2 chain}},}\ }\href {\doibase
		10.1038/s41567-021-01191-6} {\bibfield  {journal} {\bibinfo  {journal}
			{Nature Physics}\ }\textbf {\bibinfo {volume} {17}},\ \bibinfo {pages}
		{726--730} (\bibinfo {year} {2021})}\BibitemShut {NoStop}%
	\bibitem [{\citenamefont {\ifmmode \check{Z}\else
			\v{Z}\fi{}nidari\ifmmode~\check{c}\else
			\v{c}\fi{}}(2011)}]{PhysRevLett.106.220601}%
	\BibitemOpen
	\bibfield  {author} {\bibinfo {author} {\bibfnamefont {M.}~\bibnamefont
			{\ifmmode \check{Z}\else \v{Z}\fi{}nidari\ifmmode~\check{c}\else
				\v{c}\fi{}}},\ }\bibfield  {title} {\enquote {\bibinfo {title} {{Spin
					Transport in a One-Dimensional Anisotropic Heisenberg Model}},}\ }\href
	{\doibase 10.1103/PhysRevLett.106.220601} {\bibfield  {journal} {\bibinfo
			{journal} {Phys. Rev. Lett.}\ }\textbf {\bibinfo {volume} {106}},\ \bibinfo
		{pages} {220601} (\bibinfo {year} {2011})}\BibitemShut {NoStop}%
	\bibitem [{\citenamefont {Dupont}\ \emph {et~al.}(2021)\citenamefont {Dupont},
		\citenamefont {Sherman},\ and\ \citenamefont
		{Moore}}]{PhysRevLett.127.107201}%
	\BibitemOpen
	\bibfield  {author} {\bibinfo {author} {\bibfnamefont {M.}~\bibnamefont
			{Dupont}}, \bibinfo {author} {\bibfnamefont {N.~E.}\ \bibnamefont {Sherman}},
		\ and\ \bibinfo {author} {\bibfnamefont {J.~E.}\ \bibnamefont {Moore}},\
	}\bibfield  {title} {\enquote {\bibinfo {title} {{Spatiotemporal Crossover
					between Low- and High-Temperature Dynamical Regimes in the Quantum Heisenberg
					Magnet}},}\ }\href {\doibase 10.1103/PhysRevLett.127.107201} {\bibfield
		{journal} {\bibinfo  {journal} {Phys. Rev. Lett.}\ }\textbf {\bibinfo
			{volume} {127}},\ \bibinfo {pages} {107201} (\bibinfo {year}
		{2021})}\BibitemShut {NoStop}%
	\bibitem [{\citenamefont {Bertini}\ \emph {et~al.}(2021)\citenamefont
		{Bertini}, \citenamefont {Heidrich-Meisner}, \citenamefont {Karrasch},
		\citenamefont {Prosen}, \citenamefont {Steinigeweg},\ and\ \citenamefont
		{\ifmmode \check{Z}\else \v{Z}\fi{}nidari\ifmmode~\check{c}\else
			\v{c}\fi{}}}]{RevModPhys.93.025003}%
	\BibitemOpen
	\bibfield  {author} {\bibinfo {author} {\bibfnamefont {B.}~\bibnamefont
			{Bertini}}, \bibinfo {author} {\bibfnamefont {F.}~\bibnamefont
			{Heidrich-Meisner}}, \bibinfo {author} {\bibfnamefont {C.}~\bibnamefont
			{Karrasch}}, \bibinfo {author} {\bibfnamefont {T.}~\bibnamefont {Prosen}},
		\bibinfo {author} {\bibfnamefont {R.}~\bibnamefont {Steinigeweg}}, \ and\
		\bibinfo {author} {\bibfnamefont {M.}~\bibnamefont {\ifmmode \check{Z}\else
				\v{Z}\fi{}nidari\ifmmode~\check{c}\else \v{c}\fi{}}},\ }\bibfield  {title}
	{\enquote {\bibinfo {title} {{Finite-temperature transport in one-dimensional
					quantum lattice models}},}\ }\href {\doibase 10.1103/RevModPhys.93.025003}
	{\bibfield  {journal} {\bibinfo  {journal} {Rev. Mod. Phys.}\ }\textbf
		{\bibinfo {volume} {93}},\ \bibinfo {pages} {025003} (\bibinfo {year}
		{2021})}\BibitemShut {NoStop}%
	\bibitem [{\citenamefont {Eisert}\ \emph {et~al.}(2015)\citenamefont {Eisert},
		\citenamefont {Friesdorf},\ and\ \citenamefont {Gogolin}}]{Eisert:2015ws}%
	\BibitemOpen
	\bibfield  {author} {\bibinfo {author} {\bibfnamefont {J.}~\bibnamefont
			{Eisert}}, \bibinfo {author} {\bibfnamefont {M.}~\bibnamefont {Friesdorf}}, \
		and\ \bibinfo {author} {\bibfnamefont {C.}~\bibnamefont {Gogolin}},\
	}\bibfield  {title} {\enquote {\bibinfo {title} {{Quantum many-body systems
					out of equilibrium}},}\ }\href {\doibase 10.1038/nphys3215} {\bibfield
		{journal} {\bibinfo  {journal} {Nature Physics}\ }\textbf {\bibinfo {volume}
			{11}},\ \bibinfo {pages} {124--130} (\bibinfo {year} {2015})}\BibitemShut
	{NoStop}%
	\bibitem [{\citenamefont {Peng}\ \emph {et~al.}(2023)\citenamefont {Peng},
		\citenamefont {Ye}, \citenamefont {Yao},\ and\ \citenamefont
		{Cappellaro}}]{Peng:2023vn}%
	\BibitemOpen
	\bibfield  {author} {\bibinfo {author} {\bibfnamefont {P.}~\bibnamefont
			{Peng}}, \bibinfo {author} {\bibfnamefont {B.}~\bibnamefont {Ye}}, \bibinfo
		{author} {\bibfnamefont {N.~Y.}\ \bibnamefont {Yao}}, \ and\ \bibinfo
		{author} {\bibfnamefont {P.}~\bibnamefont {Cappellaro}},\ }\bibfield  {title}
	{\enquote {\bibinfo {title} {{Exploiting disorder to probe spin and energy
					hydrodynamics}},}\ }\href {https://doi.org/10.1038/s41567-023-02024-4}
	{\bibfield  {journal} {\bibinfo  {journal} {Nature Physics}\ } (\bibinfo
		{year} {2023})}\BibitemShut {NoStop}%
	\bibitem [{\citenamefont {Steinigeweg}\ \emph {et~al.}(2014)\citenamefont
		{Steinigeweg}, \citenamefont {Heidrich-Meisner}, \citenamefont {Gemmer},
		\citenamefont {Michielsen},\ and\ \citenamefont
		{De~Raedt}}]{PhysRevB.90.094417}%
	\BibitemOpen
	\bibfield  {author} {\bibinfo {author} {\bibfnamefont {R.}~\bibnamefont
			{Steinigeweg}}, \bibinfo {author} {\bibfnamefont {F.}~\bibnamefont
			{Heidrich-Meisner}}, \bibinfo {author} {\bibfnamefont {J.}~\bibnamefont
			{Gemmer}}, \bibinfo {author} {\bibfnamefont {K.}~\bibnamefont {Michielsen}},
		\ and\ \bibinfo {author} {\bibfnamefont {H.}~\bibnamefont {De~Raedt}},\
	}\bibfield  {title} {\enquote {\bibinfo {title} {{Scaling of diffusion
					constants in the spin-$\frac{1}{2}$ XX ladder}},}\ }\href {\doibase
		10.1103/PhysRevB.90.094417} {\bibfield  {journal} {\bibinfo  {journal} {Phys.
				Rev. B}\ }\textbf {\bibinfo {volume} {90}},\ \bibinfo {pages} {094417}
		(\bibinfo {year} {2014})}\BibitemShut {NoStop}%
	\bibitem [{\citenamefont {Schubert}\ \emph {et~al.}(2021)\citenamefont
		{Schubert}, \citenamefont {Richter}, \citenamefont {Jin}, \citenamefont
		{Michielsen}, \citenamefont {De~Raedt},\ and\ \citenamefont
		{Steinigeweg}}]{PhysRevB.104.054415}%
	\BibitemOpen
	\bibfield  {author} {\bibinfo {author} {\bibfnamefont {D.}~\bibnamefont
			{Schubert}}, \bibinfo {author} {\bibfnamefont {J.}~\bibnamefont {Richter}},
		\bibinfo {author} {\bibfnamefont {F.}~\bibnamefont {Jin}}, \bibinfo {author}
		{\bibfnamefont {K.}~\bibnamefont {Michielsen}}, \bibinfo {author}
		{\bibfnamefont {H.}~\bibnamefont {De~Raedt}}, \ and\ \bibinfo {author}
		{\bibfnamefont {R.}~\bibnamefont {Steinigeweg}},\ }\bibfield  {title}
	{\enquote {\bibinfo {title} {Quantum versus classical dynamics in spin
				models: Chains, ladders, and square lattices},}\ }\href {\doibase
		10.1103/PhysRevB.104.054415} {\bibfield  {journal} {\bibinfo  {journal}
			{Phys. Rev. B}\ }\textbf {\bibinfo {volume} {104}},\ \bibinfo {pages}
		{054415} (\bibinfo {year} {2021})}\BibitemShut {NoStop}%
	\bibitem [{\citenamefont {Ljubotina}\ \emph {et~al.}(2017)\citenamefont
		{Ljubotina}, \citenamefont {{\v{Z}}nidari{\v{c}}},\ and\ \citenamefont
		{Prosen}}]{Ljubotina:2017ub}%
	\BibitemOpen
	\bibfield  {author} {\bibinfo {author} {\bibfnamefont {Marko}\ \bibnamefont
			{Ljubotina}}, \bibinfo {author} {\bibfnamefont {Marko}\ \bibnamefont
			{{\v{Z}}nidari{\v{c}}}}, \ and\ \bibinfo {author} {\bibfnamefont
			{Toma{\v{z}}}\ \bibnamefont {Prosen}},\ }\bibfield  {title} {\enquote
		{\bibinfo {title} {{Spin diffusion from an inhomogeneous quench in an
					integrable system}},}\ }\href {\doibase 10.1038/ncomms16117} {\bibfield
		{journal} {\bibinfo  {journal} {Nature Communications}\ }\textbf {\bibinfo
			{volume} {8}},\ \bibinfo {pages} {16117} (\bibinfo {year}
		{2017})}\BibitemShut {NoStop}%
	\bibitem [{\citenamefont {Wei}\ \emph {et~al.}(2022)\citenamefont {Wei},
		\citenamefont {Rubio-Abadal}, \citenamefont {Ye}, \citenamefont {Machado},
		\citenamefont {Kemp}, \citenamefont {Srakaew}, \citenamefont {Hollerith},
		\citenamefont {Rui}, \citenamefont {Gopalakrishnan}, \citenamefont {Yao},
		\citenamefont {Bloch},\ and\ \citenamefont {Zeiher}}]{Wei:2022vj}%
	\BibitemOpen
	\bibfield  {author} {\bibinfo {author} {\bibfnamefont {David}\ \bibnamefont
			{Wei}}, \bibinfo {author} {\bibfnamefont {Antonio}\ \bibnamefont
			{Rubio-Abadal}}, \bibinfo {author} {\bibfnamefont {Bingtian}\ \bibnamefont
			{Ye}}, \bibinfo {author} {\bibfnamefont {Francisco}\ \bibnamefont {Machado}},
		\bibinfo {author} {\bibfnamefont {Jack}\ \bibnamefont {Kemp}}, \bibinfo
		{author} {\bibfnamefont {Kritsana}\ \bibnamefont {Srakaew}}, \bibinfo
		{author} {\bibfnamefont {Simon}\ \bibnamefont {Hollerith}}, \bibinfo {author}
		{\bibfnamefont {Jun}\ \bibnamefont {Rui}}, \bibinfo {author} {\bibfnamefont
			{Sarang}\ \bibnamefont {Gopalakrishnan}}, \bibinfo {author} {\bibfnamefont
			{Norman~Y.}\ \bibnamefont {Yao}}, \bibinfo {author} {\bibfnamefont
			{Immanuel}\ \bibnamefont {Bloch}}, \ and\ \bibinfo {author} {\bibfnamefont
			{Johannes}\ \bibnamefont {Zeiher}},\ }\bibfield  {title} {\enquote {\bibinfo
			{title} {{Quantum gas microscopy of Kardar-Parisi-Zhang superdiffusion}},}\
	}\href {\doibase 10.1126/science.abk2397} {\bibfield  {journal} {\bibinfo
			{journal} {Science}\ }\textbf {\bibinfo {volume} {376}},\ \bibinfo {pages}
		{716--720} (\bibinfo {year} {2022})}\BibitemShut {NoStop}%
	\bibitem [{\citenamefont {Joshi}\ \emph {et~al.}(2022)\citenamefont {Joshi},
		\citenamefont {Kranzl}, \citenamefont {Schuckert}, \citenamefont {Lovas},
		\citenamefont {Maier}, \citenamefont {Blatt}, \citenamefont {Knap},\ and\
		\citenamefont {Roos}}]{Joshi:2022tr}%
	\BibitemOpen
	\bibfield  {author} {\bibinfo {author} {\bibfnamefont {M.~K.}\ \bibnamefont
			{Joshi}}, \bibinfo {author} {\bibfnamefont {F.}~\bibnamefont {Kranzl}},
		\bibinfo {author} {\bibfnamefont {A.}~\bibnamefont {Schuckert}}, \bibinfo
		{author} {\bibfnamefont {I.}~\bibnamefont {Lovas}}, \bibinfo {author}
		{\bibfnamefont {C.}~\bibnamefont {Maier}}, \bibinfo {author} {\bibfnamefont
			{R.}~\bibnamefont {Blatt}}, \bibinfo {author} {\bibfnamefont
			{M.}~\bibnamefont {Knap}}, \ and\ \bibinfo {author} {\bibfnamefont {C.~F.}\
			\bibnamefont {Roos}},\ }\bibfield  {title} {\enquote {\bibinfo {title}
			{{Observing emergent hydrodynamics in a long-range quantum magnet}},}\ }\href
	{\doibase 10.1126/science.abk2400} {\bibfield  {journal} {\bibinfo  {journal}
			{Science}\ }\textbf {\bibinfo {volume} {376}},\ \bibinfo {pages} {720--724}
		(\bibinfo {year} {2022})}\BibitemShut {NoStop}%
	\bibitem [{\citenamefont {Rosenberg}\ \emph {et~al.}(2024)\citenamefont
		{Rosenberg}, \citenamefont {Andersen}, \citenamefont {Samajdar},
		\citenamefont {Petukhov}, \citenamefont {Hoke}, \citenamefont {Abanin},
		\citenamefont {Bengtsson}, \citenamefont {Drozdov}, \citenamefont {Erickson},
		\citenamefont {Klimov}, \citenamefont {Mi}, \citenamefont {Morvan},
		\citenamefont {Neeley}, \citenamefont {Neill}, \citenamefont {Acharya},
		\citenamefont {Allen}, \citenamefont {Anderson}, \citenamefont {Ansmann},
		\citenamefont {Arute}, \citenamefont {Arya}, \citenamefont {Asfaw},
		\citenamefont {Atalaya}, \citenamefont {Bardin}, \citenamefont {Bilmes},
		\citenamefont {Bortoli}, \citenamefont {Bourassa}, \citenamefont {Bovaird},
		\citenamefont {Brill}, \citenamefont {Broughton}, \citenamefont {Buckley},
		\citenamefont {Buell}, \citenamefont {Burger}, \citenamefont {Burkett},
		\citenamefont {Bushnell}, \citenamefont {Campero}, \citenamefont {Chang},
		\citenamefont {Chen}, \citenamefont {Chiaro}, \citenamefont {Chik},
		\citenamefont {Cogan}, \citenamefont {Collins}, \citenamefont {Conner},
		\citenamefont {Courtney}, \citenamefont {Crook}, \citenamefont {Curtin},
		\citenamefont {Debroy}, \citenamefont {Barba}, \citenamefont {Demura},
		\citenamefont {{Di Paolo}}, \citenamefont {Dunsworth}, \citenamefont {Earle},
		\citenamefont {Faoro}, \citenamefont {Farhi}, \citenamefont {Fatemi},
		\citenamefont {Ferreira}, \citenamefont {Burgos}, \citenamefont {Forati},
		\citenamefont {Fowler}, \citenamefont {Foxen}, \citenamefont {Garcia},
		\citenamefont {Genois}, \citenamefont {Giang}, \citenamefont {Gidney},
		\citenamefont {Gilboa}, \citenamefont {Giustina}, \citenamefont {Gosula},
		\citenamefont {Dau}, \citenamefont {Gross}, \citenamefont {Habegger},
		\citenamefont {Hamilton}, \citenamefont {Hansen}, \citenamefont {Harrigan},
		\citenamefont {Harrington}, \citenamefont {Heu}, \citenamefont {Hill},
		\citenamefont {Hoffmann}, \citenamefont {Hong}, \citenamefont {Huang},
		\citenamefont {Huff}, \citenamefont {Huggins}, \citenamefont {Ioffe},
		\citenamefont {Isakov}, \citenamefont {Iveland}, \citenamefont {Jeffrey},
		\citenamefont {Jiang}, \citenamefont {Jones}, \citenamefont {Juhas},
		\citenamefont {Kafri}, \citenamefont {Khattar}, \citenamefont {Khezri},
		\citenamefont {Kieferov{\'{a}}}, \citenamefont {Kim}, \citenamefont {Kitaev},
		\citenamefont {Klots}, \citenamefont {Korotkov}, \citenamefont {Kostritsa},
		\citenamefont {Kreikebaum}, \citenamefont {Landhuis}, \citenamefont {Laptev},
		\citenamefont {Lau}, \citenamefont {Laws}, \citenamefont {Lee}, \citenamefont
		{Lee}, \citenamefont {Lensky}, \citenamefont {Lester}, \citenamefont {Lill},
		\citenamefont {Liu}, \citenamefont {Locharla}, \citenamefont {Mandr{\`{a}}},
		\citenamefont {Martin}, \citenamefont {Martin}, \citenamefont {McClean},
		\citenamefont {McEwen}, \citenamefont {Meeks}, \citenamefont {Miao},
		\citenamefont {Mieszala}, \citenamefont {Montazeri}, \citenamefont
		{Movassagh}, \citenamefont {Mruczkiewicz}, \citenamefont {Nersisyan},
		\citenamefont {Newman}, \citenamefont {Ng}, \citenamefont {Nguyen},
		\citenamefont {Nguyen}, \citenamefont {Niu}, \citenamefont {O'Brien},
		\citenamefont {Omonije}, \citenamefont {Opremcak}, \citenamefont {Potter},
		\citenamefont {Pryadko}, \citenamefont {Quintana}, \citenamefont {Rhodes},
		\citenamefont {Rocque}, \citenamefont {Rubin}, \citenamefont {Saei},
		\citenamefont {Sank}, \citenamefont {Sankaragomathi}, \citenamefont
		{Satzinger}, \citenamefont {Schurkus}, \citenamefont {Schuster},
		\citenamefont {Shearn}, \citenamefont {Shorter}, \citenamefont {Shutty},
		\citenamefont {Shvarts}, \citenamefont {Sivak}, \citenamefont {Skruzny},
		\citenamefont {Smith}, \citenamefont {Somma}, \citenamefont {Sterling},
		\citenamefont {Strain}, \citenamefont {Szalay}, \citenamefont {Thor},
		\citenamefont {Torres}, \citenamefont {Vidal}, \citenamefont {Villalonga},
		\citenamefont {Heidweiller}, \citenamefont {White}, \citenamefont {Woo},
		\citenamefont {Xing}, \citenamefont {Yao}, \citenamefont {Yeh}, \citenamefont
		{Yoo}, \citenamefont {Young}, \citenamefont {Zalcman}, \citenamefont {Zhang},
		\citenamefont {Zhu}, \citenamefont {Zobrist}, \citenamefont {Neven},
		\citenamefont {Babbush}, \citenamefont {Bacon}, \citenamefont {Boixo},
		\citenamefont {Hilton}, \citenamefont {Lucero}, \citenamefont {Megrant},
		\citenamefont {Kelly}, \citenamefont {Chen}, \citenamefont {Smelyanskiy},
		\citenamefont {Khemani}, \citenamefont {Gopalakrishnan}, \citenamefont
		{Prosen},\ and\ \citenamefont {Roushan}}]{2023arXiv230609333R}%
	\BibitemOpen
	\bibfield  {author} {\bibinfo {author} {\bibfnamefont {E.}~\bibnamefont
			{Rosenberg}}, \bibinfo {author} {\bibfnamefont {T.~I.}\ \bibnamefont
			{Andersen}}, \bibinfo {author} {\bibfnamefont {R.}~\bibnamefont {Samajdar}},
		\bibinfo {author} {\bibfnamefont {A.}~\bibnamefont {Petukhov}}, \bibinfo
		{author} {\bibfnamefont {J.~C.}\ \bibnamefont {Hoke}}, \bibinfo {author}
		{\bibfnamefont {D.}~\bibnamefont {Abanin}}, \bibinfo {author} {\bibfnamefont
			{A.}~\bibnamefont {Bengtsson}}, \bibinfo {author} {\bibfnamefont {I.~K.}\
			\bibnamefont {Drozdov}}, \bibinfo {author} {\bibfnamefont {C.}~\bibnamefont
			{Erickson}}, \bibinfo {author} {\bibfnamefont {P.~V.}\ \bibnamefont
			{Klimov}}, \bibinfo {author} {\bibfnamefont {X.}~\bibnamefont {Mi}}, \bibinfo
		{author} {\bibfnamefont {A.}~\bibnamefont {Morvan}}, \bibinfo {author}
		{\bibfnamefont {M.}~\bibnamefont {Neeley}}, \bibinfo {author} {\bibfnamefont
			{C.}~\bibnamefont {Neill}}, \bibinfo {author} {\bibfnamefont
			{R.}~\bibnamefont {Acharya}}, \bibinfo {author} {\bibfnamefont
			{R.}~\bibnamefont {Allen}}, \bibinfo {author} {\bibfnamefont
			{K.}~\bibnamefont {Anderson}}, \bibinfo {author} {\bibfnamefont
			{M.}~\bibnamefont {Ansmann}}, \bibinfo {author} {\bibfnamefont
			{F.}~\bibnamefont {Arute}}, \bibinfo {author} {\bibfnamefont
			{K.}~\bibnamefont {Arya}}, \bibinfo {author} {\bibfnamefont {A.}~\bibnamefont
			{Asfaw}}, \bibinfo {author} {\bibfnamefont {J.}~\bibnamefont {Atalaya}},
		\bibinfo {author} {\bibfnamefont {J.~C.}\ \bibnamefont {Bardin}}, \bibinfo
		{author} {\bibfnamefont {A.}~\bibnamefont {Bilmes}}, \bibinfo {author}
		{\bibfnamefont {G.}~\bibnamefont {Bortoli}}, \bibinfo {author} {\bibfnamefont
			{A.}~\bibnamefont {Bourassa}}, \bibinfo {author} {\bibfnamefont
			{J.}~\bibnamefont {Bovaird}}, \bibinfo {author} {\bibfnamefont
			{L.}~\bibnamefont {Brill}}, \bibinfo {author} {\bibfnamefont
			{M.}~\bibnamefont {Broughton}}, \bibinfo {author} {\bibfnamefont {B.~B.}\
			\bibnamefont {Buckley}}, \bibinfo {author} {\bibfnamefont {D.~A.}\
			\bibnamefont {Buell}}, \bibinfo {author} {\bibfnamefont {T.}~\bibnamefont
			{Burger}}, \bibinfo {author} {\bibfnamefont {B.}~\bibnamefont {Burkett}},
		\bibinfo {author} {\bibfnamefont {N.}~\bibnamefont {Bushnell}}, \bibinfo
		{author} {\bibfnamefont {J.}~\bibnamefont {Campero}}, \bibinfo {author}
		{\bibfnamefont {H.-S.}\ \bibnamefont {Chang}}, \bibinfo {author}
		{\bibfnamefont {Z.}~\bibnamefont {Chen}}, \bibinfo {author} {\bibfnamefont
			{B.}~\bibnamefont {Chiaro}}, \bibinfo {author} {\bibfnamefont
			{D.}~\bibnamefont {Chik}}, \bibinfo {author} {\bibfnamefont {J.}~\bibnamefont
			{Cogan}}, \bibinfo {author} {\bibfnamefont {R.}~\bibnamefont {Collins}},
		\bibinfo {author} {\bibfnamefont {P.}~\bibnamefont {Conner}}, \bibinfo
		{author} {\bibfnamefont {W.}~\bibnamefont {Courtney}}, \bibinfo {author}
		{\bibfnamefont {A.~L.}\ \bibnamefont {Crook}}, \bibinfo {author}
		{\bibfnamefont {B.}~\bibnamefont {Curtin}}, \bibinfo {author} {\bibfnamefont
			{D.~M.}\ \bibnamefont {Debroy}}, \bibinfo {author} {\bibfnamefont
			{A.~Del~Toro}\ \bibnamefont {Barba}}, \bibinfo {author} {\bibfnamefont
			{S.}~\bibnamefont {Demura}}, \bibinfo {author} {\bibfnamefont
			{A.}~\bibnamefont {{Di Paolo}}}, \bibinfo {author} {\bibfnamefont
			{A.}~\bibnamefont {Dunsworth}}, \bibinfo {author} {\bibfnamefont
			{C.}~\bibnamefont {Earle}}, \bibinfo {author} {\bibfnamefont
			{L.}~\bibnamefont {Faoro}}, \bibinfo {author} {\bibfnamefont
			{E.}~\bibnamefont {Farhi}}, \bibinfo {author} {\bibfnamefont
			{R.}~\bibnamefont {Fatemi}}, \bibinfo {author} {\bibfnamefont {V.~S.}\
			\bibnamefont {Ferreira}}, \bibinfo {author} {\bibfnamefont {L.~Flores}\
			\bibnamefont {Burgos}}, \bibinfo {author} {\bibfnamefont {E.}~\bibnamefont
			{Forati}}, \bibinfo {author} {\bibfnamefont {A.~G.}\ \bibnamefont {Fowler}},
		\bibinfo {author} {\bibfnamefont {B.}~\bibnamefont {Foxen}}, \bibinfo
		{author} {\bibfnamefont {G.}~\bibnamefont {Garcia}}, \bibinfo {author}
		{\bibfnamefont {{\'{E}}.}~\bibnamefont {Genois}}, \bibinfo {author}
		{\bibfnamefont {W.}~\bibnamefont {Giang}}, \bibinfo {author} {\bibfnamefont
			{C.}~\bibnamefont {Gidney}}, \bibinfo {author} {\bibfnamefont
			{D.}~\bibnamefont {Gilboa}}, \bibinfo {author} {\bibfnamefont
			{M.}~\bibnamefont {Giustina}}, \bibinfo {author} {\bibfnamefont
			{R.}~\bibnamefont {Gosula}}, \bibinfo {author} {\bibfnamefont {A.~Grajales}\
			\bibnamefont {Dau}}, \bibinfo {author} {\bibfnamefont {J.~A.}\ \bibnamefont
			{Gross}}, \bibinfo {author} {\bibfnamefont {S.}~\bibnamefont {Habegger}},
		\bibinfo {author} {\bibfnamefont {M.~C.}\ \bibnamefont {Hamilton}}, \bibinfo
		{author} {\bibfnamefont {M.}~\bibnamefont {Hansen}}, \bibinfo {author}
		{\bibfnamefont {M.~P.}\ \bibnamefont {Harrigan}}, \bibinfo {author}
		{\bibfnamefont {S.~D.}\ \bibnamefont {Harrington}}, \bibinfo {author}
		{\bibfnamefont {P.}~\bibnamefont {Heu}}, \bibinfo {author} {\bibfnamefont
			{G.}~\bibnamefont {Hill}}, \bibinfo {author} {\bibfnamefont {M.~R.}\
			\bibnamefont {Hoffmann}}, \bibinfo {author} {\bibfnamefont {S.}~\bibnamefont
			{Hong}}, \bibinfo {author} {\bibfnamefont {T.}~\bibnamefont {Huang}},
		\bibinfo {author} {\bibfnamefont {A.}~\bibnamefont {Huff}}, \bibinfo {author}
		{\bibfnamefont {W.~J.}\ \bibnamefont {Huggins}}, \bibinfo {author}
		{\bibfnamefont {L.~B.}\ \bibnamefont {Ioffe}}, \bibinfo {author}
		{\bibfnamefont {S.~V.}\ \bibnamefont {Isakov}}, \bibinfo {author}
		{\bibfnamefont {J.}~\bibnamefont {Iveland}}, \bibinfo {author} {\bibfnamefont
			{E.}~\bibnamefont {Jeffrey}}, \bibinfo {author} {\bibfnamefont
			{Z.}~\bibnamefont {Jiang}}, \bibinfo {author} {\bibfnamefont
			{C.}~\bibnamefont {Jones}}, \bibinfo {author} {\bibfnamefont
			{P.}~\bibnamefont {Juhas}}, \bibinfo {author} {\bibfnamefont
			{D.}~\bibnamefont {Kafri}}, \bibinfo {author} {\bibfnamefont
			{T.}~\bibnamefont {Khattar}}, \bibinfo {author} {\bibfnamefont
			{M.}~\bibnamefont {Khezri}}, \bibinfo {author} {\bibfnamefont
			{M.}~\bibnamefont {Kieferov{\'{a}}}}, \bibinfo {author} {\bibfnamefont
			{S.}~\bibnamefont {Kim}}, \bibinfo {author} {\bibfnamefont {A.}~\bibnamefont
			{Kitaev}}, \bibinfo {author} {\bibfnamefont {A.~R.}\ \bibnamefont {Klots}},
		\bibinfo {author} {\bibfnamefont {A.~N.}\ \bibnamefont {Korotkov}}, \bibinfo
		{author} {\bibfnamefont {F.}~\bibnamefont {Kostritsa}}, \bibinfo {author}
		{\bibfnamefont {J.~M.}\ \bibnamefont {Kreikebaum}}, \bibinfo {author}
		{\bibfnamefont {D.}~\bibnamefont {Landhuis}}, \bibinfo {author}
		{\bibfnamefont {P.}~\bibnamefont {Laptev}}, \bibinfo {author} {\bibfnamefont
			{K.-M.}\ \bibnamefont {Lau}}, \bibinfo {author} {\bibfnamefont
			{L.}~\bibnamefont {Laws}}, \bibinfo {author} {\bibfnamefont {J.}~\bibnamefont
			{Lee}}, \bibinfo {author} {\bibfnamefont {K.~W.}\ \bibnamefont {Lee}},
		\bibinfo {author} {\bibfnamefont {Y.~D.}\ \bibnamefont {Lensky}}, \bibinfo
		{author} {\bibfnamefont {B.~J.}\ \bibnamefont {Lester}}, \bibinfo {author}
		{\bibfnamefont {A.~T.}\ \bibnamefont {Lill}}, \bibinfo {author}
		{\bibfnamefont {W.}~\bibnamefont {Liu}}, \bibinfo {author} {\bibfnamefont
			{A.}~\bibnamefont {Locharla}}, \bibinfo {author} {\bibfnamefont
			{S.}~\bibnamefont {Mandr{\`{a}}}}, \bibinfo {author} {\bibfnamefont
			{O.}~\bibnamefont {Martin}}, \bibinfo {author} {\bibfnamefont
			{S.}~\bibnamefont {Martin}}, \bibinfo {author} {\bibfnamefont {J.~R.}\
			\bibnamefont {McClean}}, \bibinfo {author} {\bibfnamefont {M.}~\bibnamefont
			{McEwen}}, \bibinfo {author} {\bibfnamefont {S.}~\bibnamefont {Meeks}},
		\bibinfo {author} {\bibfnamefont {K.~C.}\ \bibnamefont {Miao}}, \bibinfo
		{author} {\bibfnamefont {A.}~\bibnamefont {Mieszala}}, \bibinfo {author}
		{\bibfnamefont {S.}~\bibnamefont {Montazeri}}, \bibinfo {author}
		{\bibfnamefont {R.}~\bibnamefont {Movassagh}}, \bibinfo {author}
		{\bibfnamefont {W.}~\bibnamefont {Mruczkiewicz}}, \bibinfo {author}
		{\bibfnamefont {A.}~\bibnamefont {Nersisyan}}, \bibinfo {author}
		{\bibfnamefont {M.}~\bibnamefont {Newman}}, \bibinfo {author} {\bibfnamefont
			{J.~H.}\ \bibnamefont {Ng}}, \bibinfo {author} {\bibfnamefont
			{A.}~\bibnamefont {Nguyen}}, \bibinfo {author} {\bibfnamefont
			{M.}~\bibnamefont {Nguyen}}, \bibinfo {author} {\bibfnamefont {M.~Y.}\
			\bibnamefont {Niu}}, \bibinfo {author} {\bibfnamefont {T.~E.}\ \bibnamefont
			{O'Brien}}, \bibinfo {author} {\bibfnamefont {S.}~\bibnamefont {Omonije}},
		\bibinfo {author} {\bibfnamefont {A.}~\bibnamefont {Opremcak}}, \bibinfo
		{author} {\bibfnamefont {R.}~\bibnamefont {Potter}}, \bibinfo {author}
		{\bibfnamefont {L.~P.}\ \bibnamefont {Pryadko}}, \bibinfo {author}
		{\bibfnamefont {C.}~\bibnamefont {Quintana}}, \bibinfo {author}
		{\bibfnamefont {D.~M.}\ \bibnamefont {Rhodes}}, \bibinfo {author}
		{\bibfnamefont {C.}~\bibnamefont {Rocque}}, \bibinfo {author} {\bibfnamefont
			{N.~C.}\ \bibnamefont {Rubin}}, \bibinfo {author} {\bibfnamefont
			{N.}~\bibnamefont {Saei}}, \bibinfo {author} {\bibfnamefont {D.}~\bibnamefont
			{Sank}}, \bibinfo {author} {\bibfnamefont {K.}~\bibnamefont
			{Sankaragomathi}}, \bibinfo {author} {\bibfnamefont {K.~J.}\ \bibnamefont
			{Satzinger}}, \bibinfo {author} {\bibfnamefont {H.~F.}\ \bibnamefont
			{Schurkus}}, \bibinfo {author} {\bibfnamefont {C.}~\bibnamefont {Schuster}},
		\bibinfo {author} {\bibfnamefont {M.~J.}\ \bibnamefont {Shearn}}, \bibinfo
		{author} {\bibfnamefont {A.}~\bibnamefont {Shorter}}, \bibinfo {author}
		{\bibfnamefont {N.}~\bibnamefont {Shutty}}, \bibinfo {author} {\bibfnamefont
			{V.}~\bibnamefont {Shvarts}}, \bibinfo {author} {\bibfnamefont
			{V.}~\bibnamefont {Sivak}}, \bibinfo {author} {\bibfnamefont
			{J.}~\bibnamefont {Skruzny}}, \bibinfo {author} {\bibfnamefont {W.~Clarke}\
			\bibnamefont {Smith}}, \bibinfo {author} {\bibfnamefont {R.~D.}\ \bibnamefont
			{Somma}}, \bibinfo {author} {\bibfnamefont {G.}~\bibnamefont {Sterling}},
		\bibinfo {author} {\bibfnamefont {D.}~\bibnamefont {Strain}}, \bibinfo
		{author} {\bibfnamefont {M.}~\bibnamefont {Szalay}}, \bibinfo {author}
		{\bibfnamefont {D.}~\bibnamefont {Thor}}, \bibinfo {author} {\bibfnamefont
			{A.}~\bibnamefont {Torres}}, \bibinfo {author} {\bibfnamefont
			{G.}~\bibnamefont {Vidal}}, \bibinfo {author} {\bibfnamefont
			{B.}~\bibnamefont {Villalonga}}, \bibinfo {author} {\bibfnamefont
			{C.~Vollgraff}\ \bibnamefont {Heidweiller}}, \bibinfo {author} {\bibfnamefont
			{T.}~\bibnamefont {White}}, \bibinfo {author} {\bibfnamefont {B.~W.~K.}\
			\bibnamefont {Woo}}, \bibinfo {author} {\bibfnamefont {C.}~\bibnamefont
			{Xing}}, \bibinfo {author} {\bibfnamefont {Z.~Jamie}\ \bibnamefont {Yao}},
		\bibinfo {author} {\bibfnamefont {P.}~\bibnamefont {Yeh}}, \bibinfo {author}
		{\bibfnamefont {J.}~\bibnamefont {Yoo}}, \bibinfo {author} {\bibfnamefont
			{G.}~\bibnamefont {Young}}, \bibinfo {author} {\bibfnamefont
			{A.}~\bibnamefont {Zalcman}}, \bibinfo {author} {\bibfnamefont
			{Y.}~\bibnamefont {Zhang}}, \bibinfo {author} {\bibfnamefont
			{N.}~\bibnamefont {Zhu}}, \bibinfo {author} {\bibfnamefont {N.}~\bibnamefont
			{Zobrist}}, \bibinfo {author} {\bibfnamefont {H.}~\bibnamefont {Neven}},
		\bibinfo {author} {\bibfnamefont {R.}~\bibnamefont {Babbush}}, \bibinfo
		{author} {\bibfnamefont {D.}~\bibnamefont {Bacon}}, \bibinfo {author}
		{\bibfnamefont {S.}~\bibnamefont {Boixo}}, \bibinfo {author} {\bibfnamefont
			{J.}~\bibnamefont {Hilton}}, \bibinfo {author} {\bibfnamefont
			{E.}~\bibnamefont {Lucero}}, \bibinfo {author} {\bibfnamefont
			{A.}~\bibnamefont {Megrant}}, \bibinfo {author} {\bibfnamefont
			{J.}~\bibnamefont {Kelly}}, \bibinfo {author} {\bibfnamefont
			{Y.}~\bibnamefont {Chen}}, \bibinfo {author} {\bibfnamefont {V.}~\bibnamefont
			{Smelyanskiy}}, \bibinfo {author} {\bibfnamefont {V.}~\bibnamefont
			{Khemani}}, \bibinfo {author} {\bibfnamefont {S.}~\bibnamefont
			{Gopalakrishnan}}, \bibinfo {author} {\bibfnamefont {T.}~\bibnamefont
			{Prosen}}, \ and\ \bibinfo {author} {\bibfnamefont {P.}~\bibnamefont
			{Roushan}},\ }\bibfield  {title} {\enquote {\bibinfo {title} {{Dynamics of
					magnetization at infinite temperature in a Heisenberg spin chain}},}\ }\href
	{\doibase 10.1126/science.adi7877} {\bibfield  {journal} {\bibinfo  {journal}
			{Science}\ }\textbf {\bibinfo {volume} {384}},\ \bibinfo {pages} {48--53}
		(\bibinfo {year} {2024})}\BibitemShut {NoStop}%
	\bibitem [{\citenamefont {Feldmeier}\ \emph {et~al.}(2020)\citenamefont
		{Feldmeier}, \citenamefont {Sala}, \citenamefont {De~Tomasi}, \citenamefont
		{Pollmann},\ and\ \citenamefont {Knap}}]{PhysRevLett.125.245303}%
	\BibitemOpen
	\bibfield  {author} {\bibinfo {author} {\bibfnamefont {J.}~\bibnamefont
			{Feldmeier}}, \bibinfo {author} {\bibfnamefont {P.}~\bibnamefont {Sala}},
		\bibinfo {author} {\bibfnamefont {G.}~\bibnamefont {De~Tomasi}}, \bibinfo
		{author} {\bibfnamefont {F.}~\bibnamefont {Pollmann}}, \ and\ \bibinfo
		{author} {\bibfnamefont {M.}~\bibnamefont {Knap}},\ }\bibfield  {title}
	{\enquote {\bibinfo {title} {Anomalous diffusion in dipole- and
				higher-moment-conserving systems},}\ }\href {\doibase
		10.1103/PhysRevLett.125.245303} {\bibfield  {journal} {\bibinfo  {journal}
			{Phys. Rev. Lett.}\ }\textbf {\bibinfo {volume} {125}},\ \bibinfo {pages}
		{245303} (\bibinfo {year} {2020})}\BibitemShut {NoStop}%
	\bibitem [{\citenamefont {De~Nardis}\ \emph {et~al.}(2022)\citenamefont
		{De~Nardis}, \citenamefont {Gopalakrishnan}, \citenamefont {Vasseur},\ and\
		\citenamefont {Ware}}]{De-Nardis:2022ud}%
	\BibitemOpen
	\bibfield  {author} {\bibinfo {author} {\bibfnamefont {J.}~\bibnamefont
			{De~Nardis}}, \bibinfo {author} {\bibfnamefont {S.}~\bibnamefont
			{Gopalakrishnan}}, \bibinfo {author} {\bibfnamefont {R.}~\bibnamefont
			{Vasseur}}, \ and\ \bibinfo {author} {\bibfnamefont {B.}~\bibnamefont
			{Ware}},\ }\bibfield  {title} {\enquote {\bibinfo {title} {Subdiffusive
				hydrodynamics of nearly integrable anisotropic spin chains},}\ }\href
	{\doibase 10.1073/pnas.2202823119} {\bibfield  {journal} {\bibinfo  {journal}
			{Proceedings of the National Academy of Sciences}\ }\textbf {\bibinfo
			{volume} {119}},\ \bibinfo {pages} {e2202823119} (\bibinfo {year}
		{2022})}\BibitemShut {NoStop}%
	\bibitem [{\citenamefont {Gromov}\ \emph {et~al.}(2020)\citenamefont {Gromov},
		\citenamefont {Lucas},\ and\ \citenamefont
		{Nandkishore}}]{PhysRevResearch.2.033124}%
	\BibitemOpen
	\bibfield  {author} {\bibinfo {author} {\bibfnamefont {A.}~\bibnamefont
			{Gromov}}, \bibinfo {author} {\bibfnamefont {A.}~\bibnamefont {Lucas}}, \
		and\ \bibinfo {author} {\bibfnamefont {R.~M.}\ \bibnamefont {Nandkishore}},\
	}\bibfield  {title} {\enquote {\bibinfo {title} {{Fracton hydrodynamics}},}\
	}\href {\doibase 10.1103/PhysRevResearch.2.033124} {\bibfield  {journal}
		{\bibinfo  {journal} {Phys. Rev. Res.}\ }\textbf {\bibinfo {volume} {2}},\
		\bibinfo {pages} {033124} (\bibinfo {year} {2020})}\BibitemShut {NoStop}%
	\bibitem [{\citenamefont {Ma}\ \emph {et~al.}(2019)\citenamefont {Ma},
		\citenamefont {Saxberg}, \citenamefont {Owens}, \citenamefont {Leung},
		\citenamefont {Lu}, \citenamefont {Simon},\ and\ \citenamefont
		{Schuster}}]{Ma:2019um}%
	\BibitemOpen
	\bibfield  {author} {\bibinfo {author} {\bibfnamefont {Ruichao}\ \bibnamefont
			{Ma}}, \bibinfo {author} {\bibfnamefont {Brendan}\ \bibnamefont {Saxberg}},
		\bibinfo {author} {\bibfnamefont {Clai}\ \bibnamefont {Owens}}, \bibinfo
		{author} {\bibfnamefont {Nelson}\ \bibnamefont {Leung}}, \bibinfo {author}
		{\bibfnamefont {Yao}\ \bibnamefont {Lu}}, \bibinfo {author} {\bibfnamefont
			{Jonathan}\ \bibnamefont {Simon}}, \ and\ \bibinfo {author} {\bibfnamefont
			{David~I.}\ \bibnamefont {Schuster}},\ }\bibfield  {title} {\enquote
		{\bibinfo {title} {{A dissipatively stabilized Mott insulator of photons}},}\
	}\href {\doibase 10.1038/s41586-019-0897-9} {\bibfield  {journal} {\bibinfo
			{journal} {Nature}\ }\textbf {\bibinfo {volume} {566}},\ \bibinfo {pages}
		{51--57} (\bibinfo {year} {2019})}\BibitemShut {NoStop}%
	\bibitem [{\citenamefont {Zhang}\ \emph
		{et~al.}(2023{\natexlab{a}})\citenamefont {Zhang}, \citenamefont {Kim},
		\citenamefont {Mark}, \citenamefont {Choi},\ and\ \citenamefont
		{Painter}}]{doi:10.1126/science.ade7651}%
	\BibitemOpen
	\bibfield  {author} {\bibinfo {author} {\bibfnamefont {X.}~\bibnamefont
			{Zhang}}, \bibinfo {author} {\bibfnamefont {E.}~\bibnamefont {Kim}}, \bibinfo
		{author} {\bibfnamefont {D.~K.}\ \bibnamefont {Mark}}, \bibinfo {author}
		{\bibfnamefont {S.}~\bibnamefont {Choi}}, \ and\ \bibinfo {author}
		{\bibfnamefont {O.}~\bibnamefont {Painter}},\ }\bibfield  {title} {\enquote
		{\bibinfo {title} {{A superconducting quantum simulator based on a
					photonic-bandgap metamaterial}},}\ }\href {\doibase 10.1126/science.ade7651}
	{\bibfield  {journal} {\bibinfo  {journal} {Science}\ }\textbf {\bibinfo
			{volume} {379}},\ \bibinfo {pages} {278--283} (\bibinfo {year}
		{2023}{\natexlab{a}})}\BibitemShut {NoStop}%
	\bibitem [{\citenamefont {Xiang}\ \emph {et~al.}(2023)\citenamefont {Xiang},
		\citenamefont {Huang}, \citenamefont {Zhang}, \citenamefont {Liu},
		\citenamefont {Shi}, \citenamefont {Deng}, \citenamefont {Liu}, \citenamefont
		{Li}, \citenamefont {Liang}, \citenamefont {Mei}, \citenamefont {Yu},
		\citenamefont {Xue}, \citenamefont {Tian}, \citenamefont {Song},
		\citenamefont {Liu}, \citenamefont {Xu}, \citenamefont {Zheng}, \citenamefont
		{Nori},\ and\ \citenamefont {Fan}}]{Xiang2023}%
	\BibitemOpen
	\bibfield  {author} {\bibinfo {author} {\bibfnamefont {Zhong-Cheng}\
			\bibnamefont {Xiang}}, \bibinfo {author} {\bibfnamefont {Kaixuan}\
			\bibnamefont {Huang}}, \bibinfo {author} {\bibfnamefont {Yu-Ran}\
			\bibnamefont {Zhang}}, \bibinfo {author} {\bibfnamefont {Tao}\ \bibnamefont
			{Liu}}, \bibinfo {author} {\bibfnamefont {Yun-Hao}\ \bibnamefont {Shi}},
		\bibinfo {author} {\bibfnamefont {Cheng-Lin}\ \bibnamefont {Deng}}, \bibinfo
		{author} {\bibfnamefont {Tong}\ \bibnamefont {Liu}}, \bibinfo {author}
		{\bibfnamefont {Hao}\ \bibnamefont {Li}}, \bibinfo {author} {\bibfnamefont
			{Gui-Han}\ \bibnamefont {Liang}}, \bibinfo {author} {\bibfnamefont
			{Zheng-Yang}\ \bibnamefont {Mei}}, \bibinfo {author} {\bibfnamefont
			{Haifeng}\ \bibnamefont {Yu}}, \bibinfo {author} {\bibfnamefont {Guangming}\
			\bibnamefont {Xue}}, \bibinfo {author} {\bibfnamefont {Ye}~\bibnamefont
			{Tian}}, \bibinfo {author} {\bibfnamefont {Xiaohui}\ \bibnamefont {Song}},
		\bibinfo {author} {\bibfnamefont {Zhi-Bo}\ \bibnamefont {Liu}}, \bibinfo
		{author} {\bibfnamefont {Kai}\ \bibnamefont {Xu}}, \bibinfo {author}
		{\bibfnamefont {Dongning}\ \bibnamefont {Zheng}}, \bibinfo {author}
		{\bibfnamefont {Franco}\ \bibnamefont {Nori}}, \ and\ \bibinfo {author}
		{\bibfnamefont {Heng}\ \bibnamefont {Fan}},\ }\bibfield  {title} {\enquote
		{\bibinfo {title} {{Simulating Chern insulators on a superconducting quantum
					processor}},}\ }\href {\doibase 10.1038/s41467-023-41230-9} {\bibfield
		{journal} {\bibinfo  {journal} {Nature Communications}\ }\textbf {\bibinfo
			{volume} {14}},\ \bibinfo {pages} {5433} (\bibinfo {year}
		{2023})}\BibitemShut {NoStop}%
	\bibitem [{\citenamefont {Gu}\ \emph {et~al.}(2017)\citenamefont {Gu},
		\citenamefont {Kockum}, \citenamefont {Miranowicz}, \citenamefont {Liu},\
		and\ \citenamefont {Nori}}]{Gu:2017ut}%
	\BibitemOpen
	\bibfield  {author} {\bibinfo {author} {\bibfnamefont {Xiu}\ \bibnamefont
			{Gu}}, \bibinfo {author} {\bibfnamefont {Anton~Frisk}\ \bibnamefont
			{Kockum}}, \bibinfo {author} {\bibfnamefont {Adam}\ \bibnamefont
			{Miranowicz}}, \bibinfo {author} {\bibfnamefont {Yu-xi}\ \bibnamefont {Liu}},
		\ and\ \bibinfo {author} {\bibfnamefont {Franco}\ \bibnamefont {Nori}},\
	}\bibfield  {title} {\enquote {\bibinfo {title} {{Microwave photonics with
					superconducting quantum circuits}},}\ }\href {\doibase
		10.1016/j.physrep.2017.10.002} {\bibfield  {journal} {\bibinfo  {journal}
			{Physics Reports}\ }\textbf {\bibinfo {volume} {718-719}},\ \bibinfo {pages}
		{1--102} (\bibinfo {year} {2017})}\BibitemShut {NoStop}%
	\bibitem [{\citenamefont {Chen}\ \emph {et~al.}(2021)\citenamefont {Chen},
		\citenamefont {Sun}, \citenamefont {Gong}, \citenamefont {Zhu}, \citenamefont
		{Zhang}, \citenamefont {Wu}, \citenamefont {Ye}, \citenamefont {Zha},
		\citenamefont {Li}, \citenamefont {Guo}, \citenamefont {Qian}, \citenamefont
		{Huang}, \citenamefont {Yu}, \citenamefont {Deng}, \citenamefont {Rong},
		\citenamefont {Lin}, \citenamefont {Xu}, \citenamefont {Sun}, \citenamefont
		{Guo}, \citenamefont {Li}, \citenamefont {Liang}, \citenamefont {Peng},
		\citenamefont {Fan}, \citenamefont {Zhu},\ and\ \citenamefont
		{Pan}}]{PhysRevLett.127.020602}%
	\BibitemOpen
	\bibfield  {author} {\bibinfo {author} {\bibfnamefont {Fusheng}\ \bibnamefont
			{Chen}}, \bibinfo {author} {\bibfnamefont {Zheng-Hang}\ \bibnamefont {Sun}},
		\bibinfo {author} {\bibfnamefont {Ming}\ \bibnamefont {Gong}}, \bibinfo
		{author} {\bibfnamefont {Qingling}\ \bibnamefont {Zhu}}, \bibinfo {author}
		{\bibfnamefont {Yu-Ran}\ \bibnamefont {Zhang}}, \bibinfo {author}
		{\bibfnamefont {Yulin}\ \bibnamefont {Wu}}, \bibinfo {author} {\bibfnamefont
			{Yangsen}\ \bibnamefont {Ye}}, \bibinfo {author} {\bibfnamefont {Chen}\
			\bibnamefont {Zha}}, \bibinfo {author} {\bibfnamefont {Shaowei}\ \bibnamefont
			{Li}}, \bibinfo {author} {\bibfnamefont {Shaojun}\ \bibnamefont {Guo}},
		\bibinfo {author} {\bibfnamefont {Haoran}\ \bibnamefont {Qian}}, \bibinfo
		{author} {\bibfnamefont {He-Liang}\ \bibnamefont {Huang}}, \bibinfo {author}
		{\bibfnamefont {Jiale}\ \bibnamefont {Yu}}, \bibinfo {author} {\bibfnamefont
			{Hui}\ \bibnamefont {Deng}}, \bibinfo {author} {\bibfnamefont {Hao}\
			\bibnamefont {Rong}}, \bibinfo {author} {\bibfnamefont {Jin}\ \bibnamefont
			{Lin}}, \bibinfo {author} {\bibfnamefont {Yu}~\bibnamefont {Xu}}, \bibinfo
		{author} {\bibfnamefont {Lihua}\ \bibnamefont {Sun}}, \bibinfo {author}
		{\bibfnamefont {Cheng}\ \bibnamefont {Guo}}, \bibinfo {author} {\bibfnamefont
			{Na}~\bibnamefont {Li}}, \bibinfo {author} {\bibfnamefont {Futian}\
			\bibnamefont {Liang}}, \bibinfo {author} {\bibfnamefont {Cheng-Zhi}\
			\bibnamefont {Peng}}, \bibinfo {author} {\bibfnamefont {Heng}\ \bibnamefont
			{Fan}}, \bibinfo {author} {\bibfnamefont {Xiaobo}\ \bibnamefont {Zhu}}, \
		and\ \bibinfo {author} {\bibfnamefont {Jian-Wei}\ \bibnamefont {Pan}},\
	}\bibfield  {title} {\enquote {\bibinfo {title} {{Observation of Strong and
					Weak Thermalization in a Superconducting Quantum Processor}},}\ }\href
	{\doibase 10.1103/PhysRevLett.127.020602} {\bibfield  {journal} {\bibinfo
			{journal} {Phys. Rev. Lett.}\ }\textbf {\bibinfo {volume} {127}},\ \bibinfo
		{pages} {020602} (\bibinfo {year} {2021})}\BibitemShut {NoStop}%
	\bibitem [{\citenamefont {Zhu}\ \emph {et~al.}(2022)\citenamefont {Zhu},
		\citenamefont {Sun}, \citenamefont {Gong}, \citenamefont {Chen},
		\citenamefont {Zhang}, \citenamefont {Wu}, \citenamefont {Ye}, \citenamefont
		{Zha}, \citenamefont {Li}, \citenamefont {Guo}, \citenamefont {Qian},
		\citenamefont {Huang}, \citenamefont {Yu}, \citenamefont {Deng},
		\citenamefont {Rong}, \citenamefont {Lin}, \citenamefont {Xu}, \citenamefont
		{Sun}, \citenamefont {Guo}, \citenamefont {Li}, \citenamefont {Liang},
		\citenamefont {Peng}, \citenamefont {Fan}, \citenamefont {Zhu},\ and\
		\citenamefont {Pan}}]{PhysRevLett.128.160502}%
	\BibitemOpen
	\bibfield  {author} {\bibinfo {author} {\bibfnamefont {Qingling}\
			\bibnamefont {Zhu}}, \bibinfo {author} {\bibfnamefont {Zheng-Hang}\
			\bibnamefont {Sun}}, \bibinfo {author} {\bibfnamefont {Ming}\ \bibnamefont
			{Gong}}, \bibinfo {author} {\bibfnamefont {Fusheng}\ \bibnamefont {Chen}},
		\bibinfo {author} {\bibfnamefont {Yu-Ran}\ \bibnamefont {Zhang}}, \bibinfo
		{author} {\bibfnamefont {Yulin}\ \bibnamefont {Wu}}, \bibinfo {author}
		{\bibfnamefont {Yangsen}\ \bibnamefont {Ye}}, \bibinfo {author}
		{\bibfnamefont {Chen}\ \bibnamefont {Zha}}, \bibinfo {author} {\bibfnamefont
			{Shaowei}\ \bibnamefont {Li}}, \bibinfo {author} {\bibfnamefont {Shaojun}\
			\bibnamefont {Guo}}, \bibinfo {author} {\bibfnamefont {Haoran}\ \bibnamefont
			{Qian}}, \bibinfo {author} {\bibfnamefont {He-Liang}\ \bibnamefont {Huang}},
		\bibinfo {author} {\bibfnamefont {Jiale}\ \bibnamefont {Yu}}, \bibinfo
		{author} {\bibfnamefont {Hui}\ \bibnamefont {Deng}}, \bibinfo {author}
		{\bibfnamefont {Hao}\ \bibnamefont {Rong}}, \bibinfo {author} {\bibfnamefont
			{Jin}\ \bibnamefont {Lin}}, \bibinfo {author} {\bibfnamefont
			{Yu}~\bibnamefont {Xu}}, \bibinfo {author} {\bibfnamefont {Lihua}\
			\bibnamefont {Sun}}, \bibinfo {author} {\bibfnamefont {Cheng}\ \bibnamefont
			{Guo}}, \bibinfo {author} {\bibfnamefont {Na}~\bibnamefont {Li}}, \bibinfo
		{author} {\bibfnamefont {Futian}\ \bibnamefont {Liang}}, \bibinfo {author}
		{\bibfnamefont {Cheng-Zhi}\ \bibnamefont {Peng}}, \bibinfo {author}
		{\bibfnamefont {Heng}\ \bibnamefont {Fan}}, \bibinfo {author} {\bibfnamefont
			{Xiaobo}\ \bibnamefont {Zhu}}, \ and\ \bibinfo {author} {\bibfnamefont
			{Jian-Wei}\ \bibnamefont {Pan}},\ }\bibfield  {title} {\enquote {\bibinfo
			{title} {{Observation of Thermalization and Information Scrambling in a
					Superconducting Quantum Processor}},}\ }\href {\doibase
		10.1103/PhysRevLett.128.160502} {\bibfield  {journal} {\bibinfo  {journal}
			{Phys. Rev. Lett.}\ }\textbf {\bibinfo {volume} {128}},\ \bibinfo {pages}
		{160502} (\bibinfo {year} {2022})}\BibitemShut {NoStop}%
	\bibitem [{\citenamefont {Roushan}\ \emph {et~al.}(2017)\citenamefont
		{Roushan}, \citenamefont {Neill}, \citenamefont {Tangpanitanon},
		\citenamefont {Bastidas}, \citenamefont {Megrant}, \citenamefont {Barends},
		\citenamefont {Chen}, \citenamefont {Chen}, \citenamefont {Chiaro},
		\citenamefont {Dunsworth}, \citenamefont {Fowler}, \citenamefont {Foxen},
		\citenamefont {Giustina}, \citenamefont {Jeffrey}, \citenamefont {Kelly},
		\citenamefont {Lucero}, \citenamefont {Mutus}, \citenamefont {Neeley},
		\citenamefont {Quintana}, \citenamefont {Sank}, \citenamefont {Vainsencher},
		\citenamefont {Wenner}, \citenamefont {White}, \citenamefont {Neven},
		\citenamefont {Angelakis},\ and\ \citenamefont {Martinis}}]{Roushan:2017vp}%
	\BibitemOpen
	\bibfield  {author} {\bibinfo {author} {\bibfnamefont {P.}~\bibnamefont
			{Roushan}}, \bibinfo {author} {\bibfnamefont {C.}~\bibnamefont {Neill}},
		\bibinfo {author} {\bibfnamefont {J.}~\bibnamefont {Tangpanitanon}}, \bibinfo
		{author} {\bibfnamefont {V.~M.}\ \bibnamefont {Bastidas}}, \bibinfo {author}
		{\bibfnamefont {A.}~\bibnamefont {Megrant}}, \bibinfo {author} {\bibfnamefont
			{R.}~\bibnamefont {Barends}}, \bibinfo {author} {\bibfnamefont
			{Y.}~\bibnamefont {Chen}}, \bibinfo {author} {\bibfnamefont {Z.}~\bibnamefont
			{Chen}}, \bibinfo {author} {\bibfnamefont {B.}~\bibnamefont {Chiaro}},
		\bibinfo {author} {\bibfnamefont {A.}~\bibnamefont {Dunsworth}}, \bibinfo
		{author} {\bibfnamefont {A.}~\bibnamefont {Fowler}}, \bibinfo {author}
		{\bibfnamefont {B.}~\bibnamefont {Foxen}}, \bibinfo {author} {\bibfnamefont
			{M.}~\bibnamefont {Giustina}}, \bibinfo {author} {\bibfnamefont
			{E.}~\bibnamefont {Jeffrey}}, \bibinfo {author} {\bibfnamefont
			{J.}~\bibnamefont {Kelly}}, \bibinfo {author} {\bibfnamefont
			{E.}~\bibnamefont {Lucero}}, \bibinfo {author} {\bibfnamefont
			{J.}~\bibnamefont {Mutus}}, \bibinfo {author} {\bibfnamefont
			{M.}~\bibnamefont {Neeley}}, \bibinfo {author} {\bibfnamefont
			{C.}~\bibnamefont {Quintana}}, \bibinfo {author} {\bibfnamefont
			{D.}~\bibnamefont {Sank}}, \bibinfo {author} {\bibfnamefont {A.}~\bibnamefont
			{Vainsencher}}, \bibinfo {author} {\bibfnamefont {J.}~\bibnamefont {Wenner}},
		\bibinfo {author} {\bibfnamefont {T.}~\bibnamefont {White}}, \bibinfo
		{author} {\bibfnamefont {H.}~\bibnamefont {Neven}}, \bibinfo {author}
		{\bibfnamefont {D.~G.}\ \bibnamefont {Angelakis}}, \ and\ \bibinfo {author}
		{\bibfnamefont {J.}~\bibnamefont {Martinis}},\ }\bibfield  {title} {\enquote
		{\bibinfo {title} {{Spectroscopic signatures of localization with interacting
					photons in superconducting qubits}},}\ }\href {\doibase
		10.1126/science.aao1401} {\bibfield  {journal} {\bibinfo  {journal}
			{Science}\ }\textbf {\bibinfo {volume} {358}},\ \bibinfo {pages} {1175--1179}
		(\bibinfo {year} {2017})}\BibitemShut {NoStop}%
	\bibitem [{\citenamefont {Guo}\ \emph {et~al.}(2021{\natexlab{a}})\citenamefont
		{Guo}, \citenamefont {Cheng}, \citenamefont {Sun}, \citenamefont {Song},
		\citenamefont {Li}, \citenamefont {Wang}, \citenamefont {Ren}, \citenamefont
		{Dong}, \citenamefont {Zheng}, \citenamefont {Zhang}, \citenamefont
		{Mondaini}, \citenamefont {Fan},\ and\ \citenamefont {Wang}}]{Guo:2021wr}%
	\BibitemOpen
	\bibfield  {author} {\bibinfo {author} {\bibfnamefont {Qiujiang}\
			\bibnamefont {Guo}}, \bibinfo {author} {\bibfnamefont {Chen}\ \bibnamefont
			{Cheng}}, \bibinfo {author} {\bibfnamefont {Zheng-Hang}\ \bibnamefont {Sun}},
		\bibinfo {author} {\bibfnamefont {Zixuan}\ \bibnamefont {Song}}, \bibinfo
		{author} {\bibfnamefont {Hekang}\ \bibnamefont {Li}}, \bibinfo {author}
		{\bibfnamefont {Zhen}\ \bibnamefont {Wang}}, \bibinfo {author} {\bibfnamefont
			{Wenhui}\ \bibnamefont {Ren}}, \bibinfo {author} {\bibfnamefont {Hang}\
			\bibnamefont {Dong}}, \bibinfo {author} {\bibfnamefont {Dongning}\
			\bibnamefont {Zheng}}, \bibinfo {author} {\bibfnamefont {Yu-Ran}\
			\bibnamefont {Zhang}}, \bibinfo {author} {\bibfnamefont {Rubem}\ \bibnamefont
			{Mondaini}}, \bibinfo {author} {\bibfnamefont {Heng}\ \bibnamefont {Fan}}, \
		and\ \bibinfo {author} {\bibfnamefont {H.}~\bibnamefont {Wang}},\ }\bibfield
	{title} {\enquote {\bibinfo {title} {{Observation of energy-resolved
					many-body localization}},}\ }\href {\doibase 10.1038/s41567-020-1035-1}
	{\bibfield  {journal} {\bibinfo  {journal} {Nature Physics}\ }\textbf
		{\bibinfo {volume} {17}},\ \bibinfo {pages} {234--239} (\bibinfo {year}
		{2021}{\natexlab{a}})}\BibitemShut {NoStop}%
	\bibitem [{\citenamefont {Guo}\ \emph {et~al.}(2021{\natexlab{b}})\citenamefont
		{Guo}, \citenamefont {Cheng}, \citenamefont {Li}, \citenamefont {Xu},
		\citenamefont {Zhang}, \citenamefont {Wang}, \citenamefont {Song},
		\citenamefont {Liu}, \citenamefont {Ren}, \citenamefont {Dong}, \citenamefont
		{Mondaini},\ and\ \citenamefont {Wang}}]{PhysRevLett.127.240502}%
	\BibitemOpen
	\bibfield  {author} {\bibinfo {author} {\bibfnamefont {Qiujiang}\
			\bibnamefont {Guo}}, \bibinfo {author} {\bibfnamefont {Chen}\ \bibnamefont
			{Cheng}}, \bibinfo {author} {\bibfnamefont {Hekang}\ \bibnamefont {Li}},
		\bibinfo {author} {\bibfnamefont {Shibo}\ \bibnamefont {Xu}}, \bibinfo
		{author} {\bibfnamefont {Pengfei}\ \bibnamefont {Zhang}}, \bibinfo {author}
		{\bibfnamefont {Zhen}\ \bibnamefont {Wang}}, \bibinfo {author} {\bibfnamefont
			{Chao}\ \bibnamefont {Song}}, \bibinfo {author} {\bibfnamefont {Wuxin}\
			\bibnamefont {Liu}}, \bibinfo {author} {\bibfnamefont {Wenhui}\ \bibnamefont
			{Ren}}, \bibinfo {author} {\bibfnamefont {Hang}\ \bibnamefont {Dong}},
		\bibinfo {author} {\bibfnamefont {Rubem}\ \bibnamefont {Mondaini}}, \ and\
		\bibinfo {author} {\bibfnamefont {H.}~\bibnamefont {Wang}},\ }\bibfield
	{title} {\enquote {\bibinfo {title} {{Stark Many-Body Localization on a
					Superconducting Quantum Processor}},}\ }\href {\doibase
		10.1103/PhysRevLett.127.240502} {\bibfield  {journal} {\bibinfo  {journal}
			{Phys. Rev. Lett.}\ }\textbf {\bibinfo {volume} {127}},\ \bibinfo {pages}
		{240502} (\bibinfo {year} {2021}{\natexlab{b}})}\BibitemShut {NoStop}%
	\bibitem [{\citenamefont {Zhang}\ \emph
		{et~al.}(2023{\natexlab{b}})\citenamefont {Zhang}, \citenamefont {Dong},
		\citenamefont {Gao}, \citenamefont {Zhao}, \citenamefont {Hao}, \citenamefont
		{Desaules}, \citenamefont {Guo}, \citenamefont {Chen}, \citenamefont {Deng},
		\citenamefont {Liu}, \citenamefont {Ren}, \citenamefont {Yao}, \citenamefont
		{Zhang}, \citenamefont {Xu}, \citenamefont {Wang}, \citenamefont {Jin},
		\citenamefont {Zhu}, \citenamefont {Zhang}, \citenamefont {Li}, \citenamefont
		{Song}, \citenamefont {Wang}, \citenamefont {Liu}, \citenamefont
		{Papi{\'{c}}}, \citenamefont {Ying}, \citenamefont {Wang},\ and\
		\citenamefont {Lai}}]{Zhang:2023vv}%
	\BibitemOpen
	\bibfield  {author} {\bibinfo {author} {\bibfnamefont {Pengfei}\ \bibnamefont
			{Zhang}}, \bibinfo {author} {\bibfnamefont {Hang}\ \bibnamefont {Dong}},
		\bibinfo {author} {\bibfnamefont {Yu}~\bibnamefont {Gao}}, \bibinfo {author}
		{\bibfnamefont {Liangtian}\ \bibnamefont {Zhao}}, \bibinfo {author}
		{\bibfnamefont {Jie}\ \bibnamefont {Hao}}, \bibinfo {author} {\bibfnamefont
			{Jean-Yves}\ \bibnamefont {Desaules}}, \bibinfo {author} {\bibfnamefont
			{Qiujiang}\ \bibnamefont {Guo}}, \bibinfo {author} {\bibfnamefont {Jiachen}\
			\bibnamefont {Chen}}, \bibinfo {author} {\bibfnamefont {Jinfeng}\
			\bibnamefont {Deng}}, \bibinfo {author} {\bibfnamefont {Bobo}\ \bibnamefont
			{Liu}}, \bibinfo {author} {\bibfnamefont {Wenhui}\ \bibnamefont {Ren}},
		\bibinfo {author} {\bibfnamefont {Yunyan}\ \bibnamefont {Yao}}, \bibinfo
		{author} {\bibfnamefont {Xu}~\bibnamefont {Zhang}}, \bibinfo {author}
		{\bibfnamefont {Shibo}\ \bibnamefont {Xu}}, \bibinfo {author} {\bibfnamefont
			{Ke}~\bibnamefont {Wang}}, \bibinfo {author} {\bibfnamefont {Feitong}\
			\bibnamefont {Jin}}, \bibinfo {author} {\bibfnamefont {Xuhao}\ \bibnamefont
			{Zhu}}, \bibinfo {author} {\bibfnamefont {Bing}\ \bibnamefont {Zhang}},
		\bibinfo {author} {\bibfnamefont {Hekang}\ \bibnamefont {Li}}, \bibinfo
		{author} {\bibfnamefont {Chao}\ \bibnamefont {Song}}, \bibinfo {author}
		{\bibfnamefont {Zhen}\ \bibnamefont {Wang}}, \bibinfo {author} {\bibfnamefont
			{Fangli}\ \bibnamefont {Liu}}, \bibinfo {author} {\bibfnamefont {Zlatko}\
			\bibnamefont {Papi{\'{c}}}}, \bibinfo {author} {\bibfnamefont {Lei}\
			\bibnamefont {Ying}}, \bibinfo {author} {\bibfnamefont {H.}~\bibnamefont
			{Wang}}, \ and\ \bibinfo {author} {\bibfnamefont {Ying-Cheng}\ \bibnamefont
			{Lai}},\ }\bibfield  {title} {\enquote {\bibinfo {title} {{Many-body Hilbert
					space scarring on a superconducting processor}},}\ }\href {\doibase
		10.1038/s41567-022-01784-9} {\bibfield  {journal} {\bibinfo  {journal}
			{Nature Physics}\ }\textbf {\bibinfo {volume} {19}},\ \bibinfo {pages}
		{120--125} (\bibinfo {year} {2023}{\natexlab{b}})}\BibitemShut {NoStop}%
	\bibitem [{\citenamefont {Zhang}\ \emph {et~al.}(2022)\citenamefont {Zhang},
		\citenamefont {Jiang}, \citenamefont {Deng}, \citenamefont {Wang},
		\citenamefont {Chen}, \citenamefont {Zhang}, \citenamefont {Ren},
		\citenamefont {Dong}, \citenamefont {Xu}, \citenamefont {Gao}, \citenamefont
		{Jin}, \citenamefont {Zhu}, \citenamefont {Guo}, \citenamefont {Li},
		\citenamefont {Song}, \citenamefont {Gorshkov}, \citenamefont {Iadecola},
		\citenamefont {Liu}, \citenamefont {Gong}, \citenamefont {Wang},
		\citenamefont {Deng},\ and\ \citenamefont {Wang}}]{Zhang:2022wf}%
	\BibitemOpen
	\bibfield  {author} {\bibinfo {author} {\bibfnamefont {Xu}~\bibnamefont
			{Zhang}}, \bibinfo {author} {\bibfnamefont {Wenjie}\ \bibnamefont {Jiang}},
		\bibinfo {author} {\bibfnamefont {Jinfeng}\ \bibnamefont {Deng}}, \bibinfo
		{author} {\bibfnamefont {Ke}~\bibnamefont {Wang}}, \bibinfo {author}
		{\bibfnamefont {Jiachen}\ \bibnamefont {Chen}}, \bibinfo {author}
		{\bibfnamefont {Pengfei}\ \bibnamefont {Zhang}}, \bibinfo {author}
		{\bibfnamefont {Wenhui}\ \bibnamefont {Ren}}, \bibinfo {author}
		{\bibfnamefont {Hang}\ \bibnamefont {Dong}}, \bibinfo {author} {\bibfnamefont
			{Shibo}\ \bibnamefont {Xu}}, \bibinfo {author} {\bibfnamefont
			{Yu}~\bibnamefont {Gao}}, \bibinfo {author} {\bibfnamefont {Feitong}\
			\bibnamefont {Jin}}, \bibinfo {author} {\bibfnamefont {Xuhao}\ \bibnamefont
			{Zhu}}, \bibinfo {author} {\bibfnamefont {Qiujiang}\ \bibnamefont {Guo}},
		\bibinfo {author} {\bibfnamefont {Hekang}\ \bibnamefont {Li}}, \bibinfo
		{author} {\bibfnamefont {Chao}\ \bibnamefont {Song}}, \bibinfo {author}
		{\bibfnamefont {Alexey~V.}\ \bibnamefont {Gorshkov}}, \bibinfo {author}
		{\bibfnamefont {Thomas}\ \bibnamefont {Iadecola}}, \bibinfo {author}
		{\bibfnamefont {Fangli}\ \bibnamefont {Liu}}, \bibinfo {author}
		{\bibfnamefont {Zhe-Xuan}\ \bibnamefont {Gong}}, \bibinfo {author}
		{\bibfnamefont {Zhen}\ \bibnamefont {Wang}}, \bibinfo {author} {\bibfnamefont
			{Dong-Ling}\ \bibnamefont {Deng}}, \ and\ \bibinfo {author} {\bibfnamefont
			{H.}~\bibnamefont {Wang}},\ }\bibfield  {title} {\enquote {\bibinfo {title}
			{{Digital quantum simulation of Floquet symmetry-protected topological
					phases}},}\ }\href {\doibase 10.1038/s41586-022-04854-3} {\bibfield
		{journal} {\bibinfo  {journal} {Nature}\ }\textbf {\bibinfo {volume} {607}},\
		\bibinfo {pages} {468--473} (\bibinfo {year} {2022})}\BibitemShut {NoStop}%
	\bibitem [{\citenamefont {Mi}\ \emph {et~al.}(2022)\citenamefont {Mi},
		\citenamefont {Ippoliti}, \citenamefont {Quintana}, \citenamefont {Greene},
		\citenamefont {Chen}, \citenamefont {Gross}, \citenamefont {Arute},
		\citenamefont {Arya}, \citenamefont {Atalaya}, \citenamefont {Babbush},
		\citenamefont {Bardin}, \citenamefont {Basso}, \citenamefont {Bengtsson},
		\citenamefont {Bilmes}, \citenamefont {Bourassa}, \citenamefont {Brill},
		\citenamefont {Broughton}, \citenamefont {Buckley}, \citenamefont {Buell},
		\citenamefont {Burkett}, \citenamefont {Bushnell}, \citenamefont {Chiaro},
		\citenamefont {Collins}, \citenamefont {Courtney}, \citenamefont {Debroy},
		\citenamefont {Demura}, \citenamefont {Derk}, \citenamefont {Dunsworth},
		\citenamefont {Eppens}, \citenamefont {Erickson}, \citenamefont {Farhi},
		\citenamefont {Fowler}, \citenamefont {Foxen}, \citenamefont {Gidney},
		\citenamefont {Giustina}, \citenamefont {Harrigan}, \citenamefont
		{Harrington}, \citenamefont {Hilton}, \citenamefont {Ho}, \citenamefont
		{Hong}, \citenamefont {Huang}, \citenamefont {Huff}, \citenamefont {Huggins},
		\citenamefont {Ioffe}, \citenamefont {Isakov}, \citenamefont {Iveland},
		\citenamefont {Jeffrey}, \citenamefont {Jiang}, \citenamefont {Jones},
		\citenamefont {Kafri}, \citenamefont {Khattar}, \citenamefont {Kim},
		\citenamefont {Kitaev}, \citenamefont {Klimov}, \citenamefont {Korotkov},
		\citenamefont {Kostritsa}, \citenamefont {Landhuis}, \citenamefont {Laptev},
		\citenamefont {Lee}, \citenamefont {Lee}, \citenamefont {Locharla},
		\citenamefont {Lucero}, \citenamefont {Martin}, \citenamefont {McClean},
		\citenamefont {McCourt}, \citenamefont {McEwen}, \citenamefont {Miao},
		\citenamefont {Mohseni}, \citenamefont {Montazeri}, \citenamefont
		{Mruczkiewicz}, \citenamefont {Naaman}, \citenamefont {Neeley}, \citenamefont
		{Neill}, \citenamefont {Newman}, \citenamefont {Niu}, \citenamefont
		{O'Brien}, \citenamefont {Opremcak}, \citenamefont {Ostby}, \citenamefont
		{Pato}, \citenamefont {Petukhov}, \citenamefont {Rubin}, \citenamefont
		{Sank}, \citenamefont {Satzinger}, \citenamefont {Shvarts}, \citenamefont
		{Su}, \citenamefont {Strain}, \citenamefont {Szalay}, \citenamefont
		{Trevithick}, \citenamefont {Villalonga}, \citenamefont {White},
		\citenamefont {Yao}, \citenamefont {Yeh}, \citenamefont {Yoo}, \citenamefont
		{Zalcman}, \citenamefont {Neven}, \citenamefont {Boixo}, \citenamefont
		{Smelyanskiy}, \citenamefont {Megrant}, \citenamefont {Kelly}, \citenamefont
		{Chen}, \citenamefont {Sondhi}, \citenamefont {Moessner}, \citenamefont
		{Kechedzhi}, \citenamefont {Khemani},\ and\ \citenamefont
		{Roushan}}]{Mi:2022tv}%
	\BibitemOpen
	\bibfield  {author} {\bibinfo {author} {\bibfnamefont {Xiao}\ \bibnamefont
			{Mi}}, \bibinfo {author} {\bibfnamefont {Matteo}\ \bibnamefont {Ippoliti}},
		\bibinfo {author} {\bibfnamefont {Chris}\ \bibnamefont {Quintana}}, \bibinfo
		{author} {\bibfnamefont {Ami}\ \bibnamefont {Greene}}, \bibinfo {author}
		{\bibfnamefont {Zijun}\ \bibnamefont {Chen}}, \bibinfo {author}
		{\bibfnamefont {Jonathan}\ \bibnamefont {Gross}}, \bibinfo {author}
		{\bibfnamefont {Frank}\ \bibnamefont {Arute}}, \bibinfo {author}
		{\bibfnamefont {Kunal}\ \bibnamefont {Arya}}, \bibinfo {author}
		{\bibfnamefont {Juan}\ \bibnamefont {Atalaya}}, \bibinfo {author}
		{\bibfnamefont {Ryan}\ \bibnamefont {Babbush}}, \bibinfo {author}
		{\bibfnamefont {Joseph~C.}\ \bibnamefont {Bardin}}, \bibinfo {author}
		{\bibfnamefont {Joao}\ \bibnamefont {Basso}}, \bibinfo {author}
		{\bibfnamefont {Andreas}\ \bibnamefont {Bengtsson}}, \bibinfo {author}
		{\bibfnamefont {Alexander}\ \bibnamefont {Bilmes}}, \bibinfo {author}
		{\bibfnamefont {Alexandre}\ \bibnamefont {Bourassa}}, \bibinfo {author}
		{\bibfnamefont {Leon}\ \bibnamefont {Brill}}, \bibinfo {author}
		{\bibfnamefont {Michael}\ \bibnamefont {Broughton}}, \bibinfo {author}
		{\bibfnamefont {Bob~B.}\ \bibnamefont {Buckley}}, \bibinfo {author}
		{\bibfnamefont {David~A.}\ \bibnamefont {Buell}}, \bibinfo {author}
		{\bibfnamefont {Brian}\ \bibnamefont {Burkett}}, \bibinfo {author}
		{\bibfnamefont {Nicholas}\ \bibnamefont {Bushnell}}, \bibinfo {author}
		{\bibfnamefont {Benjamin}\ \bibnamefont {Chiaro}}, \bibinfo {author}
		{\bibfnamefont {Roberto}\ \bibnamefont {Collins}}, \bibinfo {author}
		{\bibfnamefont {William}\ \bibnamefont {Courtney}}, \bibinfo {author}
		{\bibfnamefont {Dripto}\ \bibnamefont {Debroy}}, \bibinfo {author}
		{\bibfnamefont {Sean}\ \bibnamefont {Demura}}, \bibinfo {author}
		{\bibfnamefont {Alan~R.}\ \bibnamefont {Derk}}, \bibinfo {author}
		{\bibfnamefont {Andrew}\ \bibnamefont {Dunsworth}}, \bibinfo {author}
		{\bibfnamefont {Daniel}\ \bibnamefont {Eppens}}, \bibinfo {author}
		{\bibfnamefont {Catherine}\ \bibnamefont {Erickson}}, \bibinfo {author}
		{\bibfnamefont {Edward}\ \bibnamefont {Farhi}}, \bibinfo {author}
		{\bibfnamefont {Austin~G.}\ \bibnamefont {Fowler}}, \bibinfo {author}
		{\bibfnamefont {Brooks}\ \bibnamefont {Foxen}}, \bibinfo {author}
		{\bibfnamefont {Craig}\ \bibnamefont {Gidney}}, \bibinfo {author}
		{\bibfnamefont {Marissa}\ \bibnamefont {Giustina}}, \bibinfo {author}
		{\bibfnamefont {Matthew~P.}\ \bibnamefont {Harrigan}}, \bibinfo {author}
		{\bibfnamefont {Sean~D.}\ \bibnamefont {Harrington}}, \bibinfo {author}
		{\bibfnamefont {Jeremy}\ \bibnamefont {Hilton}}, \bibinfo {author}
		{\bibfnamefont {Alan}\ \bibnamefont {Ho}}, \bibinfo {author} {\bibfnamefont
			{Sabrina}\ \bibnamefont {Hong}}, \bibinfo {author} {\bibfnamefont {Trent}\
			\bibnamefont {Huang}}, \bibinfo {author} {\bibfnamefont {Ashley}\
			\bibnamefont {Huff}}, \bibinfo {author} {\bibfnamefont {William~J.}\
			\bibnamefont {Huggins}}, \bibinfo {author} {\bibfnamefont {L.~B.}\
			\bibnamefont {Ioffe}}, \bibinfo {author} {\bibfnamefont {Sergei~V.}\
			\bibnamefont {Isakov}}, \bibinfo {author} {\bibfnamefont {Justin}\
			\bibnamefont {Iveland}}, \bibinfo {author} {\bibfnamefont {Evan}\
			\bibnamefont {Jeffrey}}, \bibinfo {author} {\bibfnamefont {Zhang}\
			\bibnamefont {Jiang}}, \bibinfo {author} {\bibfnamefont {Cody}\ \bibnamefont
			{Jones}}, \bibinfo {author} {\bibfnamefont {Dvir}\ \bibnamefont {Kafri}},
		\bibinfo {author} {\bibfnamefont {Tanuj}\ \bibnamefont {Khattar}}, \bibinfo
		{author} {\bibfnamefont {Seon}\ \bibnamefont {Kim}}, \bibinfo {author}
		{\bibfnamefont {Alexei}\ \bibnamefont {Kitaev}}, \bibinfo {author}
		{\bibfnamefont {Paul~V.}\ \bibnamefont {Klimov}}, \bibinfo {author}
		{\bibfnamefont {Alexander~N.}\ \bibnamefont {Korotkov}}, \bibinfo {author}
		{\bibfnamefont {Fedor}\ \bibnamefont {Kostritsa}}, \bibinfo {author}
		{\bibfnamefont {David}\ \bibnamefont {Landhuis}}, \bibinfo {author}
		{\bibfnamefont {Pavel}\ \bibnamefont {Laptev}}, \bibinfo {author}
		{\bibfnamefont {Joonho}\ \bibnamefont {Lee}}, \bibinfo {author}
		{\bibfnamefont {Kenny}\ \bibnamefont {Lee}}, \bibinfo {author} {\bibfnamefont
			{Aditya}\ \bibnamefont {Locharla}}, \bibinfo {author} {\bibfnamefont {Erik}\
			\bibnamefont {Lucero}}, \bibinfo {author} {\bibfnamefont {Orion}\
			\bibnamefont {Martin}}, \bibinfo {author} {\bibfnamefont {Jarrod~R.}\
			\bibnamefont {McClean}}, \bibinfo {author} {\bibfnamefont {Trevor}\
			\bibnamefont {McCourt}}, \bibinfo {author} {\bibfnamefont {Matt}\
			\bibnamefont {McEwen}}, \bibinfo {author} {\bibfnamefont {Kevin~C.}\
			\bibnamefont {Miao}}, \bibinfo {author} {\bibfnamefont {Masoud}\ \bibnamefont
			{Mohseni}}, \bibinfo {author} {\bibfnamefont {Shirin}\ \bibnamefont
			{Montazeri}}, \bibinfo {author} {\bibfnamefont {Wojciech}\ \bibnamefont
			{Mruczkiewicz}}, \bibinfo {author} {\bibfnamefont {Ofer}\ \bibnamefont
			{Naaman}}, \bibinfo {author} {\bibfnamefont {Matthew}\ \bibnamefont
			{Neeley}}, \bibinfo {author} {\bibfnamefont {Charles}\ \bibnamefont {Neill}},
		\bibinfo {author} {\bibfnamefont {Michael}\ \bibnamefont {Newman}}, \bibinfo
		{author} {\bibfnamefont {Murphy~Yuezhen}\ \bibnamefont {Niu}}, \bibinfo
		{author} {\bibfnamefont {Thomas~E.}\ \bibnamefont {O'Brien}}, \bibinfo
		{author} {\bibfnamefont {Alex}\ \bibnamefont {Opremcak}}, \bibinfo {author}
		{\bibfnamefont {Eric}\ \bibnamefont {Ostby}}, \bibinfo {author}
		{\bibfnamefont {Balint}\ \bibnamefont {Pato}}, \bibinfo {author}
		{\bibfnamefont {Andre}\ \bibnamefont {Petukhov}}, \bibinfo {author}
		{\bibfnamefont {Nicholas~C.}\ \bibnamefont {Rubin}}, \bibinfo {author}
		{\bibfnamefont {Daniel}\ \bibnamefont {Sank}}, \bibinfo {author}
		{\bibfnamefont {Kevin~J.}\ \bibnamefont {Satzinger}}, \bibinfo {author}
		{\bibfnamefont {Vladimir}\ \bibnamefont {Shvarts}}, \bibinfo {author}
		{\bibfnamefont {Yuan}\ \bibnamefont {Su}}, \bibinfo {author} {\bibfnamefont
			{Doug}\ \bibnamefont {Strain}}, \bibinfo {author} {\bibfnamefont {Marco}\
			\bibnamefont {Szalay}}, \bibinfo {author} {\bibfnamefont {Matthew~D.}\
			\bibnamefont {Trevithick}}, \bibinfo {author} {\bibfnamefont {Benjamin}\
			\bibnamefont {Villalonga}}, \bibinfo {author} {\bibfnamefont {Theodore}\
			\bibnamefont {White}}, \bibinfo {author} {\bibfnamefont {Z.~Jamie}\
			\bibnamefont {Yao}}, \bibinfo {author} {\bibfnamefont {Ping}\ \bibnamefont
			{Yeh}}, \bibinfo {author} {\bibfnamefont {Juhwan}\ \bibnamefont {Yoo}},
		\bibinfo {author} {\bibfnamefont {Adam}\ \bibnamefont {Zalcman}}, \bibinfo
		{author} {\bibfnamefont {Hartmut}\ \bibnamefont {Neven}}, \bibinfo {author}
		{\bibfnamefont {Sergio}\ \bibnamefont {Boixo}}, \bibinfo {author}
		{\bibfnamefont {Vadim}\ \bibnamefont {Smelyanskiy}}, \bibinfo {author}
		{\bibfnamefont {Anthony}\ \bibnamefont {Megrant}}, \bibinfo {author}
		{\bibfnamefont {Julian}\ \bibnamefont {Kelly}}, \bibinfo {author}
		{\bibfnamefont {Yu}~\bibnamefont {Chen}}, \bibinfo {author} {\bibfnamefont
			{S.~L.}\ \bibnamefont {Sondhi}}, \bibinfo {author} {\bibfnamefont {Roderich}\
			\bibnamefont {Moessner}}, \bibinfo {author} {\bibfnamefont {Kostyantyn}\
			\bibnamefont {Kechedzhi}}, \bibinfo {author} {\bibfnamefont {Vedika}\
			\bibnamefont {Khemani}}, \ and\ \bibinfo {author} {\bibfnamefont {Pedram}\
			\bibnamefont {Roushan}},\ }\bibfield  {title} {\enquote {\bibinfo {title}
			{{Time-crystalline eigenstate order on a quantum processor}},}\ }\href
	{\doibase 10.1038/s41586-021-04257-w} {\bibfield  {journal} {\bibinfo
			{journal} {Nature}\ }\textbf {\bibinfo {volume} {601}},\ \bibinfo {pages}
		{531--536} (\bibinfo {year} {2022})}\BibitemShut {NoStop}%
	\bibitem [{\citenamefont {Frey}\ and\ \citenamefont
		{Rachel}(2022)}]{doi:10.1126/sciadv.abm7652}%
	\BibitemOpen
	\bibfield  {author} {\bibinfo {author} {\bibfnamefont {P.}~\bibnamefont
			{Frey}}\ and\ \bibinfo {author} {\bibfnamefont {S.}~\bibnamefont {Rachel}},\
	}\bibfield  {title} {\enquote {\bibinfo {title} {{Realization of a discrete
					time crystal on 57 qubits of a quantum computer}},}\ }\href {\doibase
		10.1126/sciadv.abm7652} {\bibfield  {journal} {\bibinfo  {journal} {Science
				Advances}\ }\textbf {\bibinfo {volume} {8}},\ \bibinfo {pages} {eabm7652}
		(\bibinfo {year} {2022})}\BibitemShut {NoStop}%
	\bibitem [{\citenamefont {Mi}\ \emph {et~al.}(2021)\citenamefont {Mi},
		\citenamefont {Roushan}, \citenamefont {Quintana}, \citenamefont
		{Mandr{\`{a}}}, \citenamefont {Marshall}, \citenamefont {Neill},
		\citenamefont {Arute}, \citenamefont {Arya}, \citenamefont {Atalaya},
		\citenamefont {Babbush}, \citenamefont {Bardin}, \citenamefont {Barends},
		\citenamefont {Basso}, \citenamefont {Bengtsson}, \citenamefont {Boixo},
		\citenamefont {Bourassa}, \citenamefont {Broughton}, \citenamefont {Buckley},
		\citenamefont {Buell}, \citenamefont {Burkett}, \citenamefont {Bushnell},
		\citenamefont {Chen}, \citenamefont {Chiaro}, \citenamefont {Collins},
		\citenamefont {Courtney}, \citenamefont {Demura}, \citenamefont {Derk},
		\citenamefont {Dunsworth}, \citenamefont {Eppens}, \citenamefont {Erickson},
		\citenamefont {Farhi}, \citenamefont {Fowler}, \citenamefont {Foxen},
		\citenamefont {Gidney}, \citenamefont {Giustina}, \citenamefont {Gross},
		\citenamefont {Harrigan}, \citenamefont {Harrington}, \citenamefont {Hilton},
		\citenamefont {Ho}, \citenamefont {Hong}, \citenamefont {Huang},
		\citenamefont {Huggins}, \citenamefont {Ioffe}, \citenamefont {Isakov},
		\citenamefont {Jeffrey}, \citenamefont {Jiang}, \citenamefont {Jones},
		\citenamefont {Kafri}, \citenamefont {Kelly}, \citenamefont {Kim},
		\citenamefont {Kitaev}, \citenamefont {Klimov}, \citenamefont {Korotkov},
		\citenamefont {Kostritsa}, \citenamefont {Landhuis}, \citenamefont {Laptev},
		\citenamefont {Lucero}, \citenamefont {Martin}, \citenamefont {McClean},
		\citenamefont {McCourt}, \citenamefont {McEwen}, \citenamefont {Megrant},
		\citenamefont {Miao}, \citenamefont {Mohseni}, \citenamefont {Montazeri},
		\citenamefont {Mruczkiewicz}, \citenamefont {Mutus}, \citenamefont {Naaman},
		\citenamefont {Neeley}, \citenamefont {Newman}, \citenamefont {Niu},
		\citenamefont {O'Brien}, \citenamefont {Opremcak}, \citenamefont {Ostby},
		\citenamefont {Pato}, \citenamefont {Petukhov}, \citenamefont {Redd},
		\citenamefont {Rubin}, \citenamefont {Sank}, \citenamefont {Satzinger},
		\citenamefont {Shvarts}, \citenamefont {Strain}, \citenamefont {Szalay},
		\citenamefont {Trevithick}, \citenamefont {Villalonga}, \citenamefont
		{White}, \citenamefont {Yao}, \citenamefont {Yeh}, \citenamefont {Zalcman},
		\citenamefont {Neven}, \citenamefont {Aleiner}, \citenamefont {Kechedzhi},
		\citenamefont {Smelyanskiy},\ and\ \citenamefont {Chen}}]{Mi:2021ta}%
	\BibitemOpen
	\bibfield  {author} {\bibinfo {author} {\bibfnamefont {Xiao}\ \bibnamefont
			{Mi}}, \bibinfo {author} {\bibfnamefont {Pedram}\ \bibnamefont {Roushan}},
		\bibinfo {author} {\bibfnamefont {Chris}\ \bibnamefont {Quintana}}, \bibinfo
		{author} {\bibfnamefont {Salvatore}\ \bibnamefont {Mandr{\`{a}}}}, \bibinfo
		{author} {\bibfnamefont {Jeffrey}\ \bibnamefont {Marshall}}, \bibinfo
		{author} {\bibfnamefont {Charles}\ \bibnamefont {Neill}}, \bibinfo {author}
		{\bibfnamefont {Frank}\ \bibnamefont {Arute}}, \bibinfo {author}
		{\bibfnamefont {Kunal}\ \bibnamefont {Arya}}, \bibinfo {author}
		{\bibfnamefont {Juan}\ \bibnamefont {Atalaya}}, \bibinfo {author}
		{\bibfnamefont {Ryan}\ \bibnamefont {Babbush}}, \bibinfo {author}
		{\bibfnamefont {Joseph~C.}\ \bibnamefont {Bardin}}, \bibinfo {author}
		{\bibfnamefont {Rami}\ \bibnamefont {Barends}}, \bibinfo {author}
		{\bibfnamefont {Joao}\ \bibnamefont {Basso}}, \bibinfo {author}
		{\bibfnamefont {Andreas}\ \bibnamefont {Bengtsson}}, \bibinfo {author}
		{\bibfnamefont {Sergio}\ \bibnamefont {Boixo}}, \bibinfo {author}
		{\bibfnamefont {Alexandre}\ \bibnamefont {Bourassa}}, \bibinfo {author}
		{\bibfnamefont {Michael}\ \bibnamefont {Broughton}}, \bibinfo {author}
		{\bibfnamefont {Bob~B.}\ \bibnamefont {Buckley}}, \bibinfo {author}
		{\bibfnamefont {David~A.}\ \bibnamefont {Buell}}, \bibinfo {author}
		{\bibfnamefont {Brian}\ \bibnamefont {Burkett}}, \bibinfo {author}
		{\bibfnamefont {Nicholas}\ \bibnamefont {Bushnell}}, \bibinfo {author}
		{\bibfnamefont {Zijun}\ \bibnamefont {Chen}}, \bibinfo {author}
		{\bibfnamefont {Benjamin}\ \bibnamefont {Chiaro}}, \bibinfo {author}
		{\bibfnamefont {Roberto}\ \bibnamefont {Collins}}, \bibinfo {author}
		{\bibfnamefont {William}\ \bibnamefont {Courtney}}, \bibinfo {author}
		{\bibfnamefont {Sean}\ \bibnamefont {Demura}}, \bibinfo {author}
		{\bibfnamefont {Alan~R.}\ \bibnamefont {Derk}}, \bibinfo {author}
		{\bibfnamefont {Andrew}\ \bibnamefont {Dunsworth}}, \bibinfo {author}
		{\bibfnamefont {Daniel}\ \bibnamefont {Eppens}}, \bibinfo {author}
		{\bibfnamefont {Catherine}\ \bibnamefont {Erickson}}, \bibinfo {author}
		{\bibfnamefont {Edward}\ \bibnamefont {Farhi}}, \bibinfo {author}
		{\bibfnamefont {Austin~G.}\ \bibnamefont {Fowler}}, \bibinfo {author}
		{\bibfnamefont {Brooks}\ \bibnamefont {Foxen}}, \bibinfo {author}
		{\bibfnamefont {Craig}\ \bibnamefont {Gidney}}, \bibinfo {author}
		{\bibfnamefont {Marissa}\ \bibnamefont {Giustina}}, \bibinfo {author}
		{\bibfnamefont {Jonathan~A.}\ \bibnamefont {Gross}}, \bibinfo {author}
		{\bibfnamefont {Matthew~P.}\ \bibnamefont {Harrigan}}, \bibinfo {author}
		{\bibfnamefont {Sean~D.}\ \bibnamefont {Harrington}}, \bibinfo {author}
		{\bibfnamefont {Jeremy}\ \bibnamefont {Hilton}}, \bibinfo {author}
		{\bibfnamefont {Alan}\ \bibnamefont {Ho}}, \bibinfo {author} {\bibfnamefont
			{Sabrina}\ \bibnamefont {Hong}}, \bibinfo {author} {\bibfnamefont {Trent}\
			\bibnamefont {Huang}}, \bibinfo {author} {\bibfnamefont {William~J.}\
			\bibnamefont {Huggins}}, \bibinfo {author} {\bibfnamefont {L.~B.}\
			\bibnamefont {Ioffe}}, \bibinfo {author} {\bibfnamefont {Sergei~V.}\
			\bibnamefont {Isakov}}, \bibinfo {author} {\bibfnamefont {Evan}\ \bibnamefont
			{Jeffrey}}, \bibinfo {author} {\bibfnamefont {Zhang}\ \bibnamefont {Jiang}},
		\bibinfo {author} {\bibfnamefont {Cody}\ \bibnamefont {Jones}}, \bibinfo
		{author} {\bibfnamefont {Dvir}\ \bibnamefont {Kafri}}, \bibinfo {author}
		{\bibfnamefont {Julian}\ \bibnamefont {Kelly}}, \bibinfo {author}
		{\bibfnamefont {Seon}\ \bibnamefont {Kim}}, \bibinfo {author} {\bibfnamefont
			{Alexei}\ \bibnamefont {Kitaev}}, \bibinfo {author} {\bibfnamefont {Paul~V.}\
			\bibnamefont {Klimov}}, \bibinfo {author} {\bibfnamefont {Alexander~N.}\
			\bibnamefont {Korotkov}}, \bibinfo {author} {\bibfnamefont {Fedor}\
			\bibnamefont {Kostritsa}}, \bibinfo {author} {\bibfnamefont {David}\
			\bibnamefont {Landhuis}}, \bibinfo {author} {\bibfnamefont {Pavel}\
			\bibnamefont {Laptev}}, \bibinfo {author} {\bibfnamefont {Erik}\ \bibnamefont
			{Lucero}}, \bibinfo {author} {\bibfnamefont {Orion}\ \bibnamefont {Martin}},
		\bibinfo {author} {\bibfnamefont {Jarrod~R.}\ \bibnamefont {McClean}},
		\bibinfo {author} {\bibfnamefont {Trevor}\ \bibnamefont {McCourt}}, \bibinfo
		{author} {\bibfnamefont {Matt}\ \bibnamefont {McEwen}}, \bibinfo {author}
		{\bibfnamefont {Anthony}\ \bibnamefont {Megrant}}, \bibinfo {author}
		{\bibfnamefont {Kevin~C.}\ \bibnamefont {Miao}}, \bibinfo {author}
		{\bibfnamefont {Masoud}\ \bibnamefont {Mohseni}}, \bibinfo {author}
		{\bibfnamefont {Shirin}\ \bibnamefont {Montazeri}}, \bibinfo {author}
		{\bibfnamefont {Wojciech}\ \bibnamefont {Mruczkiewicz}}, \bibinfo {author}
		{\bibfnamefont {Josh}\ \bibnamefont {Mutus}}, \bibinfo {author}
		{\bibfnamefont {Ofer}\ \bibnamefont {Naaman}}, \bibinfo {author}
		{\bibfnamefont {Matthew}\ \bibnamefont {Neeley}}, \bibinfo {author}
		{\bibfnamefont {Michael}\ \bibnamefont {Newman}}, \bibinfo {author}
		{\bibfnamefont {Murphy~Yuezhen}\ \bibnamefont {Niu}}, \bibinfo {author}
		{\bibfnamefont {Thomas~E.}\ \bibnamefont {O'Brien}}, \bibinfo {author}
		{\bibfnamefont {Alex}\ \bibnamefont {Opremcak}}, \bibinfo {author}
		{\bibfnamefont {Eric}\ \bibnamefont {Ostby}}, \bibinfo {author}
		{\bibfnamefont {Balint}\ \bibnamefont {Pato}}, \bibinfo {author}
		{\bibfnamefont {Andre}\ \bibnamefont {Petukhov}}, \bibinfo {author}
		{\bibfnamefont {Nicholas}\ \bibnamefont {Redd}}, \bibinfo {author}
		{\bibfnamefont {Nicholas~C.}\ \bibnamefont {Rubin}}, \bibinfo {author}
		{\bibfnamefont {Daniel}\ \bibnamefont {Sank}}, \bibinfo {author}
		{\bibfnamefont {Kevin~J.}\ \bibnamefont {Satzinger}}, \bibinfo {author}
		{\bibfnamefont {Vladimir}\ \bibnamefont {Shvarts}}, \bibinfo {author}
		{\bibfnamefont {Doug}\ \bibnamefont {Strain}}, \bibinfo {author}
		{\bibfnamefont {Marco}\ \bibnamefont {Szalay}}, \bibinfo {author}
		{\bibfnamefont {Matthew~D.}\ \bibnamefont {Trevithick}}, \bibinfo {author}
		{\bibfnamefont {Benjamin}\ \bibnamefont {Villalonga}}, \bibinfo {author}
		{\bibfnamefont {Theodore}\ \bibnamefont {White}}, \bibinfo {author}
		{\bibfnamefont {Z.~Jamie}\ \bibnamefont {Yao}}, \bibinfo {author}
		{\bibfnamefont {Ping}\ \bibnamefont {Yeh}}, \bibinfo {author} {\bibfnamefont
			{Adam}\ \bibnamefont {Zalcman}}, \bibinfo {author} {\bibfnamefont {Hartmut}\
			\bibnamefont {Neven}}, \bibinfo {author} {\bibfnamefont {Igor}\ \bibnamefont
			{Aleiner}}, \bibinfo {author} {\bibfnamefont {Kostyantyn}\ \bibnamefont
			{Kechedzhi}}, \bibinfo {author} {\bibfnamefont {Vadim}\ \bibnamefont
			{Smelyanskiy}}, \ and\ \bibinfo {author} {\bibfnamefont {Yu}~\bibnamefont
			{Chen}},\ }\bibfield  {title} {\enquote {\bibinfo {title} {{Information
					scrambling in quantum circuits}},}\ }\href {\doibase 10.1126/science.abg5029}
	{\bibfield  {journal} {\bibinfo  {journal} {Science}\ }\textbf {\bibinfo
			{volume} {374}},\ \bibinfo {pages} {1479--1483} (\bibinfo {year}
		{2021})}\BibitemShut {NoStop}%
	\bibitem [{\citenamefont {Braum{\"{u}}ller}\ \emph {et~al.}(2022)\citenamefont
		{Braum{\"{u}}ller}, \citenamefont {Karamlou}, \citenamefont {Yanay},
		\citenamefont {Kannan}, \citenamefont {Kim}, \citenamefont {Kjaergaard},
		\citenamefont {Melville}, \citenamefont {Niedzielski}, \citenamefont {Sung},
		\citenamefont {Veps{\"{a}}l{\"{a}}inen}, \citenamefont {Winik}, \citenamefont
		{Yoder}, \citenamefont {Orlando}, \citenamefont {Gustavsson}, \citenamefont
		{Tahan},\ and\ \citenamefont {Oliver}}]{Braumuller:2022wj}%
	\BibitemOpen
	\bibfield  {author} {\bibinfo {author} {\bibfnamefont {Jochen}\ \bibnamefont
			{Braum{\"{u}}ller}}, \bibinfo {author} {\bibfnamefont {Amir~H.}\ \bibnamefont
			{Karamlou}}, \bibinfo {author} {\bibfnamefont {Yariv}\ \bibnamefont {Yanay}},
		\bibinfo {author} {\bibfnamefont {Bharath}\ \bibnamefont {Kannan}}, \bibinfo
		{author} {\bibfnamefont {David}\ \bibnamefont {Kim}}, \bibinfo {author}
		{\bibfnamefont {Morten}\ \bibnamefont {Kjaergaard}}, \bibinfo {author}
		{\bibfnamefont {Alexander}\ \bibnamefont {Melville}}, \bibinfo {author}
		{\bibfnamefont {Bethany~M.}\ \bibnamefont {Niedzielski}}, \bibinfo {author}
		{\bibfnamefont {Youngkyu}\ \bibnamefont {Sung}}, \bibinfo {author}
		{\bibfnamefont {Antti}\ \bibnamefont {Veps{\"{a}}l{\"{a}}inen}}, \bibinfo
		{author} {\bibfnamefont {Roni}\ \bibnamefont {Winik}}, \bibinfo {author}
		{\bibfnamefont {Jonilyn~L.}\ \bibnamefont {Yoder}}, \bibinfo {author}
		{\bibfnamefont {Terry~P.}\ \bibnamefont {Orlando}}, \bibinfo {author}
		{\bibfnamefont {Simon}\ \bibnamefont {Gustavsson}}, \bibinfo {author}
		{\bibfnamefont {Charles}\ \bibnamefont {Tahan}}, \ and\ \bibinfo {author}
		{\bibfnamefont {William~D.}\ \bibnamefont {Oliver}},\ }\bibfield  {title}
	{\enquote {\bibinfo {title} {{Probing quantum information propagation with
					out-of-time-ordered correlators}},}\ }\href {\doibase
		10.1038/s41567-021-01430-w} {\bibfield  {journal} {\bibinfo  {journal}
			{Nature Physics}\ }\textbf {\bibinfo {volume} {18}},\ \bibinfo {pages}
		{172--178} (\bibinfo {year} {2022})}\BibitemShut {NoStop}%
	\bibitem [{\citenamefont {Neill}\ \emph {et~al.}(2018)\citenamefont {Neill},
		\citenamefont {Roushan}, \citenamefont {Kechedzhi}, \citenamefont {Boixo},
		\citenamefont {Isakov}, \citenamefont {Smelyanskiy}, \citenamefont {Megrant},
		\citenamefont {Chiaro}, \citenamefont {Dunsworth}, \citenamefont {Arya},
		\citenamefont {Barends}, \citenamefont {Burkett}, \citenamefont {Chen},
		\citenamefont {Chen}, \citenamefont {Fowler}, \citenamefont {Foxen},
		\citenamefont {Giustina}, \citenamefont {Graff}, \citenamefont {Jeffrey},
		\citenamefont {Huang}, \citenamefont {Kelly}, \citenamefont {Klimov},
		\citenamefont {Lucero}, \citenamefont {Mutus}, \citenamefont {Neeley},
		\citenamefont {Quintana}, \citenamefont {Sank}, \citenamefont {Vainsencher},
		\citenamefont {Wenner}, \citenamefont {White}, \citenamefont {Neven},\ and\
		\citenamefont {Martinis}}]{Neill:2018wy}%
	\BibitemOpen
	\bibfield  {author} {\bibinfo {author} {\bibfnamefont {C.}~\bibnamefont
			{Neill}}, \bibinfo {author} {\bibfnamefont {P.}~\bibnamefont {Roushan}},
		\bibinfo {author} {\bibfnamefont {K.}~\bibnamefont {Kechedzhi}}, \bibinfo
		{author} {\bibfnamefont {S.}~\bibnamefont {Boixo}}, \bibinfo {author}
		{\bibfnamefont {S.~V.}\ \bibnamefont {Isakov}}, \bibinfo {author}
		{\bibfnamefont {V.}~\bibnamefont {Smelyanskiy}}, \bibinfo {author}
		{\bibfnamefont {A.}~\bibnamefont {Megrant}}, \bibinfo {author} {\bibfnamefont
			{B.}~\bibnamefont {Chiaro}}, \bibinfo {author} {\bibfnamefont
			{A.}~\bibnamefont {Dunsworth}}, \bibinfo {author} {\bibfnamefont
			{K.}~\bibnamefont {Arya}}, \bibinfo {author} {\bibfnamefont {R.}~\bibnamefont
			{Barends}}, \bibinfo {author} {\bibfnamefont {B.}~\bibnamefont {Burkett}},
		\bibinfo {author} {\bibfnamefont {Y.}~\bibnamefont {Chen}}, \bibinfo {author}
		{\bibfnamefont {Z.}~\bibnamefont {Chen}}, \bibinfo {author} {\bibfnamefont
			{A.}~\bibnamefont {Fowler}}, \bibinfo {author} {\bibfnamefont
			{B.}~\bibnamefont {Foxen}}, \bibinfo {author} {\bibfnamefont
			{M.}~\bibnamefont {Giustina}}, \bibinfo {author} {\bibfnamefont
			{R.}~\bibnamefont {Graff}}, \bibinfo {author} {\bibfnamefont
			{E.}~\bibnamefont {Jeffrey}}, \bibinfo {author} {\bibfnamefont
			{T.}~\bibnamefont {Huang}}, \bibinfo {author} {\bibfnamefont
			{J.}~\bibnamefont {Kelly}}, \bibinfo {author} {\bibfnamefont
			{P.}~\bibnamefont {Klimov}}, \bibinfo {author} {\bibfnamefont
			{E.}~\bibnamefont {Lucero}}, \bibinfo {author} {\bibfnamefont
			{J.}~\bibnamefont {Mutus}}, \bibinfo {author} {\bibfnamefont
			{M.}~\bibnamefont {Neeley}}, \bibinfo {author} {\bibfnamefont
			{C.}~\bibnamefont {Quintana}}, \bibinfo {author} {\bibfnamefont
			{D.}~\bibnamefont {Sank}}, \bibinfo {author} {\bibfnamefont {A.}~\bibnamefont
			{Vainsencher}}, \bibinfo {author} {\bibfnamefont {J.}~\bibnamefont {Wenner}},
		\bibinfo {author} {\bibfnamefont {T.~C.}\ \bibnamefont {White}}, \bibinfo
		{author} {\bibfnamefont {H.}~\bibnamefont {Neven}}, \ and\ \bibinfo {author}
		{\bibfnamefont {J.~M.}\ \bibnamefont {Martinis}},\ }\bibfield  {title}
	{\enquote {\bibinfo {title} {{A blueprint for demonstrating quantum supremacy
					with superconducting qubits}},}\ }\href {\doibase 10.1126/science.aao4309}
	{\bibfield  {journal} {\bibinfo  {journal} {Science}\ }\textbf {\bibinfo
			{volume} {360}},\ \bibinfo {pages} {195--199} (\bibinfo {year}
		{2018})}\BibitemShut {NoStop}%
	\bibitem [{\citenamefont {Boixo}\ \emph {et~al.}(2018)\citenamefont {Boixo},
		\citenamefont {Isakov}, \citenamefont {Smelyanskiy}, \citenamefont {Babbush},
		\citenamefont {Ding}, \citenamefont {Jiang}, \citenamefont {Bremner},
		\citenamefont {Martinis},\ and\ \citenamefont {Neven}}]{Boixo:2018un}%
	\BibitemOpen
	\bibfield  {author} {\bibinfo {author} {\bibfnamefont {Sergio}\ \bibnamefont
			{Boixo}}, \bibinfo {author} {\bibfnamefont {Sergei~V.}\ \bibnamefont
			{Isakov}}, \bibinfo {author} {\bibfnamefont {Vadim~N.}\ \bibnamefont
			{Smelyanskiy}}, \bibinfo {author} {\bibfnamefont {Ryan}\ \bibnamefont
			{Babbush}}, \bibinfo {author} {\bibfnamefont {Nan}\ \bibnamefont {Ding}},
		\bibinfo {author} {\bibfnamefont {Zhang}\ \bibnamefont {Jiang}}, \bibinfo
		{author} {\bibfnamefont {Michael~J.}\ \bibnamefont {Bremner}}, \bibinfo
		{author} {\bibfnamefont {John~M.}\ \bibnamefont {Martinis}}, \ and\ \bibinfo
		{author} {\bibfnamefont {Hartmut}\ \bibnamefont {Neven}},\ }\bibfield
	{title} {\enquote {\bibinfo {title} {{Characterizing quantum supremacy in
					near-term devices}},}\ }\href {\doibase 10.1038/s41567-018-0124-x} {\bibfield
		{journal} {\bibinfo  {journal} {Nature Physics}\ }\textbf {\bibinfo {volume}
			{14}},\ \bibinfo {pages} {595--600} (\bibinfo {year} {2018})}\BibitemShut
	{NoStop}%
	\bibitem [{\citenamefont {Arute~\emph {et~al.}}(2019)}]{Arute:2019ts}%
	\BibitemOpen
	\bibfield  {author} {\bibinfo {author} {\bibfnamefont {F.}~\bibnamefont
			{Arute~\emph {et~al.}}},\ }\bibfield  {title} {\enquote {\bibinfo {title}
			{Quantum supremacy using a programmable superconducting processor},}\ }\href
	{\doibase 10.1038/s41586-019-1666-5} {\bibfield  {journal} {\bibinfo
			{journal} {Nature}\ }\textbf {\bibinfo {volume} {574}},\ \bibinfo {pages}
		{505--510} (\bibinfo {year} {2019})}\BibitemShut {NoStop}%
	\bibitem [{\citenamefont {Wu}\ \emph {et~al.}(2021)\citenamefont {Wu},
		\citenamefont {Bao}, \citenamefont {Cao}, \citenamefont {Chen}, \citenamefont
		{Chen}, \citenamefont {Chen}, \citenamefont {Chung}, \citenamefont {Deng},
		\citenamefont {Du}, \citenamefont {Fan}, \citenamefont {Gong}, \citenamefont
		{Guo}, \citenamefont {Guo}, \citenamefont {Guo}, \citenamefont {Han},
		\citenamefont {Hong}, \citenamefont {Huang}, \citenamefont {Huo},
		\citenamefont {Li}, \citenamefont {Li}, \citenamefont {Li}, \citenamefont
		{Li}, \citenamefont {Liang}, \citenamefont {Lin}, \citenamefont {Lin},
		\citenamefont {Qian}, \citenamefont {Qiao}, \citenamefont {Rong},
		\citenamefont {Su}, \citenamefont {Sun}, \citenamefont {Wang}, \citenamefont
		{Wang}, \citenamefont {Wu}, \citenamefont {Xu}, \citenamefont {Yan},
		\citenamefont {Yang}, \citenamefont {Yang}, \citenamefont {Ye}, \citenamefont
		{Yin}, \citenamefont {Ying}, \citenamefont {Yu}, \citenamefont {Zha},
		\citenamefont {Zhang}, \citenamefont {Zhang}, \citenamefont {Zhang},
		\citenamefont {Zhang}, \citenamefont {Zhao}, \citenamefont {Zhao},
		\citenamefont {Zhou}, \citenamefont {Zhu}, \citenamefont {Lu}, \citenamefont
		{Peng}, \citenamefont {Zhu},\ and\ \citenamefont
		{Pan}}]{PhysRevLett.127.180501}%
	\BibitemOpen
	\bibfield  {author} {\bibinfo {author} {\bibfnamefont {Yulin}\ \bibnamefont
			{Wu}}, \bibinfo {author} {\bibfnamefont {Wan-Su}\ \bibnamefont {Bao}},
		\bibinfo {author} {\bibfnamefont {Sirui}\ \bibnamefont {Cao}}, \bibinfo
		{author} {\bibfnamefont {Fusheng}\ \bibnamefont {Chen}}, \bibinfo {author}
		{\bibfnamefont {Ming-Cheng}\ \bibnamefont {Chen}}, \bibinfo {author}
		{\bibfnamefont {Xiawei}\ \bibnamefont {Chen}}, \bibinfo {author}
		{\bibfnamefont {Tung-Hsun}\ \bibnamefont {Chung}}, \bibinfo {author}
		{\bibfnamefont {Hui}\ \bibnamefont {Deng}}, \bibinfo {author} {\bibfnamefont
			{Yajie}\ \bibnamefont {Du}}, \bibinfo {author} {\bibfnamefont {Daojin}\
			\bibnamefont {Fan}}, \bibinfo {author} {\bibfnamefont {Ming}\ \bibnamefont
			{Gong}}, \bibinfo {author} {\bibfnamefont {Cheng}\ \bibnamefont {Guo}},
		\bibinfo {author} {\bibfnamefont {Chu}\ \bibnamefont {Guo}}, \bibinfo
		{author} {\bibfnamefont {Shaojun}\ \bibnamefont {Guo}}, \bibinfo {author}
		{\bibfnamefont {Lianchen}\ \bibnamefont {Han}}, \bibinfo {author}
		{\bibfnamefont {Linyin}\ \bibnamefont {Hong}}, \bibinfo {author}
		{\bibfnamefont {He-Liang}\ \bibnamefont {Huang}}, \bibinfo {author}
		{\bibfnamefont {Yong-Heng}\ \bibnamefont {Huo}}, \bibinfo {author}
		{\bibfnamefont {Liping}\ \bibnamefont {Li}}, \bibinfo {author} {\bibfnamefont
			{Na}~\bibnamefont {Li}}, \bibinfo {author} {\bibfnamefont {Shaowei}\
			\bibnamefont {Li}}, \bibinfo {author} {\bibfnamefont {Yuan}\ \bibnamefont
			{Li}}, \bibinfo {author} {\bibfnamefont {Futian}\ \bibnamefont {Liang}},
		\bibinfo {author} {\bibfnamefont {Chun}\ \bibnamefont {Lin}}, \bibinfo
		{author} {\bibfnamefont {Jin}\ \bibnamefont {Lin}}, \bibinfo {author}
		{\bibfnamefont {Haoran}\ \bibnamefont {Qian}}, \bibinfo {author}
		{\bibfnamefont {Dan}\ \bibnamefont {Qiao}}, \bibinfo {author} {\bibfnamefont
			{Hao}\ \bibnamefont {Rong}}, \bibinfo {author} {\bibfnamefont {Hong}\
			\bibnamefont {Su}}, \bibinfo {author} {\bibfnamefont {Lihua}\ \bibnamefont
			{Sun}}, \bibinfo {author} {\bibfnamefont {Liangyuan}\ \bibnamefont {Wang}},
		\bibinfo {author} {\bibfnamefont {Shiyu}\ \bibnamefont {Wang}}, \bibinfo
		{author} {\bibfnamefont {Dachao}\ \bibnamefont {Wu}}, \bibinfo {author}
		{\bibfnamefont {Yu}~\bibnamefont {Xu}}, \bibinfo {author} {\bibfnamefont
			{Kai}\ \bibnamefont {Yan}}, \bibinfo {author} {\bibfnamefont {Weifeng}\
			\bibnamefont {Yang}}, \bibinfo {author} {\bibfnamefont {Yang}\ \bibnamefont
			{Yang}}, \bibinfo {author} {\bibfnamefont {Yangsen}\ \bibnamefont {Ye}},
		\bibinfo {author} {\bibfnamefont {Jianghan}\ \bibnamefont {Yin}}, \bibinfo
		{author} {\bibfnamefont {Chong}\ \bibnamefont {Ying}}, \bibinfo {author}
		{\bibfnamefont {Jiale}\ \bibnamefont {Yu}}, \bibinfo {author} {\bibfnamefont
			{Chen}\ \bibnamefont {Zha}}, \bibinfo {author} {\bibfnamefont {Cha}\
			\bibnamefont {Zhang}}, \bibinfo {author} {\bibfnamefont {Haibin}\
			\bibnamefont {Zhang}}, \bibinfo {author} {\bibfnamefont {Kaili}\ \bibnamefont
			{Zhang}}, \bibinfo {author} {\bibfnamefont {Yiming}\ \bibnamefont {Zhang}},
		\bibinfo {author} {\bibfnamefont {Han}\ \bibnamefont {Zhao}}, \bibinfo
		{author} {\bibfnamefont {Youwei}\ \bibnamefont {Zhao}}, \bibinfo {author}
		{\bibfnamefont {Liang}\ \bibnamefont {Zhou}}, \bibinfo {author}
		{\bibfnamefont {Qingling}\ \bibnamefont {Zhu}}, \bibinfo {author}
		{\bibfnamefont {Chao-Yang}\ \bibnamefont {Lu}}, \bibinfo {author}
		{\bibfnamefont {Cheng-Zhi}\ \bibnamefont {Peng}}, \bibinfo {author}
		{\bibfnamefont {Xiaobo}\ \bibnamefont {Zhu}}, \ and\ \bibinfo {author}
		{\bibfnamefont {Jian-Wei}\ \bibnamefont {Pan}},\ }\bibfield  {title}
	{\enquote {\bibinfo {title} {{Strong Quantum Computational Advantage Using a
					Superconducting Quantum Processor}},}\ }\href {\doibase
		10.1103/PhysRevLett.127.180501} {\bibfield  {journal} {\bibinfo  {journal}
			{Phys. Rev. Lett.}\ }\textbf {\bibinfo {volume} {127}},\ \bibinfo {pages}
		{180501} (\bibinfo {year} {2021})}\BibitemShut {NoStop}%
	\bibitem [{\citenamefont {Morvan}\ \emph {et~al.}(2023)\citenamefont {Morvan},
		\citenamefont {Villalonga}, \citenamefont {Mi}, \citenamefont {Mandr{\`{a}}},
		\citenamefont {Bengtsson}, \citenamefont {Klimov}, \citenamefont {Chen},
		\citenamefont {Hong}, \citenamefont {Erickson}, \citenamefont {Drozdov},
		\citenamefont {Chau}, \citenamefont {Laun}, \citenamefont {Movassagh},
		\citenamefont {Asfaw}, \citenamefont {Brand{\~{a}}o}, \citenamefont
		{Peralta}, \citenamefont {Abanin}, \citenamefont {Acharya}, \citenamefont
		{Allen}, \citenamefont {Andersen}, \citenamefont {Anderson}, \citenamefont
		{Ansmann}, \citenamefont {Arute}, \citenamefont {Arya}, \citenamefont
		{Atalaya}, \citenamefont {Bardin}, \citenamefont {Bilmes}, \citenamefont
		{Bortoli}, \citenamefont {Bourassa}, \citenamefont {Bovaird}, \citenamefont
		{Brill}, \citenamefont {Broughton}, \citenamefont {Buckley}, \citenamefont
		{Buell}, \citenamefont {Burger}, \citenamefont {Burkett}, \citenamefont
		{Bushnell}, \citenamefont {Campero}, \citenamefont {Chang}, \citenamefont
		{Chiaro}, \citenamefont {Chik}, \citenamefont {Chou}, \citenamefont {Cogan},
		\citenamefont {Collins}, \citenamefont {Conner}, \citenamefont {Courtney},
		\citenamefont {Crook}, \citenamefont {Curtin}, \citenamefont {Debroy},
		\citenamefont {Barba}, \citenamefont {Demura}, \citenamefont {{Di Paolo}},
		\citenamefont {Dunsworth}, \citenamefont {Faoro}, \citenamefont {Farhi},
		\citenamefont {Fatemi}, \citenamefont {Ferreira}, \citenamefont {Burgos},
		\citenamefont {Forati}, \citenamefont {Fowler}, \citenamefont {Foxen},
		\citenamefont {Garcia}, \citenamefont {Genois}, \citenamefont {Giang},
		\citenamefont {Gidney}, \citenamefont {Gilboa}, \citenamefont {Giustina},
		\citenamefont {Gosula}, \citenamefont {Dau}, \citenamefont {Gross},
		\citenamefont {Habegger}, \citenamefont {Hamilton}, \citenamefont {Hansen},
		\citenamefont {Harrigan}, \citenamefont {Harrington}, \citenamefont {Heu},
		\citenamefont {Hoffmann}, \citenamefont {Huang}, \citenamefont {Huff},
		\citenamefont {Huggins}, \citenamefont {Ioffe}, \citenamefont {Isakov},
		\citenamefont {Iveland}, \citenamefont {Jeffrey}, \citenamefont {Jiang},
		\citenamefont {Jones}, \citenamefont {Juhas}, \citenamefont {Kafri},
		\citenamefont {Khattar}, \citenamefont {Khezri}, \citenamefont
		{Kieferov{\'{a}}}, \citenamefont {Kim}, \citenamefont {Kitaev}, \citenamefont
		{Klots}, \citenamefont {Korotkov}, \citenamefont {Kostritsa}, \citenamefont
		{Kreikebaum}, \citenamefont {Landhuis}, \citenamefont {Laptev}, \citenamefont
		{Lau}, \citenamefont {Laws}, \citenamefont {Lee}, \citenamefont {Lee},
		\citenamefont {Lensky}, \citenamefont {Lester}, \citenamefont {Lill},
		\citenamefont {Liu}, \citenamefont {Livingston}, \citenamefont {Locharla},
		\citenamefont {Malone}, \citenamefont {Martin}, \citenamefont {Martin},
		\citenamefont {McClean}, \citenamefont {McEwen}, \citenamefont {Miao},
		\citenamefont {Mieszala}, \citenamefont {Montazeri}, \citenamefont
		{Mruczkiewicz}, \citenamefont {Naaman}, \citenamefont {Neeley}, \citenamefont
		{Neill}, \citenamefont {Nersisyan}, \citenamefont {Newman}, \citenamefont
		{Ng}, \citenamefont {Nguyen}, \citenamefont {Nguyen}, \citenamefont {Niu},
		\citenamefont {O'Brien}, \citenamefont {Omonije}, \citenamefont {Opremcak},
		\citenamefont {Petukhov}, \citenamefont {Potter}, \citenamefont {Pryadko},
		\citenamefont {Quintana}, \citenamefont {Rhodes}, \citenamefont {Rosenberg},
		\citenamefont {Rocque}, \citenamefont {Roushan}, \citenamefont {Rubin},
		\citenamefont {Saei}, \citenamefont {Sank}, \citenamefont {Sankaragomathi},
		\citenamefont {Satzinger}, \citenamefont {Schurkus}, \citenamefont
		{Schuster}, \citenamefont {Shearn}, \citenamefont {Shorter}, \citenamefont
		{Shutty}, \citenamefont {Shvarts}, \citenamefont {Sivak}, \citenamefont
		{Skruzny}, \citenamefont {Smith}, \citenamefont {Somma}, \citenamefont
		{Sterling}, \citenamefont {Strain}, \citenamefont {Szalay}, \citenamefont
		{Thor}, \citenamefont {Torres}, \citenamefont {Vidal}, \citenamefont
		{Heidweiller}, \citenamefont {White}, \citenamefont {Woo}, \citenamefont
		{Xing}, \citenamefont {Yao}, \citenamefont {Yeh}, \citenamefont {Yoo},
		\citenamefont {Young}, \citenamefont {Zalcman}, \citenamefont {Zhang},
		\citenamefont {Zhu}, \citenamefont {Zobrist}, \citenamefont {Rieffel},
		\citenamefont {Biswas}, \citenamefont {Babbush}, \citenamefont {Bacon},
		\citenamefont {Hilton}, \citenamefont {Lucero}, \citenamefont {Neven},
		\citenamefont {Megrant}, \citenamefont {Kelly}, \citenamefont {Aleiner},
		\citenamefont {Smelyanskiy}, \citenamefont {Kechedzhi}, \citenamefont
		{Chen},\ and\ \citenamefont {Boixo}}]{2023arXiv230411119M}%
	\BibitemOpen
	\bibfield  {author} {\bibinfo {author} {\bibfnamefont {A.}~\bibnamefont
			{Morvan}}, \bibinfo {author} {\bibfnamefont {B.}~\bibnamefont {Villalonga}},
		\bibinfo {author} {\bibfnamefont {X.}~\bibnamefont {Mi}}, \bibinfo {author}
		{\bibfnamefont {S.}~\bibnamefont {Mandr{\`{a}}}}, \bibinfo {author}
		{\bibfnamefont {A.}~\bibnamefont {Bengtsson}}, \bibinfo {author}
		{\bibfnamefont {P.~V.}\ \bibnamefont {Klimov}}, \bibinfo {author}
		{\bibfnamefont {Z.}~\bibnamefont {Chen}}, \bibinfo {author} {\bibfnamefont
			{S.}~\bibnamefont {Hong}}, \bibinfo {author} {\bibfnamefont {C.}~\bibnamefont
			{Erickson}}, \bibinfo {author} {\bibfnamefont {I.~K.}\ \bibnamefont
			{Drozdov}}, \bibinfo {author} {\bibfnamefont {J.}~\bibnamefont {Chau}},
		\bibinfo {author} {\bibfnamefont {G.}~\bibnamefont {Laun}}, \bibinfo {author}
		{\bibfnamefont {R.}~\bibnamefont {Movassagh}}, \bibinfo {author}
		{\bibfnamefont {A.}~\bibnamefont {Asfaw}}, \bibinfo {author} {\bibfnamefont
			{L.~T. A.~N.}\ \bibnamefont {Brand{\~{a}}o}}, \bibinfo {author}
		{\bibfnamefont {R.}~\bibnamefont {Peralta}}, \bibinfo {author} {\bibfnamefont
			{D.}~\bibnamefont {Abanin}}, \bibinfo {author} {\bibfnamefont
			{R.}~\bibnamefont {Acharya}}, \bibinfo {author} {\bibfnamefont
			{R.}~\bibnamefont {Allen}}, \bibinfo {author} {\bibfnamefont {T.~I.}\
			\bibnamefont {Andersen}}, \bibinfo {author} {\bibfnamefont {K.}~\bibnamefont
			{Anderson}}, \bibinfo {author} {\bibfnamefont {M.}~\bibnamefont {Ansmann}},
		\bibinfo {author} {\bibfnamefont {F.}~\bibnamefont {Arute}}, \bibinfo
		{author} {\bibfnamefont {K.}~\bibnamefont {Arya}}, \bibinfo {author}
		{\bibfnamefont {J.}~\bibnamefont {Atalaya}}, \bibinfo {author} {\bibfnamefont
			{J.~C.}\ \bibnamefont {Bardin}}, \bibinfo {author} {\bibfnamefont
			{A.}~\bibnamefont {Bilmes}}, \bibinfo {author} {\bibfnamefont
			{G.}~\bibnamefont {Bortoli}}, \bibinfo {author} {\bibfnamefont
			{A.}~\bibnamefont {Bourassa}}, \bibinfo {author} {\bibfnamefont
			{J.}~\bibnamefont {Bovaird}}, \bibinfo {author} {\bibfnamefont
			{L.}~\bibnamefont {Brill}}, \bibinfo {author} {\bibfnamefont
			{M.}~\bibnamefont {Broughton}}, \bibinfo {author} {\bibfnamefont {B.~B.}\
			\bibnamefont {Buckley}}, \bibinfo {author} {\bibfnamefont {D.~A.}\
			\bibnamefont {Buell}}, \bibinfo {author} {\bibfnamefont {T.}~\bibnamefont
			{Burger}}, \bibinfo {author} {\bibfnamefont {B.}~\bibnamefont {Burkett}},
		\bibinfo {author} {\bibfnamefont {N.}~\bibnamefont {Bushnell}}, \bibinfo
		{author} {\bibfnamefont {J.}~\bibnamefont {Campero}}, \bibinfo {author}
		{\bibfnamefont {H.~S.}\ \bibnamefont {Chang}}, \bibinfo {author}
		{\bibfnamefont {B.}~\bibnamefont {Chiaro}}, \bibinfo {author} {\bibfnamefont
			{D.}~\bibnamefont {Chik}}, \bibinfo {author} {\bibfnamefont {C.}~\bibnamefont
			{Chou}}, \bibinfo {author} {\bibfnamefont {J.}~\bibnamefont {Cogan}},
		\bibinfo {author} {\bibfnamefont {R.}~\bibnamefont {Collins}}, \bibinfo
		{author} {\bibfnamefont {P.}~\bibnamefont {Conner}}, \bibinfo {author}
		{\bibfnamefont {W.}~\bibnamefont {Courtney}}, \bibinfo {author}
		{\bibfnamefont {A.~L.}\ \bibnamefont {Crook}}, \bibinfo {author}
		{\bibfnamefont {B.}~\bibnamefont {Curtin}}, \bibinfo {author} {\bibfnamefont
			{D.~M.}\ \bibnamefont {Debroy}}, \bibinfo {author} {\bibfnamefont
			{A.~Del~Toro}\ \bibnamefont {Barba}}, \bibinfo {author} {\bibfnamefont
			{S.}~\bibnamefont {Demura}}, \bibinfo {author} {\bibfnamefont
			{A.}~\bibnamefont {{Di Paolo}}}, \bibinfo {author} {\bibfnamefont
			{A.}~\bibnamefont {Dunsworth}}, \bibinfo {author} {\bibfnamefont
			{L.}~\bibnamefont {Faoro}}, \bibinfo {author} {\bibfnamefont
			{E.}~\bibnamefont {Farhi}}, \bibinfo {author} {\bibfnamefont
			{R.}~\bibnamefont {Fatemi}}, \bibinfo {author} {\bibfnamefont {V.~S.}\
			\bibnamefont {Ferreira}}, \bibinfo {author} {\bibfnamefont {L.~Flores}\
			\bibnamefont {Burgos}}, \bibinfo {author} {\bibfnamefont {E.}~\bibnamefont
			{Forati}}, \bibinfo {author} {\bibfnamefont {A.~G.}\ \bibnamefont {Fowler}},
		\bibinfo {author} {\bibfnamefont {B.}~\bibnamefont {Foxen}}, \bibinfo
		{author} {\bibfnamefont {G.}~\bibnamefont {Garcia}}, \bibinfo {author}
		{\bibfnamefont {E.}~\bibnamefont {Genois}}, \bibinfo {author} {\bibfnamefont
			{W.}~\bibnamefont {Giang}}, \bibinfo {author} {\bibfnamefont
			{C.}~\bibnamefont {Gidney}}, \bibinfo {author} {\bibfnamefont
			{D.}~\bibnamefont {Gilboa}}, \bibinfo {author} {\bibfnamefont
			{M.}~\bibnamefont {Giustina}}, \bibinfo {author} {\bibfnamefont
			{R.}~\bibnamefont {Gosula}}, \bibinfo {author} {\bibfnamefont {A.~Grajales}\
			\bibnamefont {Dau}}, \bibinfo {author} {\bibfnamefont {J.~A.}\ \bibnamefont
			{Gross}}, \bibinfo {author} {\bibfnamefont {S.}~\bibnamefont {Habegger}},
		\bibinfo {author} {\bibfnamefont {M.~C.}\ \bibnamefont {Hamilton}}, \bibinfo
		{author} {\bibfnamefont {M.}~\bibnamefont {Hansen}}, \bibinfo {author}
		{\bibfnamefont {M.~P.}\ \bibnamefont {Harrigan}}, \bibinfo {author}
		{\bibfnamefont {S.~D.}\ \bibnamefont {Harrington}}, \bibinfo {author}
		{\bibfnamefont {P.}~\bibnamefont {Heu}}, \bibinfo {author} {\bibfnamefont
			{M.~R.}\ \bibnamefont {Hoffmann}}, \bibinfo {author} {\bibfnamefont
			{T.}~\bibnamefont {Huang}}, \bibinfo {author} {\bibfnamefont
			{A.}~\bibnamefont {Huff}}, \bibinfo {author} {\bibfnamefont {W.~J.}\
			\bibnamefont {Huggins}}, \bibinfo {author} {\bibfnamefont {L.~B.}\
			\bibnamefont {Ioffe}}, \bibinfo {author} {\bibfnamefont {S.~V.}\ \bibnamefont
			{Isakov}}, \bibinfo {author} {\bibfnamefont {J.}~\bibnamefont {Iveland}},
		\bibinfo {author} {\bibfnamefont {E.}~\bibnamefont {Jeffrey}}, \bibinfo
		{author} {\bibfnamefont {Z.}~\bibnamefont {Jiang}}, \bibinfo {author}
		{\bibfnamefont {C.}~\bibnamefont {Jones}}, \bibinfo {author} {\bibfnamefont
			{P.}~\bibnamefont {Juhas}}, \bibinfo {author} {\bibfnamefont
			{D.}~\bibnamefont {Kafri}}, \bibinfo {author} {\bibfnamefont
			{T.}~\bibnamefont {Khattar}}, \bibinfo {author} {\bibfnamefont
			{M.}~\bibnamefont {Khezri}}, \bibinfo {author} {\bibfnamefont
			{M.}~\bibnamefont {Kieferov{\'{a}}}}, \bibinfo {author} {\bibfnamefont
			{S.}~\bibnamefont {Kim}}, \bibinfo {author} {\bibfnamefont {A.}~\bibnamefont
			{Kitaev}}, \bibinfo {author} {\bibfnamefont {A.~R.}\ \bibnamefont {Klots}},
		\bibinfo {author} {\bibfnamefont {A.~N.}\ \bibnamefont {Korotkov}}, \bibinfo
		{author} {\bibfnamefont {F.}~\bibnamefont {Kostritsa}}, \bibinfo {author}
		{\bibfnamefont {J.~M.}\ \bibnamefont {Kreikebaum}}, \bibinfo {author}
		{\bibfnamefont {D.}~\bibnamefont {Landhuis}}, \bibinfo {author}
		{\bibfnamefont {P.}~\bibnamefont {Laptev}}, \bibinfo {author} {\bibfnamefont
			{K.~M.}\ \bibnamefont {Lau}}, \bibinfo {author} {\bibfnamefont
			{L.}~\bibnamefont {Laws}}, \bibinfo {author} {\bibfnamefont {J.}~\bibnamefont
			{Lee}}, \bibinfo {author} {\bibfnamefont {K.~W.}\ \bibnamefont {Lee}},
		\bibinfo {author} {\bibfnamefont {Y.~D.}\ \bibnamefont {Lensky}}, \bibinfo
		{author} {\bibfnamefont {B.~J.}\ \bibnamefont {Lester}}, \bibinfo {author}
		{\bibfnamefont {A.~T.}\ \bibnamefont {Lill}}, \bibinfo {author}
		{\bibfnamefont {W.}~\bibnamefont {Liu}}, \bibinfo {author} {\bibfnamefont
			{W.~P.}\ \bibnamefont {Livingston}}, \bibinfo {author} {\bibfnamefont
			{A.}~\bibnamefont {Locharla}}, \bibinfo {author} {\bibfnamefont {F.~D.}\
			\bibnamefont {Malone}}, \bibinfo {author} {\bibfnamefont {O.}~\bibnamefont
			{Martin}}, \bibinfo {author} {\bibfnamefont {S.}~\bibnamefont {Martin}},
		\bibinfo {author} {\bibfnamefont {J.~R.}\ \bibnamefont {McClean}}, \bibinfo
		{author} {\bibfnamefont {M.}~\bibnamefont {McEwen}}, \bibinfo {author}
		{\bibfnamefont {K.~C.}\ \bibnamefont {Miao}}, \bibinfo {author}
		{\bibfnamefont {A.}~\bibnamefont {Mieszala}}, \bibinfo {author}
		{\bibfnamefont {S.}~\bibnamefont {Montazeri}}, \bibinfo {author}
		{\bibfnamefont {W.}~\bibnamefont {Mruczkiewicz}}, \bibinfo {author}
		{\bibfnamefont {O.}~\bibnamefont {Naaman}}, \bibinfo {author} {\bibfnamefont
			{M.}~\bibnamefont {Neeley}}, \bibinfo {author} {\bibfnamefont
			{C.}~\bibnamefont {Neill}}, \bibinfo {author} {\bibfnamefont
			{A.}~\bibnamefont {Nersisyan}}, \bibinfo {author} {\bibfnamefont
			{M.}~\bibnamefont {Newman}}, \bibinfo {author} {\bibfnamefont {J.~H.}\
			\bibnamefont {Ng}}, \bibinfo {author} {\bibfnamefont {A.}~\bibnamefont
			{Nguyen}}, \bibinfo {author} {\bibfnamefont {M.}~\bibnamefont {Nguyen}},
		\bibinfo {author} {\bibfnamefont {M.~Yuezhen}\ \bibnamefont {Niu}}, \bibinfo
		{author} {\bibfnamefont {T.~E.}\ \bibnamefont {O'Brien}}, \bibinfo {author}
		{\bibfnamefont {S.}~\bibnamefont {Omonije}}, \bibinfo {author} {\bibfnamefont
			{A.}~\bibnamefont {Opremcak}}, \bibinfo {author} {\bibfnamefont
			{A.}~\bibnamefont {Petukhov}}, \bibinfo {author} {\bibfnamefont
			{R.}~\bibnamefont {Potter}}, \bibinfo {author} {\bibfnamefont {L.~P.}\
			\bibnamefont {Pryadko}}, \bibinfo {author} {\bibfnamefont {C.}~\bibnamefont
			{Quintana}}, \bibinfo {author} {\bibfnamefont {D.~M.}\ \bibnamefont
			{Rhodes}}, \bibinfo {author} {\bibfnamefont {E.}~\bibnamefont {Rosenberg}},
		\bibinfo {author} {\bibfnamefont {C.}~\bibnamefont {Rocque}}, \bibinfo
		{author} {\bibfnamefont {P.}~\bibnamefont {Roushan}}, \bibinfo {author}
		{\bibfnamefont {N.~C.}\ \bibnamefont {Rubin}}, \bibinfo {author}
		{\bibfnamefont {N.}~\bibnamefont {Saei}}, \bibinfo {author} {\bibfnamefont
			{D.}~\bibnamefont {Sank}}, \bibinfo {author} {\bibfnamefont {K.}~\bibnamefont
			{Sankaragomathi}}, \bibinfo {author} {\bibfnamefont {K.~J.}\ \bibnamefont
			{Satzinger}}, \bibinfo {author} {\bibfnamefont {H.~F.}\ \bibnamefont
			{Schurkus}}, \bibinfo {author} {\bibfnamefont {C.}~\bibnamefont {Schuster}},
		\bibinfo {author} {\bibfnamefont {M.~J.}\ \bibnamefont {Shearn}}, \bibinfo
		{author} {\bibfnamefont {A.}~\bibnamefont {Shorter}}, \bibinfo {author}
		{\bibfnamefont {N.}~\bibnamefont {Shutty}}, \bibinfo {author} {\bibfnamefont
			{V.}~\bibnamefont {Shvarts}}, \bibinfo {author} {\bibfnamefont
			{V.}~\bibnamefont {Sivak}}, \bibinfo {author} {\bibfnamefont
			{J.}~\bibnamefont {Skruzny}}, \bibinfo {author} {\bibfnamefont {W.~C.}\
			\bibnamefont {Smith}}, \bibinfo {author} {\bibfnamefont {R.~D.}\ \bibnamefont
			{Somma}}, \bibinfo {author} {\bibfnamefont {G.}~\bibnamefont {Sterling}},
		\bibinfo {author} {\bibfnamefont {D.}~\bibnamefont {Strain}}, \bibinfo
		{author} {\bibfnamefont {M.}~\bibnamefont {Szalay}}, \bibinfo {author}
		{\bibfnamefont {D.}~\bibnamefont {Thor}}, \bibinfo {author} {\bibfnamefont
			{A.}~\bibnamefont {Torres}}, \bibinfo {author} {\bibfnamefont
			{G.}~\bibnamefont {Vidal}}, \bibinfo {author} {\bibfnamefont {C.~Vollgraff}\
			\bibnamefont {Heidweiller}}, \bibinfo {author} {\bibfnamefont
			{T.}~\bibnamefont {White}}, \bibinfo {author} {\bibfnamefont {B.~W.~K.}\
			\bibnamefont {Woo}}, \bibinfo {author} {\bibfnamefont {C.}~\bibnamefont
			{Xing}}, \bibinfo {author} {\bibfnamefont {Z.~J.}\ \bibnamefont {Yao}},
		\bibinfo {author} {\bibfnamefont {P.}~\bibnamefont {Yeh}}, \bibinfo {author}
		{\bibfnamefont {J.}~\bibnamefont {Yoo}}, \bibinfo {author} {\bibfnamefont
			{G.}~\bibnamefont {Young}}, \bibinfo {author} {\bibfnamefont
			{A.}~\bibnamefont {Zalcman}}, \bibinfo {author} {\bibfnamefont
			{Y.}~\bibnamefont {Zhang}}, \bibinfo {author} {\bibfnamefont
			{N.}~\bibnamefont {Zhu}}, \bibinfo {author} {\bibfnamefont {N.}~\bibnamefont
			{Zobrist}}, \bibinfo {author} {\bibfnamefont {E.~G.}\ \bibnamefont
			{Rieffel}}, \bibinfo {author} {\bibfnamefont {R.}~\bibnamefont {Biswas}},
		\bibinfo {author} {\bibfnamefont {R.}~\bibnamefont {Babbush}}, \bibinfo
		{author} {\bibfnamefont {D.}~\bibnamefont {Bacon}}, \bibinfo {author}
		{\bibfnamefont {J.}~\bibnamefont {Hilton}}, \bibinfo {author} {\bibfnamefont
			{E.}~\bibnamefont {Lucero}}, \bibinfo {author} {\bibfnamefont
			{H.}~\bibnamefont {Neven}}, \bibinfo {author} {\bibfnamefont
			{A.}~\bibnamefont {Megrant}}, \bibinfo {author} {\bibfnamefont
			{J.}~\bibnamefont {Kelly}}, \bibinfo {author} {\bibfnamefont
			{I.}~\bibnamefont {Aleiner}}, \bibinfo {author} {\bibfnamefont
			{V.}~\bibnamefont {Smelyanskiy}}, \bibinfo {author} {\bibfnamefont
			{K.}~\bibnamefont {Kechedzhi}}, \bibinfo {author} {\bibfnamefont
			{Y.}~\bibnamefont {Chen}}, \ and\ \bibinfo {author} {\bibfnamefont
			{S.}~\bibnamefont {Boixo}},\ }\bibfield  {title} {\enquote {\bibinfo {title}
			{{Phase transition in Random Circuit Sampling}},}\ }\href
	{http://arxiv.org/abs/2304.11119} {\  (\bibinfo {year} {2023})},\ \Eprint
	{http://arxiv.org/abs/2304.11119} {arXiv:2304.11119} \BibitemShut {NoStop}%
	\bibitem [{\citenamefont {Richter}\ and\ \citenamefont
		{Pal}(2021)}]{PhysRevLett.126.230501}%
	\BibitemOpen
	\bibfield  {author} {\bibinfo {author} {\bibfnamefont {J.}~\bibnamefont
			{Richter}}\ and\ \bibinfo {author} {\bibfnamefont {A.}~\bibnamefont {Pal}},\
	}\bibfield  {title} {\enquote {\bibinfo {title} {Simulating hydrodynamics on
				noisy intermediate-scale quantum devices with random circuits},}\ }\href
	{\doibase 10.1103/PhysRevLett.126.230501} {\bibfield  {journal} {\bibinfo
			{journal} {Phys. Rev. Lett.}\ }\textbf {\bibinfo {volume} {126}},\ \bibinfo
		{pages} {230501} (\bibinfo {year} {2021})}\BibitemShut {NoStop}%
	\bibitem [{\citenamefont {Keenan}\ \emph {et~al.}(2023)\citenamefont {Keenan},
		\citenamefont {Robertson}, \citenamefont {Murphy}, \citenamefont {Zhuk},\
		and\ \citenamefont {Goold}}]{2022arXiv220812243K}%
	\BibitemOpen
	\bibfield  {author} {\bibinfo {author} {\bibfnamefont {N.}~\bibnamefont
			{Keenan}}, \bibinfo {author} {\bibfnamefont {N.~F.}\ \bibnamefont
			{Robertson}}, \bibinfo {author} {\bibfnamefont {T.}~\bibnamefont {Murphy}},
		\bibinfo {author} {\bibfnamefont {S.}~\bibnamefont {Zhuk}}, \ and\ \bibinfo
		{author} {\bibfnamefont {J.}~\bibnamefont {Goold}},\ }\bibfield  {title}
	{\enquote {\bibinfo {title} {{Evidence of Kardar-Parisi-Zhang scaling on a
					digital quantum simulator}},}\ }\href {\doibase 10.1038/s41534-023-00742-4}
	{\bibfield  {journal} {\bibinfo  {journal} {npj Quantum Information}\
		}\textbf {\bibinfo {volume} {9}},\ \bibinfo {pages} {72} (\bibinfo {year}
		{2023})}\BibitemShut {NoStop}%
	\bibitem [{\citenamefont {Choi}\ \emph {et~al.}(2023)\citenamefont {Choi},
		\citenamefont {Shaw}, \citenamefont {Madjarov}, \citenamefont {Xie},
		\citenamefont {Finkelstein}, \citenamefont {Covey}, \citenamefont {Cotler},
		\citenamefont {Mark}, \citenamefont {Huang}, \citenamefont {Kale},
		\citenamefont {Pichler}, \citenamefont {Brand{\~{a}}o}, \citenamefont
		{Choi},\ and\ \citenamefont {Endres}}]{Choi:2023tt}%
	\BibitemOpen
	\bibfield  {author} {\bibinfo {author} {\bibfnamefont {Joonhee}\ \bibnamefont
			{Choi}}, \bibinfo {author} {\bibfnamefont {Adam~L.}\ \bibnamefont {Shaw}},
		\bibinfo {author} {\bibfnamefont {Ivaylo~S.}\ \bibnamefont {Madjarov}},
		\bibinfo {author} {\bibfnamefont {Xin}\ \bibnamefont {Xie}}, \bibinfo
		{author} {\bibfnamefont {Ran}\ \bibnamefont {Finkelstein}}, \bibinfo {author}
		{\bibfnamefont {Jacob~P.}\ \bibnamefont {Covey}}, \bibinfo {author}
		{\bibfnamefont {Jordan~S.}\ \bibnamefont {Cotler}}, \bibinfo {author}
		{\bibfnamefont {Daniel~K.}\ \bibnamefont {Mark}}, \bibinfo {author}
		{\bibfnamefont {Hsin-Yuan}\ \bibnamefont {Huang}}, \bibinfo {author}
		{\bibfnamefont {Anant}\ \bibnamefont {Kale}}, \bibinfo {author}
		{\bibfnamefont {Hannes}\ \bibnamefont {Pichler}}, \bibinfo {author}
		{\bibfnamefont {Fernando G. S.~L.}\ \bibnamefont {Brand{\~{a}}o}}, \bibinfo
		{author} {\bibfnamefont {Soonwon}\ \bibnamefont {Choi}}, \ and\ \bibinfo
		{author} {\bibfnamefont {Manuel}\ \bibnamefont {Endres}},\ }\bibfield
	{title} {\enquote {\bibinfo {title} {{Preparing random states and
					benchmarking with many-body quantum chaos}},}\ }\href {\doibase
		10.1038/s41586-022-05442-1} {\bibfield  {journal} {\bibinfo  {journal}
			{Nature}\ }\textbf {\bibinfo {volume} {613}},\ \bibinfo {pages} {468--473}
		(\bibinfo {year} {2023})}\BibitemShut {NoStop}%
	\bibitem [{\citenamefont {Karamlou}\ \emph {et~al.}(2024)\citenamefont
		{Karamlou}, \citenamefont {Rosen}, \citenamefont {Muschinske}, \citenamefont
		{Barrett}, \citenamefont {{Di Paolo}}, \citenamefont {Ding}, \citenamefont
		{Harrington}, \citenamefont {Hays}, \citenamefont {Das}, \citenamefont {Kim},
		\citenamefont {Niedzielski}, \citenamefont {Schuldt}, \citenamefont
		{Serniak}, \citenamefont {Schwartz}, \citenamefont {Yoder}, \citenamefont
		{Gustavsson}, \citenamefont {Yanay}, \citenamefont {Grover},\ and\
		\citenamefont {Oliver}}]{Karamlou:2024ua}%
	\BibitemOpen
	\bibfield  {author} {\bibinfo {author} {\bibfnamefont {Amir~H.}\ \bibnamefont
			{Karamlou}}, \bibinfo {author} {\bibfnamefont {Ilan~T.}\ \bibnamefont
			{Rosen}}, \bibinfo {author} {\bibfnamefont {Sarah~E.}\ \bibnamefont
			{Muschinske}}, \bibinfo {author} {\bibfnamefont {Cora~N.}\ \bibnamefont
			{Barrett}}, \bibinfo {author} {\bibfnamefont {Agustin}\ \bibnamefont {{Di
					Paolo}}}, \bibinfo {author} {\bibfnamefont {Leon}\ \bibnamefont {Ding}},
		\bibinfo {author} {\bibfnamefont {Patrick~M.}\ \bibnamefont {Harrington}},
		\bibinfo {author} {\bibfnamefont {Max}\ \bibnamefont {Hays}}, \bibinfo
		{author} {\bibfnamefont {Rabindra}\ \bibnamefont {Das}}, \bibinfo {author}
		{\bibfnamefont {David~K.}\ \bibnamefont {Kim}}, \bibinfo {author}
		{\bibfnamefont {Bethany~M.}\ \bibnamefont {Niedzielski}}, \bibinfo {author}
		{\bibfnamefont {Meghan}\ \bibnamefont {Schuldt}}, \bibinfo {author}
		{\bibfnamefont {Kyle}\ \bibnamefont {Serniak}}, \bibinfo {author}
		{\bibfnamefont {Mollie~E.}\ \bibnamefont {Schwartz}}, \bibinfo {author}
		{\bibfnamefont {Jonilyn~L.}\ \bibnamefont {Yoder}}, \bibinfo {author}
		{\bibfnamefont {Simon}\ \bibnamefont {Gustavsson}}, \bibinfo {author}
		{\bibfnamefont {Yariv}\ \bibnamefont {Yanay}}, \bibinfo {author}
		{\bibfnamefont {Jeffrey~A.}\ \bibnamefont {Grover}}, \ and\ \bibinfo {author}
		{\bibfnamefont {William~D.}\ \bibnamefont {Oliver}},\ }\bibfield  {title}
	{\enquote {\bibinfo {title} {{Probing entanglement in a 2D hard-core
					Bose–Hubbard lattice}},}\ }\href {\doibase 10.1038/s41586-024-07325-z}
	{\bibfield  {journal} {\bibinfo  {journal} {Nature}\ }\textbf {\bibinfo
			{volume} {629}},\ \bibinfo {pages} {561--566} (\bibinfo {year}
		{2024})}\BibitemShut {NoStop}%
	\bibitem [{\citenamefont {Yanay}\ \emph {et~al.}(2020)\citenamefont {Yanay},
		\citenamefont {Braum{\"u}ller}, \citenamefont {Gustavsson}, \citenamefont
		{Oliver},\ and\ \citenamefont {Tahan}}]{Yanay:2020tb}%
	\BibitemOpen
	\bibfield  {author} {\bibinfo {author} {\bibfnamefont {Yariv}\ \bibnamefont
			{Yanay}}, \bibinfo {author} {\bibfnamefont {Jochen}\ \bibnamefont
			{Braum{\"u}ller}}, \bibinfo {author} {\bibfnamefont {Simon}\ \bibnamefont
			{Gustavsson}}, \bibinfo {author} {\bibfnamefont {William~D.}\ \bibnamefont
			{Oliver}}, \ and\ \bibinfo {author} {\bibfnamefont {Charles}\ \bibnamefont
			{Tahan}},\ }\bibfield  {title} {\enquote {\bibinfo {title} {{Two-dimensional
					hard-core Bose--Hubbard model with superconducting qubits}},}\ }\href
	{\doibase 10.1038/s41534-020-0269-1} {\bibfield  {journal} {\bibinfo
			{journal} {npj Quantum Information}\ }\textbf {\bibinfo {volume} {6}},\
		\bibinfo {pages} {58} (\bibinfo {year} {2020})}\BibitemShut {NoStop}%
	\bibitem [{\citenamefont {Sun}\ \emph {et~al.}(2020)\citenamefont {Sun},
		\citenamefont {Cui},\ and\ \citenamefont {Fan}}]{PhysRevResearch.2.013163}%
	\BibitemOpen
	\bibfield  {author} {\bibinfo {author} {\bibfnamefont {Z.-H.}\ \bibnamefont
			{Sun}}, \bibinfo {author} {\bibfnamefont {J.}~\bibnamefont {Cui}}, \ and\
		\bibinfo {author} {\bibfnamefont {H.}~\bibnamefont {Fan}},\ }\bibfield
	{title} {\enquote {\bibinfo {title} {Characterizing the many-body
				localization transition by the dynamics of diagonal entropy},}\ }\href
	{\doibase 10.1103/PhysRevResearch.2.013163} {\bibfield  {journal} {\bibinfo
			{journal} {Phys. Rev. Res.}\ }\textbf {\bibinfo {volume} {2}},\ \bibinfo
		{pages} {013163} (\bibinfo {year} {2020})}\BibitemShut {NoStop}%
	\bibitem [{\citenamefont {Khait}\ \emph {et~al.}(2016)\citenamefont {Khait},
		\citenamefont {Gazit}, \citenamefont {Yao},\ and\ \citenamefont
		{Auerbach}}]{PhysRevB.93.224205}%
	\BibitemOpen
	\bibfield  {author} {\bibinfo {author} {\bibfnamefont {I.}~\bibnamefont
			{Khait}}, \bibinfo {author} {\bibfnamefont {S.}~\bibnamefont {Gazit}},
		\bibinfo {author} {\bibfnamefont {N.~Y.}\ \bibnamefont {Yao}}, \ and\
		\bibinfo {author} {\bibfnamefont {A.}~\bibnamefont {Auerbach}},\ }\bibfield
	{title} {\enquote {\bibinfo {title} {Spin transport of weakly disordered
				heisenberg chain at infinite temperature},}\ }\href {\doibase
		10.1103/PhysRevB.93.224205} {\bibfield  {journal} {\bibinfo  {journal} {Phys.
				Rev. B}\ }\textbf {\bibinfo {volume} {93}},\ \bibinfo {pages} {224205}
		(\bibinfo {year} {2016})}\BibitemShut {NoStop}%
	\bibitem [{\citenamefont {Gopalakrishnan}\ \emph {et~al.}(2016)\citenamefont
		{Gopalakrishnan}, \citenamefont {Agarwal}, \citenamefont {Demler},
		\citenamefont {Huse},\ and\ \citenamefont {Knap}}]{PhysRevB.93.134206}%
	\BibitemOpen
	\bibfield  {author} {\bibinfo {author} {\bibfnamefont {S.}~\bibnamefont
			{Gopalakrishnan}}, \bibinfo {author} {\bibfnamefont {K.}~\bibnamefont
			{Agarwal}}, \bibinfo {author} {\bibfnamefont {E.~A.}\ \bibnamefont {Demler}},
		\bibinfo {author} {\bibfnamefont {D.~A.}\ \bibnamefont {Huse}}, \ and\
		\bibinfo {author} {\bibfnamefont {M.}~\bibnamefont {Knap}},\ }\bibfield
	{title} {\enquote {\bibinfo {title} {{Griffiths effects and slow dynamics in
					nearly many-body localized systems}},}\ }\href {\doibase
		10.1103/PhysRevB.93.134206} {\bibfield  {journal} {\bibinfo  {journal} {Phys.
				Rev. B}\ }\textbf {\bibinfo {volume} {93}},\ \bibinfo {pages} {134206}
		(\bibinfo {year} {2016})}\BibitemShut {NoStop}%
	\bibitem [{\citenamefont {Setiawan}\ \emph {et~al.}(2017)\citenamefont
		{Setiawan}, \citenamefont {Deng},\ and\ \citenamefont
		{Pixley}}]{PhysRevB.96.104205}%
	\BibitemOpen
	\bibfield  {author} {\bibinfo {author} {\bibfnamefont {F.}~\bibnamefont
			{Setiawan}}, \bibinfo {author} {\bibfnamefont {D.-L.}\ \bibnamefont {Deng}},
		\ and\ \bibinfo {author} {\bibfnamefont {J.~H.}\ \bibnamefont {Pixley}},\
	}\bibfield  {title} {\enquote {\bibinfo {title} {Transport properties across
				the many-body localization transition in quasiperiodic and random systems},}\
	}\href {\doibase 10.1103/PhysRevB.96.104205} {\bibfield  {journal} {\bibinfo
			{journal} {Phys. Rev. B}\ }\textbf {\bibinfo {volume} {96}},\ \bibinfo
		{pages} {104205} (\bibinfo {year} {2017})}\BibitemShut {NoStop}%
	\bibitem [{\citenamefont {Luitz}\ and\ \citenamefont
		{Lev}(2017)}]{Luitz:2017uz}%
	\BibitemOpen
	\bibfield  {author} {\bibinfo {author} {\bibfnamefont {D.~J.}\ \bibnamefont
			{Luitz}}\ and\ \bibinfo {author} {\bibfnamefont {Y.~B.}\ \bibnamefont
			{Lev}},\ }\bibfield  {title} {\enquote {\bibinfo {title} {The ergodic side of
				the many-body localization transition},}\ }\href {\doibase
		https://doi.org/10.1002/andp.201600350} {\bibfield  {journal} {\bibinfo
			{journal} {Annalen der Physik}\ }\textbf {\bibinfo {volume} {529}},\ \bibinfo
		{pages} {1600350} (\bibinfo {year} {2017})}\BibitemShut {NoStop}%
	\bibitem [{\citenamefont {Morong}\ \emph {et~al.}(2021)\citenamefont {Morong},
		\citenamefont {Liu}, \citenamefont {Becker}, \citenamefont {Collins},
		\citenamefont {Feng}, \citenamefont {Kyprianidis}, \citenamefont {Pagano},
		\citenamefont {You}, \citenamefont {Gorshkov},\ and\ \citenamefont
		{Monroe}}]{Morong:2021ul}%
	\BibitemOpen
	\bibfield  {author} {\bibinfo {author} {\bibfnamefont {W.}~\bibnamefont
			{Morong}}, \bibinfo {author} {\bibfnamefont {F.}~\bibnamefont {Liu}},
		\bibinfo {author} {\bibfnamefont {P.}~\bibnamefont {Becker}}, \bibinfo
		{author} {\bibfnamefont {K.~S.}\ \bibnamefont {Collins}}, \bibinfo {author}
		{\bibfnamefont {L.}~\bibnamefont {Feng}}, \bibinfo {author} {\bibfnamefont
			{A.}~\bibnamefont {Kyprianidis}}, \bibinfo {author} {\bibfnamefont
			{G.}~\bibnamefont {Pagano}}, \bibinfo {author} {\bibfnamefont
			{T.}~\bibnamefont {You}}, \bibinfo {author} {\bibfnamefont {A.~V.}\
			\bibnamefont {Gorshkov}}, \ and\ \bibinfo {author} {\bibfnamefont
			{C.}~\bibnamefont {Monroe}},\ }\bibfield  {title} {\enquote {\bibinfo {title}
			{{Observation of Stark many-body localization without disorder}},}\ }\href
	{\doibase 10.1038/s41586-021-03988-0} {\bibfield  {journal} {\bibinfo
			{journal} {Nature}\ }\textbf {\bibinfo {volume} {599}},\ \bibinfo {pages}
		{393--398} (\bibinfo {year} {2021})}\BibitemShut {NoStop}%
	\bibitem [{\citenamefont {Schulz}\ \emph {et~al.}(2019)\citenamefont {Schulz},
		\citenamefont {Hooley}, \citenamefont {Moessner},\ and\ \citenamefont
		{Pollmann}}]{PhysRevLett.122.040606}%
	\BibitemOpen
	\bibfield  {author} {\bibinfo {author} {\bibfnamefont {M.}~\bibnamefont
			{Schulz}}, \bibinfo {author} {\bibfnamefont {C.~A.}\ \bibnamefont {Hooley}},
		\bibinfo {author} {\bibfnamefont {R.}~\bibnamefont {Moessner}}, \ and\
		\bibinfo {author} {\bibfnamefont {F.}~\bibnamefont {Pollmann}},\ }\bibfield
	{title} {\enquote {\bibinfo {title} {{Stark Many-Body Localization}},}\
	}\href {\doibase 10.1103/PhysRevLett.122.040606} {\bibfield  {journal}
		{\bibinfo  {journal} {Phys. Rev. Lett.}\ }\textbf {\bibinfo {volume} {122}},\
		\bibinfo {pages} {040606} (\bibinfo {year} {2019})}\BibitemShut {NoStop}%
	\bibitem [{\citenamefont {van Nieuwenburg}\ \emph {et~al.}(2019)\citenamefont
		{van Nieuwenburg}, \citenamefont {Baum},\ and\ \citenamefont
		{Refael}}]{Nieuwenburg:2019vb}%
	\BibitemOpen
	\bibfield  {author} {\bibinfo {author} {\bibfnamefont {E.}~\bibnamefont {van
				Nieuwenburg}}, \bibinfo {author} {\bibfnamefont {Y.}~\bibnamefont {Baum}}, \
		and\ \bibinfo {author} {\bibfnamefont {G.}~\bibnamefont {Refael}},\
	}\bibfield  {title} {\enquote {\bibinfo {title} {From bloch oscillations to
				many-body localization in clean interacting systems},}\ }\href {\doibase
		10.1073/pnas.1819316116} {\bibfield  {journal} {\bibinfo  {journal}
			{Proceedings of the National Academy of Sciences}\ }\textbf {\bibinfo
			{volume} {116}},\ \bibinfo {pages} {9269--9274} (\bibinfo {year}
		{2019})}\BibitemShut {NoStop}%
	\bibitem [{\citenamefont {Wang}\ \emph {et~al.}(2021)\citenamefont {Wang},
		\citenamefont {Sun},\ and\ \citenamefont {Fan}}]{PhysRevB.104.205122}%
	\BibitemOpen
	\bibfield  {author} {\bibinfo {author} {\bibfnamefont {Y.-Y.}\ \bibnamefont
			{Wang}}, \bibinfo {author} {\bibfnamefont {Z.-H.}\ \bibnamefont {Sun}}, \
		and\ \bibinfo {author} {\bibfnamefont {H.}~\bibnamefont {Fan}},\ }\bibfield
	{title} {\enquote {\bibinfo {title} {{Stark many-body localization
					transitions in superconducting circuits}},}\ }\href {\doibase
		10.1103/PhysRevB.104.205122} {\bibfield  {journal} {\bibinfo  {journal}
			{Phys. Rev. B}\ }\textbf {\bibinfo {volume} {104}},\ \bibinfo {pages}
		{205122} (\bibinfo {year} {2021})}\BibitemShut {NoStop}%
	\bibitem [{\citenamefont {Taylor}\ \emph {et~al.}(2020)\citenamefont {Taylor},
		\citenamefont {Schulz}, \citenamefont {Pollmann},\ and\ \citenamefont
		{Moessner}}]{PhysRevB.102.054206}%
	\BibitemOpen
	\bibfield  {author} {\bibinfo {author} {\bibfnamefont {S.~R.}\ \bibnamefont
			{Taylor}}, \bibinfo {author} {\bibfnamefont {M.}~\bibnamefont {Schulz}},
		\bibinfo {author} {\bibfnamefont {F.}~\bibnamefont {Pollmann}}, \ and\
		\bibinfo {author} {\bibfnamefont {R.}~\bibnamefont {Moessner}},\ }\bibfield
	{title} {\enquote {\bibinfo {title} {{Experimental probes of Stark many-body
					localization}},}\ }\href {\doibase 10.1103/PhysRevB.102.054206} {\bibfield
		{journal} {\bibinfo  {journal} {Phys. Rev. B}\ }\textbf {\bibinfo {volume}
			{102}},\ \bibinfo {pages} {054206} (\bibinfo {year} {2020})}\BibitemShut
	{NoStop}%
	\bibitem [{\citenamefont {Doggen}\ \emph {et~al.}(2021)\citenamefont {Doggen},
		\citenamefont {Gornyi},\ and\ \citenamefont
		{Polyakov}}]{PhysRevB.103.L100202}%
	\BibitemOpen
	\bibfield  {author} {\bibinfo {author} {\bibfnamefont {E.~V.~H.}\
			\bibnamefont {Doggen}}, \bibinfo {author} {\bibfnamefont {I.~V.}\
			\bibnamefont {Gornyi}}, \ and\ \bibinfo {author} {\bibfnamefont {D.~G.}\
			\bibnamefont {Polyakov}},\ }\bibfield  {title} {\enquote {\bibinfo {title}
			{{Stark many-body localization: Evidence for Hilbert-space shattering}},}\
	}\href {\doibase 10.1103/PhysRevB.103.L100202} {\bibfield  {journal}
		{\bibinfo  {journal} {Phys. Rev. B}\ }\textbf {\bibinfo {volume} {103}},\
		\bibinfo {pages} {L100202} (\bibinfo {year} {2021})}\BibitemShut {NoStop}%
	\bibitem [{\citenamefont {Khemani}\ \emph {et~al.}(2020)\citenamefont
		{Khemani}, \citenamefont {Hermele},\ and\ \citenamefont
		{Nandkishore}}]{PhysRevB.101.174204}%
	\BibitemOpen
	\bibfield  {author} {\bibinfo {author} {\bibfnamefont {V.}~\bibnamefont
			{Khemani}}, \bibinfo {author} {\bibfnamefont {M.}~\bibnamefont {Hermele}}, \
		and\ \bibinfo {author} {\bibfnamefont {R.}~\bibnamefont {Nandkishore}},\
	}\bibfield  {title} {\enquote {\bibinfo {title} {Localization from hilbert
				space shattering: From theory to physical realizations},}\ }\href {\doibase
		10.1103/PhysRevB.101.174204} {\bibfield  {journal} {\bibinfo  {journal}
			{Phys. Rev. B}\ }\textbf {\bibinfo {volume} {101}},\ \bibinfo {pages}
		{174204} (\bibinfo {year} {2020})}\BibitemShut {NoStop}%
	\bibitem [{\citenamefont {Sala}\ \emph {et~al.}(2020)\citenamefont {Sala},
		\citenamefont {Rakovszky}, \citenamefont {Verresen}, \citenamefont {Knap},\
		and\ \citenamefont {Pollmann}}]{PhysRevX.10.011047}%
	\BibitemOpen
	\bibfield  {author} {\bibinfo {author} {\bibfnamefont {P.}~\bibnamefont
			{Sala}}, \bibinfo {author} {\bibfnamefont {T.}~\bibnamefont {Rakovszky}},
		\bibinfo {author} {\bibfnamefont {R.}~\bibnamefont {Verresen}}, \bibinfo
		{author} {\bibfnamefont {M.}~\bibnamefont {Knap}}, \ and\ \bibinfo {author}
		{\bibfnamefont {F.}~\bibnamefont {Pollmann}},\ }\bibfield  {title} {\enquote
		{\bibinfo {title} {Ergodicity breaking arising from hilbert space
				fragmentation in dipole-conserving hamiltonians},}\ }\href {\doibase
		10.1103/PhysRevX.10.011047} {\bibfield  {journal} {\bibinfo  {journal} {Phys.
				Rev. X}\ }\textbf {\bibinfo {volume} {10}},\ \bibinfo {pages} {011047}
		(\bibinfo {year} {2020})}\BibitemShut {NoStop}%
	\bibitem [{\citenamefont {Serbyn}\ \emph {et~al.}(2013)\citenamefont {Serbyn},
		\citenamefont {Papi\ifmmode~\acute{c}\else \'{c}\fi{}},\ and\ \citenamefont
		{Abanin}}]{PhysRevLett.111.127201}%
	\BibitemOpen
	\bibfield  {author} {\bibinfo {author} {\bibfnamefont {Maksym}\ \bibnamefont
			{Serbyn}}, \bibinfo {author} {\bibfnamefont {Z.}~\bibnamefont
			{Papi\ifmmode~\acute{c}\else \'{c}\fi{}}}, \ and\ \bibinfo {author}
		{\bibfnamefont {Dmitry~A.}\ \bibnamefont {Abanin}},\ }\bibfield  {title}
	{\enquote {\bibinfo {title} {{Local Conservation Laws and the Structure of
					the Many-Body Localized States}},}\ }\href {\doibase
		10.1103/PhysRevLett.111.127201} {\bibfield  {journal} {\bibinfo  {journal}
			{Phys. Rev. Lett.}\ }\textbf {\bibinfo {volume} {111}},\ \bibinfo {pages}
		{127201} (\bibinfo {year} {2013})}\BibitemShut {NoStop}%
	\bibitem [{\citenamefont {Scherg}\ \emph {et~al.}(2021)\citenamefont {Scherg},
		\citenamefont {Kohlert}, \citenamefont {Sala}, \citenamefont {Pollmann},
		\citenamefont {Hebbe~Madhusudhana}, \citenamefont {Bloch},\ and\
		\citenamefont {Aidelsburger}}]{Scherg:2021tl}%
	\BibitemOpen
	\bibfield  {author} {\bibinfo {author} {\bibfnamefont {Sebastian}\
			\bibnamefont {Scherg}}, \bibinfo {author} {\bibfnamefont {Thomas}\
			\bibnamefont {Kohlert}}, \bibinfo {author} {\bibfnamefont {Pablo}\
			\bibnamefont {Sala}}, \bibinfo {author} {\bibfnamefont {Frank}\ \bibnamefont
			{Pollmann}}, \bibinfo {author} {\bibfnamefont {Bharath}\ \bibnamefont
			{Hebbe~Madhusudhana}}, \bibinfo {author} {\bibfnamefont {Immanuel}\
			\bibnamefont {Bloch}}, \ and\ \bibinfo {author} {\bibfnamefont {Monika}\
			\bibnamefont {Aidelsburger}},\ }\bibfield  {title} {\enquote {\bibinfo
			{title} {{Observing non-ergodicity due to kinetic constraints in tilted
					Fermi-Hubbard chains}},}\ }\href {\doibase 10.1038/s41467-021-24726-0}
	{\bibfield  {journal} {\bibinfo  {journal} {Nature Communications}\ }\textbf
		{\bibinfo {volume} {12}},\ \bibinfo {pages} {4490} (\bibinfo {year}
		{2021})}\BibitemShut {NoStop}%
	\bibitem [{\citenamefont {Kohlert}\ \emph {et~al.}(2023)\citenamefont
		{Kohlert}, \citenamefont {Scherg}, \citenamefont {Sala}, \citenamefont
		{Pollmann}, \citenamefont {Hebbe~Madhusudhana}, \citenamefont {Bloch},\ and\
		\citenamefont {Aidelsburger}}]{PhysRevLett.130.010201}%
	\BibitemOpen
	\bibfield  {author} {\bibinfo {author} {\bibfnamefont {Thomas}\ \bibnamefont
			{Kohlert}}, \bibinfo {author} {\bibfnamefont {Sebastian}\ \bibnamefont
			{Scherg}}, \bibinfo {author} {\bibfnamefont {Pablo}\ \bibnamefont {Sala}},
		\bibinfo {author} {\bibfnamefont {Frank}\ \bibnamefont {Pollmann}}, \bibinfo
		{author} {\bibfnamefont {Bharath}\ \bibnamefont {Hebbe~Madhusudhana}},
		\bibinfo {author} {\bibfnamefont {Immanuel}\ \bibnamefont {Bloch}}, \ and\
		\bibinfo {author} {\bibfnamefont {Monika}\ \bibnamefont {Aidelsburger}},\
	}\bibfield  {title} {\enquote {\bibinfo {title} {Exploring the regime of
				fragmentation in strongly tilted fermi-hubbard chains},}\ }\href {\doibase
		10.1103/PhysRevLett.130.010201} {\bibfield  {journal} {\bibinfo  {journal}
			{Phys. Rev. Lett.}\ }\textbf {\bibinfo {volume} {130}},\ \bibinfo {pages}
		{010201} (\bibinfo {year} {2023})}\BibitemShut {NoStop}%
	\bibitem [{\citenamefont {Bordia}\ \emph {et~al.}(2017)\citenamefont {Bordia},
		\citenamefont {L\"uschen}, \citenamefont {Scherg}, \citenamefont
		{Gopalakrishnan}, \citenamefont {Knap}, \citenamefont {Schneider},\ and\
		\citenamefont {Bloch}}]{PhysRevX.7.041047}%
	\BibitemOpen
	\bibfield  {author} {\bibinfo {author} {\bibfnamefont {Pranjal}\ \bibnamefont
			{Bordia}}, \bibinfo {author} {\bibfnamefont {Henrik}\ \bibnamefont
			{L\"uschen}}, \bibinfo {author} {\bibfnamefont {Sebastian}\ \bibnamefont
			{Scherg}}, \bibinfo {author} {\bibfnamefont {Sarang}\ \bibnamefont
			{Gopalakrishnan}}, \bibinfo {author} {\bibfnamefont {Michael}\ \bibnamefont
			{Knap}}, \bibinfo {author} {\bibfnamefont {Ulrich}\ \bibnamefont
			{Schneider}}, \ and\ \bibinfo {author} {\bibfnamefont {Immanuel}\
			\bibnamefont {Bloch}},\ }\bibfield  {title} {\enquote {\bibinfo {title}
			{{Probing Slow Relaxation and Many-Body Localization in Two-Dimensional
					Quasiperiodic Systems}},}\ }\href {\doibase 10.1103/PhysRevX.7.041047}
	{\bibfield  {journal} {\bibinfo  {journal} {Phys. Rev. X}\ }\textbf {\bibinfo
			{volume} {7}},\ \bibinfo {pages} {041047} (\bibinfo {year}
		{2017})}\BibitemShut {NoStop}%
	\bibitem [{\citenamefont {Nandy}\ \emph {et~al.}(2024)\citenamefont {Nandy},
		\citenamefont {Herbrych}, \citenamefont {Lenar{\v{c}}i{\v{c}}}, \citenamefont
		{G{\l}{\'{o}}dkowski}, \citenamefont {Prelov{\v{s}}ek},\ and\ \citenamefont
		{Mierzejewski}}]{PhysRevB.109.115120}%
	\BibitemOpen
	\bibfield  {author} {\bibinfo {author} {\bibfnamefont {S.}~\bibnamefont
			{Nandy}}, \bibinfo {author} {\bibfnamefont {J.}~\bibnamefont {Herbrych}},
		\bibinfo {author} {\bibfnamefont {Z.}~\bibnamefont {Lenar{\v{c}}i{\v{c}}}},
		\bibinfo {author} {\bibfnamefont {A.}~\bibnamefont {G{\l}{\'{o}}dkowski}},
		\bibinfo {author} {\bibfnamefont {P.}~\bibnamefont {Prelov{\v{s}}ek}}, \ and\
		\bibinfo {author} {\bibfnamefont {M.}~\bibnamefont {Mierzejewski}},\
	}\bibfield  {title} {\enquote {\bibinfo {title} {{Emergent dipole moment
					conservation and subdiffusion in tilted chains}},}\ }\href {\doibase
		10.1103/PhysRevB.109.115120} {\bibfield  {journal} {\bibinfo  {journal}
			{Phys. Rev. B}\ }\textbf {\bibinfo {volume} {109}},\ \bibinfo {pages}
		{115120} (\bibinfo {year} {2024})}\BibitemShut {NoStop}%
	\bibitem [{\citenamefont {Guardado-Sanchez}\ \emph {et~al.}(2020)\citenamefont
		{Guardado-Sanchez}, \citenamefont {Morningstar}, \citenamefont {Spar},
		\citenamefont {Brown}, \citenamefont {Huse},\ and\ \citenamefont
		{Bakr}}]{PhysRevX.10.011042}%
	\BibitemOpen
	\bibfield  {author} {\bibinfo {author} {\bibfnamefont {Elmer}\ \bibnamefont
			{Guardado-Sanchez}}, \bibinfo {author} {\bibfnamefont {Alan}\ \bibnamefont
			{Morningstar}}, \bibinfo {author} {\bibfnamefont {Benjamin~M.}\ \bibnamefont
			{Spar}}, \bibinfo {author} {\bibfnamefont {Peter~T.}\ \bibnamefont {Brown}},
		\bibinfo {author} {\bibfnamefont {David~A.}\ \bibnamefont {Huse}}, \ and\
		\bibinfo {author} {\bibfnamefont {Waseem~S.}\ \bibnamefont {Bakr}},\
	}\bibfield  {title} {\enquote {\bibinfo {title} {{Subdiffusion and Heat
					Transport in a Tilted Two-Dimensional Fermi-Hubbard System}},}\ }\href
	{\doibase 10.1103/PhysRevX.10.011042} {\bibfield  {journal} {\bibinfo
			{journal} {Phys. Rev. X}\ }\textbf {\bibinfo {volume} {10}},\ \bibinfo
		{pages} {011042} (\bibinfo {year} {2020})}\BibitemShut {NoStop}%
	\bibitem [{\citenamefont {Cross}\ \emph {et~al.}(2019)\citenamefont {Cross},
		\citenamefont {Bishop}, \citenamefont {Sheldon}, \citenamefont {Nation},\
		and\ \citenamefont {Gambetta}}]{PhysRevA.100.032328}%
	\BibitemOpen
	\bibfield  {author} {\bibinfo {author} {\bibfnamefont {A.~W.}\ \bibnamefont
			{Cross}}, \bibinfo {author} {\bibfnamefont {L.~S.}\ \bibnamefont {Bishop}},
		\bibinfo {author} {\bibfnamefont {S.}~\bibnamefont {Sheldon}}, \bibinfo
		{author} {\bibfnamefont {P.~D.}\ \bibnamefont {Nation}}, \ and\ \bibinfo
		{author} {\bibfnamefont {J.~M.}\ \bibnamefont {Gambetta}},\ }\bibfield
	{title} {\enquote {\bibinfo {title} {Validating quantum computers using
				randomized model circuits},}\ }\href {\doibase 10.1103/PhysRevA.100.032328}
	{\bibfield  {journal} {\bibinfo  {journal} {Phys. Rev. A}\ }\textbf {\bibinfo
			{volume} {100}},\ \bibinfo {pages} {032328} (\bibinfo {year}
		{2019})}\BibitemShut {NoStop}%
	\bibitem [{\citenamefont {Saffman}\ \emph {et~al.}(2010)\citenamefont
		{Saffman}, \citenamefont {Walker},\ and\ \citenamefont
		{M\o{}lmer}}]{RevModPhys.82.2313}%
	\BibitemOpen
	\bibfield  {author} {\bibinfo {author} {\bibfnamefont {M.}~\bibnamefont
			{Saffman}}, \bibinfo {author} {\bibfnamefont {T.~G.}\ \bibnamefont {Walker}},
		\ and\ \bibinfo {author} {\bibfnamefont {K.}~\bibnamefont {M\o{}lmer}},\
	}\bibfield  {title} {\enquote {\bibinfo {title} {Quantum information with
				rydberg atoms},}\ }\href {\doibase 10.1103/RevModPhys.82.2313} {\bibfield
		{journal} {\bibinfo  {journal} {Rev. Mod. Phys.}\ }\textbf {\bibinfo {volume}
			{82}},\ \bibinfo {pages} {2313--2363} (\bibinfo {year} {2010})}\BibitemShut
	{NoStop}%
	\bibitem [{\citenamefont {Browaeys}\ and\ \citenamefont
		{Lahaye}(2020)}]{Browaeys:2020tl}%
	\BibitemOpen
	\bibfield  {author} {\bibinfo {author} {\bibfnamefont {A.}~\bibnamefont
			{Browaeys}}\ and\ \bibinfo {author} {\bibfnamefont {T.}~\bibnamefont
			{Lahaye}},\ }\bibfield  {title} {\enquote {\bibinfo {title} {Many-body
				physics with individually controlled rydberg atoms},}\ }\href {\doibase
		10.1038/s41567-019-0733-z} {\bibfield  {journal} {\bibinfo  {journal} {Nature
				Physics}\ }\textbf {\bibinfo {volume} {16}},\ \bibinfo {pages} {132--142}
		(\bibinfo {year} {2020})}\BibitemShut {NoStop}%
	\bibitem [{\citenamefont {Henriet}\ \emph {et~al.}(2020)\citenamefont
		{Henriet}, \citenamefont {Beguin}, \citenamefont {Signoles}, \citenamefont
		{Lahaye}, \citenamefont {Browaeys}, \citenamefont {Reymond},\ and\
		\citenamefont {Jurczak}}]{Henriet2020quantumcomputing}%
	\BibitemOpen
	\bibfield  {author} {\bibinfo {author} {\bibfnamefont {L.}~\bibnamefont
			{Henriet}}, \bibinfo {author} {\bibfnamefont {L.}~\bibnamefont {Beguin}},
		\bibinfo {author} {\bibfnamefont {A.}~\bibnamefont {Signoles}}, \bibinfo
		{author} {\bibfnamefont {T.}~\bibnamefont {Lahaye}}, \bibinfo {author}
		{\bibfnamefont {A.}~\bibnamefont {Browaeys}}, \bibinfo {author}
		{\bibfnamefont {G.-O.}\ \bibnamefont {Reymond}}, \ and\ \bibinfo {author}
		{\bibfnamefont {C.}~\bibnamefont {Jurczak}},\ }\bibfield  {title} {\enquote
		{\bibinfo {title} {Quantum computing with neutral atoms},}\ }\href {\doibase
		10.22331/q-2020-09-21-327} {\bibfield  {journal} {\bibinfo  {journal}
			{{Quantum}}\ }\textbf {\bibinfo {volume} {4}},\ \bibinfo {pages} {327}
		(\bibinfo {year} {2020})}\BibitemShut {NoStop}%
	\bibitem [{\citenamefont {Kaufman}\ \emph {et~al.}(2016)\citenamefont
		{Kaufman}, \citenamefont {Tai}, \citenamefont {Lukin}, \citenamefont
		{Rispoli}, \citenamefont {Schittko}, \citenamefont {Preiss},\ and\
		\citenamefont {Greiner}}]{doi:10.1126/science.aaf6725}%
	\BibitemOpen
	\bibfield  {author} {\bibinfo {author} {\bibfnamefont {Adam~M.}\ \bibnamefont
			{Kaufman}}, \bibinfo {author} {\bibfnamefont {M.~Eric}\ \bibnamefont {Tai}},
		\bibinfo {author} {\bibfnamefont {Alexander}\ \bibnamefont {Lukin}}, \bibinfo
		{author} {\bibfnamefont {Matthew}\ \bibnamefont {Rispoli}}, \bibinfo {author}
		{\bibfnamefont {Robert}\ \bibnamefont {Schittko}}, \bibinfo {author}
		{\bibfnamefont {Philipp~M.}\ \bibnamefont {Preiss}}, \ and\ \bibinfo {author}
		{\bibfnamefont {Markus}\ \bibnamefont {Greiner}},\ }\bibfield  {title}
	{\enquote {\bibinfo {title} {{Quantum thermalization through entanglement in
					an isolated many-body system}},}\ }\href {\doibase 10.1126/science.aaf6725}
	{\bibfield  {journal} {\bibinfo  {journal} {Science}\ }\textbf {\bibinfo
			{volume} {353}},\ \bibinfo {pages} {794--800} (\bibinfo {year}
		{2016})}\BibitemShut {NoStop}%
	\bibitem [{\citenamefont {Gross}\ and\ \citenamefont
		{Bloch}(2017)}]{doi:10.1126/science.aal3837}%
	\BibitemOpen
	\bibfield  {author} {\bibinfo {author} {\bibfnamefont {C.}~\bibnamefont
			{Gross}}\ and\ \bibinfo {author} {\bibfnamefont {I.}~\bibnamefont {Bloch}},\
	}\bibfield  {title} {\enquote {\bibinfo {title} {{Quantum simulations with
					ultracold atoms in optical lattices}},}\ }\href {\doibase
		10.1126/science.aal3837} {\bibfield  {journal} {\bibinfo  {journal}
			{Science}\ }\textbf {\bibinfo {volume} {357}},\ \bibinfo {pages} {995--1001}
		(\bibinfo {year} {2017})}\BibitemShut {NoStop}%
	\bibitem [{\citenamefont {Xu}\ \emph {et~al.}(2020)\citenamefont {Xu},
		\citenamefont {Sun}, \citenamefont {Liu}, \citenamefont {Zhang},
		\citenamefont {Li}, \citenamefont {Dong}, \citenamefont {Ren}, \citenamefont
		{Zhang}, \citenamefont {Nori}, \citenamefont {Zheng}, \citenamefont {Fan},\
		and\ \citenamefont {Wang}}]{qc_all_2}%
	\BibitemOpen
	\bibfield  {author} {\bibinfo {author} {\bibfnamefont {Kai}\ \bibnamefont
			{Xu}}, \bibinfo {author} {\bibfnamefont {Zheng-Hang}\ \bibnamefont {Sun}},
		\bibinfo {author} {\bibfnamefont {Wuxin}\ \bibnamefont {Liu}}, \bibinfo
		{author} {\bibfnamefont {Yu-Ran}\ \bibnamefont {Zhang}}, \bibinfo {author}
		{\bibfnamefont {Hekang}\ \bibnamefont {Li}}, \bibinfo {author} {\bibfnamefont
			{Hang}\ \bibnamefont {Dong}}, \bibinfo {author} {\bibfnamefont {Wenhui}\
			\bibnamefont {Ren}}, \bibinfo {author} {\bibfnamefont {Pengfei}\ \bibnamefont
			{Zhang}}, \bibinfo {author} {\bibfnamefont {Franco}\ \bibnamefont {Nori}},
		\bibinfo {author} {\bibfnamefont {Dongning}\ \bibnamefont {Zheng}}, \bibinfo
		{author} {\bibfnamefont {Heng}\ \bibnamefont {Fan}}, \ and\ \bibinfo {author}
		{\bibfnamefont {H.}~\bibnamefont {Wang}},\ }\bibfield  {title} {\enquote
		{\bibinfo {title} {{Probing dynamical phase transitions with a
					superconducting quantum simulator}},}\ }\href {\doibase
		10.1126/sciadv.aba4935} {\bibfield  {journal} {\bibinfo  {journal} {Science
				Advances}\ }\textbf {\bibinfo {volume} {6}} (\bibinfo {year} {2020}),\
		10.1126/sciadv.aba4935}\BibitemShut {NoStop}%
	\bibitem [{\citenamefont {Xu}\ \emph {et~al.}(2022)\citenamefont {Xu},
		\citenamefont {Zhang}, \citenamefont {Sun}, \citenamefont {Li}, \citenamefont
		{Song}, \citenamefont {Xiang}, \citenamefont {Huang}, \citenamefont {Li},
		\citenamefont {Shi}, \citenamefont {Chen}, \citenamefont {Song},
		\citenamefont {Zheng}, \citenamefont {Nori}, \citenamefont {Wang},\ and\
		\citenamefont {Fan}}]{PhysRevLett.128.150501}%
	\BibitemOpen
	\bibfield  {author} {\bibinfo {author} {\bibfnamefont {Kai}\ \bibnamefont
			{Xu}}, \bibinfo {author} {\bibfnamefont {Yu-Ran}\ \bibnamefont {Zhang}},
		\bibinfo {author} {\bibfnamefont {Zheng-Hang}\ \bibnamefont {Sun}}, \bibinfo
		{author} {\bibfnamefont {Hekang}\ \bibnamefont {Li}}, \bibinfo {author}
		{\bibfnamefont {Pengtao}\ \bibnamefont {Song}}, \bibinfo {author}
		{\bibfnamefont {Zhongcheng}\ \bibnamefont {Xiang}}, \bibinfo {author}
		{\bibfnamefont {Kaixuan}\ \bibnamefont {Huang}}, \bibinfo {author}
		{\bibfnamefont {Hao}\ \bibnamefont {Li}}, \bibinfo {author} {\bibfnamefont
			{Yun-Hao}\ \bibnamefont {Shi}}, \bibinfo {author} {\bibfnamefont {Chi-Tong}\
			\bibnamefont {Chen}}, \bibinfo {author} {\bibfnamefont {Xiaohui}\
			\bibnamefont {Song}}, \bibinfo {author} {\bibfnamefont {Dongning}\
			\bibnamefont {Zheng}}, \bibinfo {author} {\bibfnamefont {Franco}\
			\bibnamefont {Nori}}, \bibinfo {author} {\bibfnamefont {H.}~\bibnamefont
			{Wang}}, \ and\ \bibinfo {author} {\bibfnamefont {Heng}\ \bibnamefont
			{Fan}},\ }\bibfield  {title} {\enquote {\bibinfo {title} {{Metrological
					Characterization of Non-Gaussian Entangled States of Superconducting
					Qubits}},}\ }\href {\doibase 10.1103/PhysRevLett.128.150501} {\bibfield
		{journal} {\bibinfo  {journal} {Phys. Rev. Lett.}\ }\textbf {\bibinfo
			{volume} {128}},\ \bibinfo {pages} {150501} (\bibinfo {year}
		{2022})}\BibitemShut {NoStop}%
	\bibitem [{\citenamefont {Jin}\ \emph {et~al.}(2020)\citenamefont {Jin},
		\citenamefont {Willsch}, \citenamefont {Willsch}, \citenamefont {Lagemann},
		\citenamefont {Michielsen},\ and\ \citenamefont {De~Raedt}}]{Jin:2020ue}%
	\BibitemOpen
	\bibfield  {author} {\bibinfo {author} {\bibfnamefont {F.}~\bibnamefont
			{Jin}}, \bibinfo {author} {\bibfnamefont {D.}~\bibnamefont {Willsch}},
		\bibinfo {author} {\bibfnamefont {M.}~\bibnamefont {Willsch}}, \bibinfo
		{author} {\bibfnamefont {H.}~\bibnamefont {Lagemann}}, \bibinfo {author}
		{\bibfnamefont {K.}~\bibnamefont {Michielsen}}, \ and\ \bibinfo {author}
		{\bibfnamefont {H.}~\bibnamefont {De~Raedt}},\ }\bibfield  {title} {\enquote
		{\bibinfo {title} {{Random State Technology}},}\ }\href {\doibase
		10.7566/JPSJ.90.012001} {\bibfield  {journal} {\bibinfo  {journal} {Journal
				of the Physical Society of Japan}\ }\textbf {\bibinfo {volume} {90}},\
		\bibinfo {pages} {012001} (\bibinfo {year} {2020})}\BibitemShut {NoStop}%
\end{thebibliography}
%

~\\
\noindent \textbf{Data availability}
\\	
\noindent
The authors declare that the data supporting the findings of this study are available within the paper and its Supplementary Information files. Should any raw data files be needed in another format they are available from the corresponding author upon reasonable request. Source data are provided with this paper.
\\
\\
\noindent \textbf{Acknowledgments}
\\	
\noindent
We thank Hai-Long Shi and H. S. Yan for helpful discussions. Z.X., D.Z., K.X. and H.F. are supported by Beijing Natural Science Foundation (Grant No. Z200009), National Natural Science Foundation of China (Grants Nos. 92265207, T2121001, 12122504, 12247168, 11934018, T2322030), Innovation Program for Quantum Science and Technology (Grant No. 2021ZD0301800), Beijing Nova Program (Nos. 20220484121, 2022000216). Y.-H.S. acknowledges the support of Postdoctoral Fellowship Program of CPSF (Grant No. GZB20240815). Z.-A.W. acknowledges the support of China Postdoctoral Science Foundation (Grant No. 2022TQ0036).\\
\\
\\
\noindent \textbf{Author contributions}
\\	
\noindent
H.F. supervised the project. Z.-H.S. proposed the idea. Y.-H.S. conducted the experiment with the help of K.H. and K.X.. Z.-H.S., Y.-Y.W., and Y.-H.S. performed the numerical simulations.  Z.X. and D.Z. fabricated the ladder-type sample. X.S., G.X., and H.Y. provided the Josephson parametric amplifiers. W.-G.M., H.-T.L., K.Z., J.-C.S., G.-H.L., Z.-Y.M., J.-C.Z., H.L., and C.-T.C. helped the experimental setup. Z.-A.W., Y.-R.Z., J.W., K.X., and H.F. discussed and commented on the manuscript. Z.-H.S., Y.-H.S., Y.-Y.W., Y.-R.Z., and H.F. co-wrote the manuscript. All authors contributed to the discussions of the results and development of the manuscript.
\\
\\
\noindent \textbf{Competing interests}
\\	
\noindent
The authors declare no competing interests.

\clearpage

\end{document}


\title{Supplementary Information for \\Probing spin hydrodynamics on a superconducting quantum simulator}

\author{Yun-Hao Shi}
\thanks{These authors contributed equally to this work.}
\affiliation{Institute of Physics, Chinese Academy of Sciences, Beijing 100190, China}
\affiliation{School of Physical Sciences, University of Chinese Academy of Sciences, Beijing 100049, China}
\affiliation{Beijing Academy of Quantum Information Sciences, Beijing 100193, China}

\author{Zheng-Hang Sun}
\thanks{These authors contributed equally to this work.}
\affiliation{Institute of Physics, Chinese Academy of Sciences, Beijing 100190, China}
\affiliation{School of Physical Sciences, University of Chinese Academy of Sciences, Beijing 100049, China}

\author{Yong-Yi Wang}
\thanks{These authors contributed equally to this work.}
\affiliation{Institute of Physics, Chinese Academy of Sciences, Beijing 100190, China}
\affiliation{School of Physical Sciences, University of Chinese Academy of Sciences, Beijing 100049, China}

\author{Zheng-An Wang}
\affiliation{Beijing Academy of Quantum Information Sciences, Beijing 100193, China}
\affiliation{Hefei National Laboratory, Hefei 230088, China}

\author{Yu-Ran Zhang}
\affiliation{School of Physics and Optoelectronics, South China University of Technology, Guangzhou 510640, China}

\author{Wei-Guo Ma}
\affiliation{Institute of Physics, Chinese Academy of Sciences, Beijing 100190, China}
\affiliation{School of Physical Sciences, University of Chinese Academy of Sciences, Beijing 100049, China}

\author{Hao-Tian Liu}
\affiliation{Institute of Physics, Chinese Academy of Sciences, Beijing 100190, China}
\affiliation{School of Physical Sciences, University of Chinese Academy of Sciences, Beijing 100049, China}

\author{Kui Zhao}
\affiliation{Beijing Academy of Quantum Information Sciences, Beijing 100193, China}

\author{Jia-Cheng Song}
\affiliation{Institute of Physics, Chinese Academy of Sciences, Beijing 100190, China}
\affiliation{School of Physical Sciences, University of Chinese Academy of Sciences, Beijing 100049, China}

\author{Gui-Han Liang}
\affiliation{Institute of Physics, Chinese Academy of Sciences, Beijing 100190, China}
\affiliation{School of Physical Sciences, University of Chinese Academy of Sciences, Beijing 100049, China}

\author{Zheng-Yang Mei}
\affiliation{Institute of Physics, Chinese Academy of Sciences, Beijing 100190, China}
\affiliation{School of Physical Sciences, University of Chinese Academy of Sciences, Beijing 100049, China}

\author{Jia-Chi Zhang}
\affiliation{Institute of Physics, Chinese Academy of Sciences, Beijing 100190, China}
\affiliation{School of Physical Sciences, University of Chinese Academy of Sciences, Beijing 100049, China}

\author{Hao Li}
\affiliation{Beijing Academy of Quantum Information Sciences, Beijing 100193, China}

\author{Chi-Tong Chen}
\affiliation{Institute of Physics, Chinese Academy of Sciences, Beijing 100190, China}
\affiliation{School of Physical Sciences, University of Chinese Academy of Sciences, Beijing 100049, China}

\author{Xiaohui Song}
\affiliation{Institute of Physics, Chinese Academy of Sciences, Beijing 100190, China}

\author{Jieci Wang}
\affiliation{Department of Physics and Key Laboratory of Low Dimensional Quantum Structures and Quantum Control of Ministry of Education, Hunan Normal University, Changsha, Hunan 410081, China}

\author{Guangming Xue}
\affiliation{Beijing Academy of Quantum Information Sciences, Beijing 100193, China}

\author{Haifeng Yu}
\affiliation{Beijing Academy of Quantum Information Sciences, Beijing 100193, China}

\author{Kaixuan Huang}
\email{huangkx@baqis.ac.cn}
\affiliation{Beijing Academy of Quantum Information Sciences, Beijing 100193, China}

\author{Zhongcheng Xiang}
\email{zcxiang@iphy.ac.cn}
\affiliation{Institute of Physics, Chinese Academy of Sciences, Beijing 100190, China}
\affiliation{School of Physical Sciences, University of Chinese Academy of Sciences, Beijing 100049, China}
\affiliation{Hefei National Laboratory, Hefei 230088, China}

\author{Kai Xu}
\email{kaixu@iphy.ac.cn}
\affiliation{Institute of Physics, Chinese Academy of Sciences, Beijing 100190, China}
\affiliation{School of Physical Sciences, University of Chinese Academy of Sciences, Beijing 100049, China}
\affiliation{Beijing Academy of Quantum Information Sciences, Beijing 100193, China}
\affiliation{Hefei National Laboratory, Hefei 230088, China}
\affiliation{Songshan Lake Materials Laboratory, Dongguan, Guangdong 523808, China}
\affiliation{CAS Center for Excellence in Topological Quantum Computation, UCAS, Beijing 100190, China}

\author{Dongning Zheng}
\affiliation{Institute of Physics, Chinese Academy of Sciences, Beijing 100190, China}
\affiliation{School of Physical Sciences, University of Chinese Academy of Sciences, Beijing 100049, China}
\affiliation{Hefei National Laboratory, Hefei 230088, China}
\affiliation{Songshan Lake Materials Laboratory, Dongguan, Guangdong 523808, China}
\affiliation{CAS Center for Excellence in Topological Quantum Computation, UCAS, Beijing 100190, China}

\author{Heng Fan}
\email{hfan@iphy.ac.cn}
\affiliation{Institute of Physics, Chinese Academy of Sciences, Beijing 100190, China}
\affiliation{School of Physical Sciences, University of Chinese Academy of Sciences, Beijing 100049, China}
\affiliation{Beijing Academy of Quantum Information Sciences, Beijing 100193, China}
\affiliation{Hefei National Laboratory, Hefei 230088, China}
\affiliation{Songshan Lake Materials Laboratory, Dongguan, Guangdong 523808, China}
\affiliation{CAS Center for Excellence in Topological Quantum Computation, UCAS, Beijing 100190, China}

\maketitle
\beginsupplement
\tableofcontents
\newpage

\section{\label{sec:model} Model and Hamiltonian}

\begin{figure}[t]
	\includegraphics[width=0.99\textwidth]{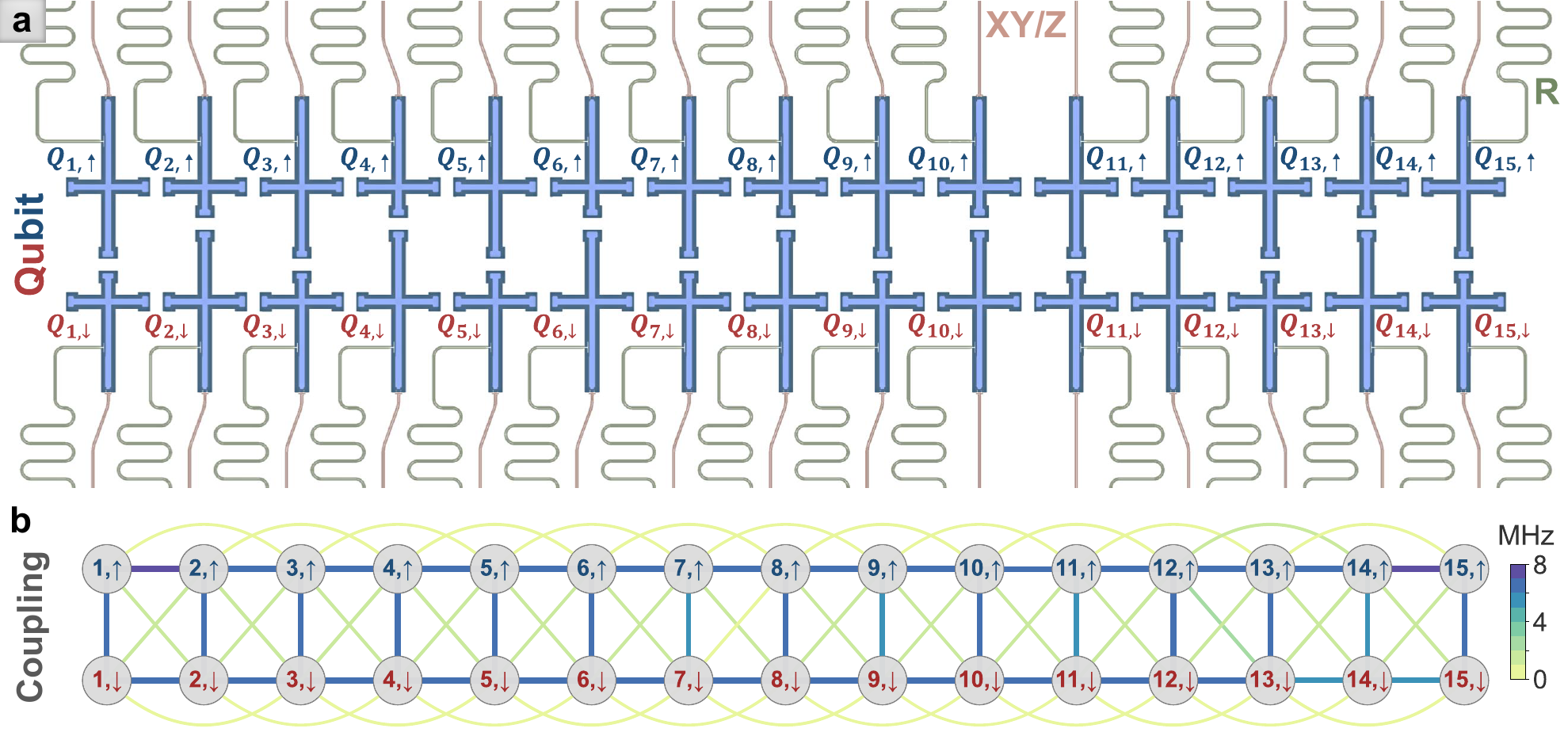}
	\caption{\textbf{{Schematic} and coupling strengths of the chip.} \textbf{a}, The ladder-type chip with 30 superconducting qubits arranged in two coupled chains. Each qubit, coupled to an independent readout resonator R, has an independent microwave line for XY and Z controls. \textbf{b}, Coupling strengths including the NN and NNN hopping couplings, which are measured by swapping experiments at the resonant frequency $\omega_{\text{ref}}\approx4.534~\!\mathrm{GHz}$.}\label{fig:chip_coupling}
\end{figure}

In this experiment, we use a ladder-type superconducting quantum processor with 30 programmable superconducting transmon qubits, which is identical to the device in ref.~[22]. The optical micrograph and coupling strengths of the chip are shown in Fig.~\ref{fig:chip_coupling}, and the device parameters are listed in Table~\ref{tab:paras}. The Hamiltonian of the total system can be essentially described by a Bose-Hubbard model of a ladder
\begin{equation}\label{eq:BH}
	\hat{H}_{\text{BH}}=\sum_{i=1}^{N}{\hbar}h_{i}\hat{a}^{\dagger}_{j}\hat{a}_{j} - \frac{E_{\mathrm{C},j}}{2}\hat{a}^{\dagger}_{j}\hat{a}^{\dagger}_{j}\hat{a}_{j}\hat{a}_{j}+\hat{H}_I,
\end{equation}
where {$\hbar$ is the reduced Planck constant}, $N$ is the total number of qubits, $\hat{a}^{\dagger}$ ($\hat{a}$) denotes the bosonic creation (annihilation) operator, $h_j$ is the tunable on-site potential, $E_{\mathrm{C},j}$ denotes the on-site charge energy, representing the magnitude of anharmonicity, and $\hat{H}_I$ is the Hamiltonian for the interactions between qubits. For qubits connected in a ladder-type with two coupled chains (`$\uparrow$' and `$\downarrow$'), the interaction Hamiltonian $\hat{H}_I$ is mainly derived from the nearest-neighbor (NN) rung (vertical, `$\bot$') and intrachain (parallel, `$\parallel$') hopping couplings, namely
\begin{eqnarray}
	\hat{H}_{\bot} &=& \sum_{j=1}^{L}{\hbar} J_{j}^{\bot}(\hat{a}^{\dagger}_{j,\uparrow}\hat{a}_{j,\downarrow}+\mathrm{H.c.}),\\
	\hat{H}_{\parallel} &=& \sum_{m\in\{\uparrow,\downarrow\}}\sum_{j=1}^{L-1}{\hbar} J_{j,m}^{\parallel}(\hat{a}^{\dagger}_{j,m}\hat{a}_{j+1,m}+\mathrm{H.c.}),
\end{eqnarray}
where $L=N/2$ is the length of each chain, $J_{j}^{\bot}$ and $J_{j,m}^{\parallel}$ are the NN rung and intrachain coupling strengths. The mean values of $J_{j}^{\bot}/2\pi$ and $J_{j,m}^{\parallel}/2\pi$ are 6.6 MHz and 7.3 MHz, respectively. In addition, it is inevitable that small next-nearest-neighbor (NNN) interactions are present, including the hopping interactions between the diagonal qubits of the upper and lower chains (`$\times$', diagonal down `$\scriptstyle{\diagdown}$' and diagonal up `$\scriptstyle{\diagup}$') and between NNN qubits on each chain (`$\cap$'), and the corresponding Hamiltonians are expressed as 
\begin{eqnarray}
	\hat{H}_{\times} &=& \sum_{j=1}^{L-1}\hbar J_{j}^{\scriptscriptstyle{\diagdown}}(\hat{a}^{\dagger}_{j,\uparrow}\hat{a}_{j+1,\downarrow} + \mathrm{H.c.})+\hbar J_{j}^{\scriptscriptstyle{\diagup}}(\hat{a}^{\dagger}_{j,\downarrow}\hat{a}_{j+1,\uparrow}+\mathrm{H.c.}),\\
	\hat{H}_{\cap} &=& \sum_{m\in\{\uparrow,\downarrow\}}\sum_{j=1}^{L-2}\hbar J_{j,m}^{\cap}(\hat{a}^{\dagger}_{j,m}\hat{a}_{j+2,m}+\mathrm{H.c.}),
\end{eqnarray}
where $J_{j}^{\scriptscriptstyle{\diagdown}}$, $J_{j}^{\scriptscriptstyle{\diagup}}$ and $J_{j,m}^{\cap}$ are the strengths of diagonal down, diagonal up and parallel NNN hopping interactions, respectively. In short, for numerical simulations, we consider $\hat{H}_I = \hat{H}_{\bot}  + \hat{H}_{\parallel} + \hat{H}_{\times} + \hat{H}_{\cap}$.

\begin{figure*}[h!]
	\centering
	\includegraphics[width=0.97\linewidth]{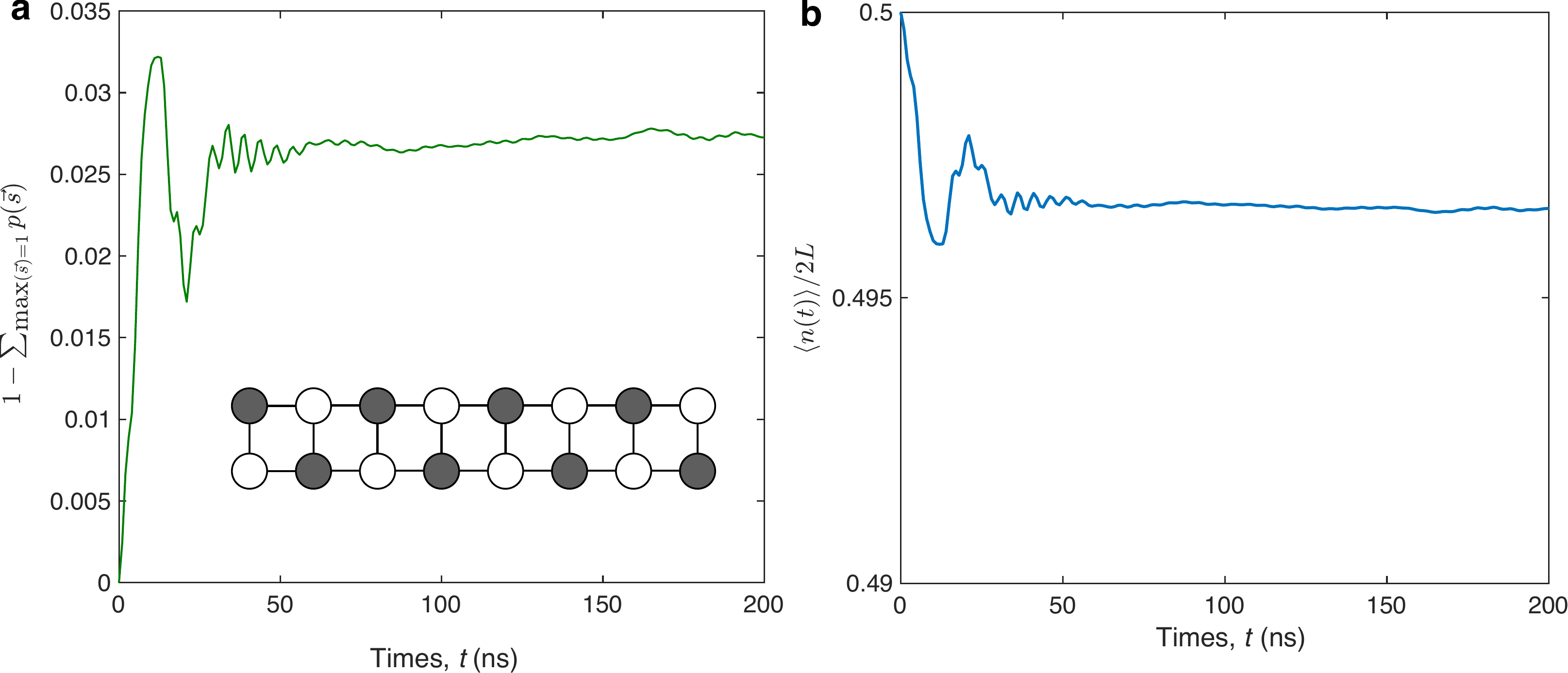}\\
	\caption{\textbf{Demonstrate of hard-core bosonic model.} \textbf{a}, Time evolution of $\sum_{\text{max}(\vec{s})=1} p(\vec{s})$  for the Hamiltonian of the superconducting circuit described by the Bose-Hubbard model (\ref{eq:BH}), with a system size $L=8$. The inset shows a schematic of the chosen initial state, where the sites represented with solid black circuits are initialized by the state $|1\rangle$, and the remainder sites are initialized by $|0\rangle$. \textbf{b}, The dynamics of the particle number $\langle n(t)\rangle$ for the Hamiltonian of the superconducting circuit with a system size $L=8$. }\label{fig_high_p}
\end{figure*}

In our quantum processor, the anharmonicity ($\ge200~\!\mathrm{MHz}$) is much greater than the coupling interaction and the model can be viewed as a ladder-type lattice of hard-core bosons~[52], i.e., the Eq.~(1) in the main text. However, in principle, the leakage to higher occupation states can be possibly induced by the finite value of the ratio between the averaged anharmonicity and coupling strength, i.e., $\overline{E_{C}}/\overline{J}$. To qualitatively characterize whether the Bose-Hubbard model (\ref{eq:BH}) can be approximate as the hard-core bosons, we consider the dynamics of the summation of the probability $\sum_{\text{max}(\vec{s})=1} p(\vec{s})$ with $\vec{s}$ denoting a configuration of product state. For instance, $\vec{s} = (1,0,1,0,...,1,0)$ corresponds to the N\'{e}el state $|\vec{s}\rangle = |1010...10\rangle$. If the system exactly becomes a hard-core bosonic model,  $\sum_{\text{max}(\vec{s})=1} p(\vec{s})=1$. Here, we numerically simulate the dynamics of $\sum_{\text{max}(\vec{s})=1} p(\vec{s})$ for the Hamiltonian of the superconducting circuit with experimentally measured hopping interactions and anharmonicity. As an example, we adopt the system size $L=16$ and a half-filling product state as the initial state $|\psi_{0}\rangle$ (see the inset of Fig.~\ref{fig_high_p}\textbf{a}). The results are plotted in Fig.~\ref{fig_high_p}\textbf{a}. One can see that the summation of the probabilities for the states with higher occupations, i.e., $\sum_{\text{max}(\vec{s})>1} p(\vec{s}) = 1 - \sum_{\text{max}(\vec{s})=1} p(\vec{s})$, only reach a relatively small value $\sim 0.03$, with the evolved time $t\simeq 200$ ns. Moreover, we numerically simulate the time evolution of the particle number $\langle n(t) \rangle \equiv \langle\psi(t) | \hat{n} |\psi(t)\rangle =   \langle\psi(t) | \sum_{i} \hat{n}_{i} |\psi(t)\rangle $, with $\hat{n}_{i}\equiv |0\rangle_{i}\langle 0 | + | 1\rangle_{i}\langle 1| $, up to the experimental time scales $t\simeq 200$ ns. The results are displayed in Fig.~\ref{fig_high_p}\textbf{b}. We emphasize that only the occupations of the states $|0\rangle$ and $|1\rangle$ are considered in the definition of $\hat{n}_{i}$, while the finite $\overline{E_{C}}/\overline{J}$ allows the possibility of the leakage to the states with higher occupations, such as $|2\rangle$. Consequently, the decay of $\langle n(t)\rangle$ shown in Fig.~\ref{fig_high_p} quantifies the leakage induced by 
the finite $\overline{E_{C}}/\overline{J}$. The stable value of $n(t)/2L$ with $t\simeq 200$ ns is about 0.4966,  indicating a moderate impact of the leakage on the conservation of the particle number. In short, the results in Fig.~\ref{fig_high_p} suggest that  hard-core bosonic Hamiltonian (1) in the main text, with a conservation of the particle number, can efficiently describe our superconducting quantum simulator.


\begin{table}[t]
		\centering
		\begin{tabular}{lcccc}
			Parameter & Median & Mean & Stdev. & Units\\[.05cm] \hline
			\\[-.2cm]
			Qubit maximum frequency & $5.025$ & $5.032$ & $0.240$ & GHz \\[.09cm]
			Qubit idle frequency & $4.723$ & $4.728$ & $0.346$ & GHz \\[.09cm]
			Qubit anharmonicity $-E_{\mathrm{C}}/(2\pi\hbar)$ & $-0.222$ & $-0.222$ & $0.022$ & GHz\\[.09cm]
			Readout frequency & $6.715$ & $6.714$ & $0.061$ & GHz\\[.09cm]
			Mean energy relaxation time $\overline{T}_1$ & $33.2$ & $32.1$ & $7.5$ & $\mu$s \\[.09cm]
			Pure dephasing time at idle frequency $T_2^{\ast}$ & $1.0$ & $2.4$ & $4.2$ & $\mu$s \\[.09cm]
			{Mean NN hopping coupling strength (vertical) $\overline{J^{\bot}}$} & {$6.7$} & {$6.6$} & {$0.2$} & {MHz}\\[.09cm]
			{Mean NN hopping coupling strength (parallel) $\overline{J^{\parallel}}$} & {$7.2$} & {$7.3$} & {$0.1$} & {MHz}\\[.09cm]
			{Mean NNN hopping coupling strength (diagonal) $\overline{J^{\times}}$} & {$1.5$} & {$1.5$} & {$0.3$} & {MHz}\\[.09cm]
			{Mean NNN hopping coupling strength (parallel) $\overline{J^{\cap}}$} & {$0.6$} & {$0.7$} & {$0.2$} & {MHz}\\[.09cm]
			{Readout fidelity of state $\ket{0}$} & {$95.2$} & {$91.4$} & {$9.6$} & {$\%$} \\[.09cm]
			{Readout fidelity of state $\ket{1}$} & {$88.5$} & {$84.7$} & {$9.3$} & {$\%$} \\[.09cm]
		\end{tabular}
	\caption{\textbf{List of device parameters.}}\label{tab:paras}
\end{table}


\section{\label{sec:wi} Wiring information}
\begin{figure*}[t]
	\includegraphics[width=0.97\textwidth]{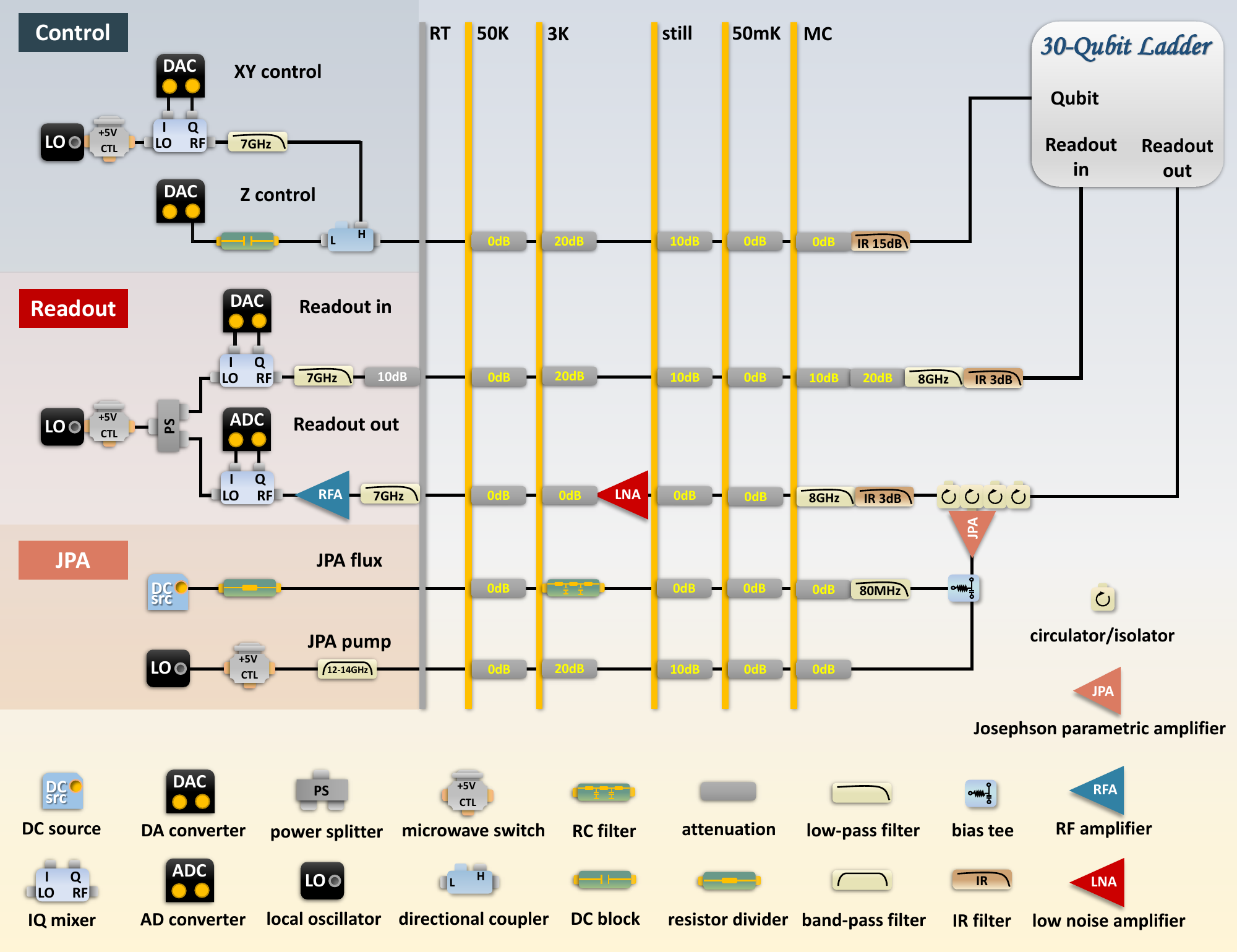}
	\caption{\textbf{Schematic diagram of the experimental system and wiring information.}}\label{fig:wiring}
\end{figure*}

The typical wiring information is shown in Fig.~\ref{fig:wiring}, in which from up to down are the control lines of qubit (XY and Z), readout, and Josephson parametric amplifier (JPA), respectively. From left to right, the ambient temperature decreases from room temperature to $12~\!\mathrm{mK}$ in a BlueFors XLD-1000 dilution refrigerator. We combine the high-frequency XY signal with the low-frequency Z bias by using directional couplers at room temperature. The XY signals are generated via frequency mixing. In detail, we use the IQ mixer to mix the intrinsic local oscillation (LO) from a microwave signal source and the IQ signals generated from two channels of arbitrary waveform generator (AWG). The output microwave signal is programmable, which depends on the pulses written into IQ signals. The joint readout signals are sent through the transmission line and amplified by the JPA, a cryo low-noise amplifier (LNA) and a room-temperature RF amplifier (RFA), and finally demodulated by the analog-digital converter (ADC).

{
\section{\label{sec:xy}{XY drive in superconducting circuits}}

\subsection{{Single-qubit XY drive}}
A transmon qubit is composed of a capacitance $C$ and a nonlinear inductance $L$ (Josephson
junction or SQUID). Its Lagrangian $\mathcal{L}_0$ and Hamiltonian $H_0$ can be written as
\begin{eqnarray}
	\mathcal{L}_0 &=& \frac{Q^2}{2C}-\frac{\Phi^2}{2L}\\
	H_0 &=& \frac{Q^2}{2C}+\frac{\Phi^2}{2L}\label{eq:HamiltonianLC},
\end{eqnarray}
where $Q=\partial\mathcal{L}_0/\partial{\dot{\Phi}}=C\dot{\Phi}$ denotes the charge, and $\Phi$ is the magnetic flux of the circuit. Here, the nonlinear inductance of the Josepshon junction with energy $E_{\text{J}}$ can be written as $L=L_{\text{c}}/{\cos{\left(2\pi{\Phi}/{\Phi_0}\right)}}$, where $\Phi_0=\hbar\pi/\mathrm{e}$ is the superconducting flux quantum, $\mathrm{e}\approx 1.602\times10^{-19}$C is the electron charge, and $L_{\text{c}}=\Phi_0^2/(4\pi^2E_{\mathrm{J}})$ is the constant inductance. This nonlinear inductance can be easily derived from the definition $L=\mathrm{d}\Phi/\mathrm{d}I$ and the Josephson equation $I=I_{\text{c}}\sin{(2\pi{\Phi}/{\Phi_0})}$ with $I_{\text{c}}=2\pi{E_{\mathrm{J}}}/\Phi_0$ being the Josephson critical current. 

\begin{figure}[t]
	\includegraphics[width=0.45\textwidth]{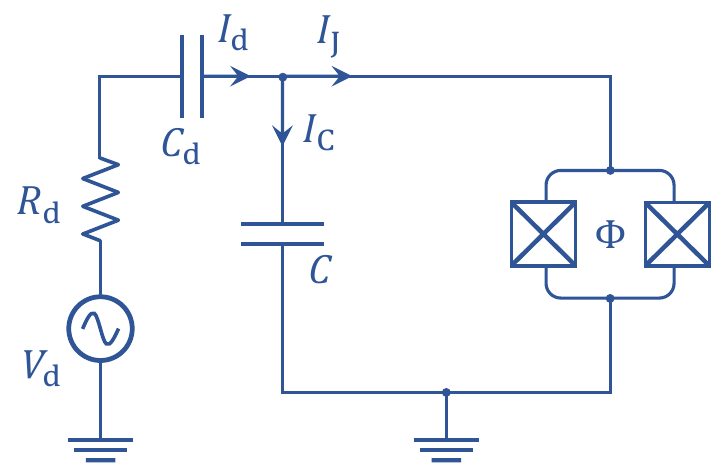}
	\caption{\textbf{Circuit diagram of a driven transmon qubit.} The qubit is coupled to a time-dependent driving voltage $V_\text{d}$. The capacitances of the qubit and the drive are labeled as $C$ and $C_\text{d}$, respectively. The magnetic flux threading the loop is denoted as $\Phi$. The driving current $I_{\text{d}}$ is split into $I_{\mathrm{C}}$ and $I_{\mathrm{J}}$.}\label{fig:xydrive1}
\end{figure}

Considering the weak flux $\Phi$, one can use the approximation $\cos{(2\pi{\Phi}/{\Phi_0})}\simeq{1-(2\pi{\Phi}/{\Phi_0})^2/2}$ and reduce the Hamiltonian Eq.~\eqref{eq:HamiltonianLC} into $H_0 \simeq \frac{Q^2}{2C}+\frac{\Phi^2}{2L_{\text{c}}}-\frac{\pi^2\Phi^4}{4L_{\text{c}}\Phi_0^2}$, which can be viewed as a harmonic oscillator with $o(\Phi^4)$ perturbation. Using canonical quantization, one can introduce
\begin{eqnarray}\label{eq:canonical_quantization}
	\begin{cases}
		&\!\!\!\!\hat{Q}=\mathrm{i}Q_{\text{zpf}}(\hat{a}^{\dagger}-\hat{a}) \cr
		&\!\!\!\!\hat{\Phi}=\Phi_{\text{zpf}}(\hat{a}^{\dagger}+\hat{a})
	\end{cases}
\end{eqnarray}
with $Q_{\text{zpf}}=\sqrt{\hbar(C/L_{\text{c}})^{\frac{1}{2}}/2}$ and $\Phi_{\text{zpf}}=\sqrt{\hbar(L_{\text{c}}/C)^{\frac{1}{2}}/2}$ being the zero point fluctuation of the charge and flux operators, respectively. The quantized Hamiltonian thus is (the constant term is omitted):
\begin{equation}
	\hat{H}_0 = \hbar\omega\hat{a}^{\dagger}\hat{a}-\frac{E_{\mathrm{C}}}{2}\hat{a}^{\dagger}\hat{a}^{\dagger}\hat{a}\hat{a},
\end{equation}
where $\omega=(\sqrt{8E_{\mathrm{C}}E_{\mathrm{J}}}-E_{\mathrm{C}})/\hbar$ denotes the qubit frequency, and $E_{\mathrm{C}}=\mathrm{e}^2/(2C)$ is the charging energy that represents the magnitude of anharmonicity. For a single Josephson junction, $E_{\mathrm{J}}$ is not tunable, while for a SQUID with two junctions, it depends on the external flux $\Phi_{\text{ext}}$ applied to the junction region. In the experiments, we can adjust the qubit frequency $\omega$ via the external fast flux bias applied to the Z control line.

When a time-dependent driving voltage $V_\text{d}(t)$ is added into a transmon qubit (Fig.~\ref{fig:xydrive1}), the driving current $I_\text{d}$ can be split into the qubit capacitance term $I_{\mathrm{C}}$ and the Josephson junction term $I_{\mathrm{J}}$. Meanwhile, according to Kirchhoff voltage law, the total voltage reduction through either of the two branches must be zero. Thus, one can obtain the following motion equation
\begin{eqnarray}
\begin{cases}
	&\!\!\!\!I_\text{d}=I_{\mathrm{C}}+I_{\mathrm{J}} \cr
	&\!\!\!\!-\dot{V}_\text{d}+\frac{I_\text{d}}{C_\text{d}}+\frac{I_{\mathrm{C}}}{C}=0 \cr
	&\!\!\!\!-\dot{V}_\text{d}+\frac{I_\text{d}}{C_\text{d}}+L\ddot{I}_{\mathrm{J}}=0
\end{cases}
~\Rightarrow~~
\ddot{\Phi}+\frac{1}{C_{\Sigma}L}\Phi-\frac{C_{\text{d}}\dot{V}_{\text{d}}(t)}{C_{\Sigma}}=0,
\end{eqnarray}
where $C_{\Sigma}=C+C_\text{d}$, $\Phi=LI_{\mathrm{J}}$. Here $C$, $C_{\text{d}}$ and $L$ are the qubit capacitance, the driving capacitance, and the nonlinear inductance, respectively. The above equation can be viewed as the Euler-Lagrange equation:  $\frac{\partial \mathcal{L}_{\text{driven}}}{\partial \Phi}-\frac{\mathrm{d}}{\mathrm{d} t}\frac{\partial \mathcal{L}_{\text{driven}}}{\partial \dot{\Phi}}=0$, where the Lagrangian of this driven qubit can be constructed as
\begin{equation}\label{eq:LagrangianLCdriving}
	\mathcal{L}_{\text{driven}}=\frac{1}{2}C\dot{\Phi}^2+\frac{1}{2}C_\text{d}\left(V_\text{d}(t)-\dot{\Phi}\right)^2-\frac{\Phi^2}{2L},
\end{equation}
where $C_\text{d}$ is the driving capacitance. In Eq.~\eqref{eq:LagrangianLCdriving}, the first term represents the charge energy of $C$, the second term denotes the charge energy of $C_{\text{d}}$ caused by induced electromotive force, and the last term is the inductance energy of $L$.

To obtain the Hamiltonian, we first calculate the conjugate to the position (flux) $\Phi$, namely the canonical momentum (charge) $\tilde{Q}={\partial \mathcal{L}_{\text{driven}}}/{\partial \dot{\Phi}}=C_{\Sigma}\dot{\Phi}-C_\text{d}V_\text{d}(t)$, and thus
\begin{equation}
	H_{\text{driven}}=\tilde{Q}\dot{\Phi}-\mathcal{L}_{\text{driven}}=\frac{\tilde{Q}}{2C_{\Sigma}}+\frac{\Phi^2}{2L}+\frac{\tilde{Q}C_\text{d}V_\text{d}(t)}{C_{\Sigma}}.
\end{equation}
Using the canonical quantization procedure like Eq.~\eqref{eq:canonical_quantization}, we introduce $\hat{\tilde{Q}}=\text{i}\tilde{Q}_{\text{zpf}}(\hat{a}^{\dagger}-\hat{a})$ and $\hat{\Phi}=\Phi_{\text{zpf}}(\hat{a}^{\dagger}+\hat{a})$ to quantize the driven system, where $\tilde{Q}_{\text{zpf}}=\sqrt{\hbar(C_{\Sigma}/L_{\text{c}})^{\frac{1}{2}}/2}$ and $\Phi_{\text{zpf}}=\sqrt{\hbar(L_{\text{c}}/C_{\Sigma})^{\frac{1}{2}}/2}$. Hence, the Hamiltonian becomes
\begin{equation}~\label{eq:ham_driven}
	\hat{H}_{\text{driven}}=\hbar\omega \hat{a}^{\dagger}\hat{a}-\frac{E_{\mathrm{C}}}{2}\hat{a}^{\dagger}\hat{a}^{\dagger}\hat{a}\hat{a}+\mathrm{i}\hbar\Omega(t)(\hat{a}^{\dagger}-\hat{a}),
\end{equation}
where $E_{\text{C}}=\mathrm{e}^2/(2C_{\Sigma})$, $E_{\mathrm{J}}={\Phi_0^2}/{(4\pi^2L_{\text{c}})}$, $\omega=(\sqrt{8E_{\mathrm{C}}E_{\mathrm{J}}}-E_{\mathrm{C}})/\hbar$, $\Omega(t)=\epsilon V_{\text{d}}(t)$, $\epsilon={\tilde{Q}_{\text{zpf}}C_\text{d}}/{(\hbar C_{\Sigma})}$. Here, we set the time-dependent driving $V_{\text{d}}(t)=-V_{\text{d}}\sin\left(\omega_{\text{d}}t+\phi\right)=\mathrm{Im}\{V_{\text{d}}e^{-\mathrm{i}(\omega_{\text{d}}t+\phi)}\}$, thus $\Omega(t)=\mathrm{i}\Omega\left(e^{\mathrm{i}(\omega_{\text{d}}t+\phi)}-e^{-\mathrm{i}(\omega_{\text{d}}t+\phi)}\right)/2$, where $\Omega=\epsilon V_{\text{d}}$ is so-called Rabi frequency. The parameter $\epsilon$ represents the Rabi frequency corresponding to the unit amplitude of the drive.

To solve the time evolution governed by the above time-dependent Hamiltonian, we consider the rotating frame which is generated by $\hat{U}_d(t)=e^{\mathrm{i}\omega_{\text{d}}t\hat{a}^{\dagger}\hat{a}}$
\begin{eqnarray}
	\hat{H}_{d}&=&\hat{U}_{d}(t)\hat{H}_{\text{driven}}(t)\hat{U}^{\dagger}_{d}(t)+\mathrm{i}\hbar\left(\frac{\mathrm{d}}{\mathrm{d}t}\hat{U}_{d}(t)\right)\hat{U}^{\dagger}_{d}(t)\nonumber\\
	&\simeq&\hbar\Delta \hat{a}^{\dagger}\hat{a}-\frac{E_{\mathrm{C}}}{2}\hat{a}^{\dagger}\hat{a}^{\dagger}\hat{a}\hat{a}+\frac{\hbar\Omega}{2}\!\left(\hat{a}^{\dagger}e^{-\mathrm{i}\phi}\!+\!\hat{a}e^{\mathrm{i}\phi}\right),
\end{eqnarray}
where $\Delta=\omega-\omega_{\text{d}}$ is the frequency detuning, and the rotating-wave approximation is adopted by ignoring high frequency oscillation $\pm2\omega_{\text{d}}$.

With $\Delta=0$ and $E_{\mathrm{C}}\gg\Omega$, the large anharmonicity results in the resonant drive acting almost exclusively between the first two energy levels $\ket{0}$ and $\ket{1}$ without leakage to higher levels. Hence, considering the two-level qubit, we have
\begin{equation}\label{eq:H_d}
	\hat{H}_{d}=\frac{\hbar\Omega}{2}\left(\hat{\sigma}^{+}e^{-\mathrm{i}\phi}+\hat{\sigma}^{-}e^{\mathrm{i}\phi}\right),
\end{equation}
where $\hat{\sigma}^{+}_j$ ($\hat{\sigma}^{-}_j$) is the raising (lowering) operator. If the qubit begins in the ground state $\ket{0}$, its time-dependent state during the unitary evolution is
\begin{equation}
	\ket{\psi_{d}(t)}=e^{-\frac{\mathrm{i}}{\hbar}\hat{H}_{d}t}\ket{0}=\cos{\frac{\Omega t}{2}}\ket{0}-\mathrm{i}e^{\mathrm{i}\phi}\sin{\frac{\Omega t}{2}}\ket{1},
\end{equation}
and the probability of qubit in $\ket{1}$ is given by $P_{1}(t)=\sin^2{({\Omega t}/{2})}=\left[1-\cos{\left(\Omega t\right)}\right]/2$. Considering the energy relaxation, the envelope of $P_{1}(t)$ will decay in a dissipative evolution and thus
\begin{equation}~\label{eq:rabi_prob1}
	P_{1}(t)=\frac{1}{2}\left[1-e^{-\frac{t}{T_1}}\cos{\left(\Omega t\right)}\right],
\end{equation}
where $T_1$ is the energy relaxation time that depends on the qubit frequency $\omega$. In order to obtain the Rabi frequency $\Omega$, one can fit the data of $P_{1}(t)$ by using the form of function $A\exp{(-{t}/{T_1})}\cos{\left(\Omega t\right)}+B$. Typical experimental data of calibrating XY drive with different driving amplitudes are displayed in Fig.~\ref{fig:xyamp2rabifreq}.

\begin{figure}[t]
	\includegraphics[width=0.7\textwidth]{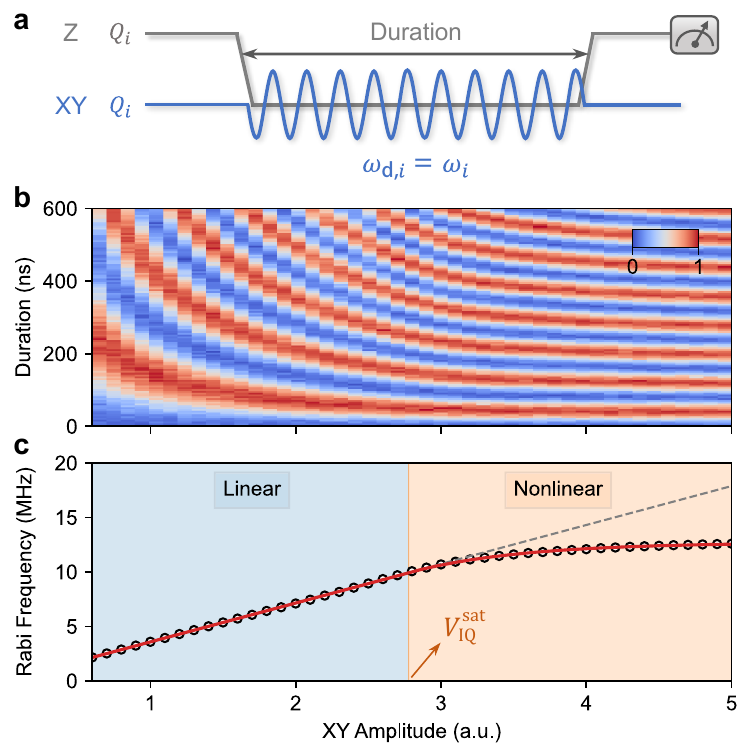}
	\caption{\textbf{Typical experimental data of measuring the relationship between Rabi frequency and XY drive amplitude.} \textbf{a}, Experimental pulse sequence. Qubit is detuned from its idle frequency to the operating $\omega_i$. Meanwhile, we apply resonant microwave drives on this qubit with scanning XY amplitude $V_{\text{IQ}}$ and measure the vacuum Rabi oscillations shown in \textbf{b}. \textbf{b}, The heatmap of the probabilities of qubit in the state $\ket{1}$ as a function of duration and XY amplitude. \textbf{c}, For each XY drive amplitude, we fit the curve of vacuum Rabi oscillation by using Eq.~\eqref{eq:rabi_prob1} to obtain the experimental Rabi frequency, denoted as black hollow circle. The red solid line is the result of fitting the experimental Rabi frequencies by using a smooth piecewise function and the grey dashed line implies the linear relationship between Rabi frequency and XY drive amplitude when the drive amplitude is less than $V^{\text{sat}}_{\text{IQ}}$.}\label{fig:xyamp2rabifreq}
\end{figure}

The above results are based on the resonance condition $\omega=\omega_{\text{d}}$. If the detuning $\Delta=\omega-\omega_{\text{d}}\neq0$, the effective Rabi frequency will be
\begin{equation}
\Omega_{\text{R}}=\sqrt{\Delta^2 + \Omega^2}.
\end{equation}
Therefore, to obtain the correct Rabi frequency when $\omega=\omega_{\text{d}}$, we should find the corresponding Z pulse amplitude that makes the qubit resonate with the microwave before calibrating XY drive. This step can be easily achieved via spectroscopy experiment or Rabi oscillation by scanning the Z pulse amplitude of the qubit.

\subsection{\label{sec:iqmixer}{Generation and manipulation}}
As shown in Fig.~\ref{fig:iqmixer}, we generate XY drive pulse by using IQ mixer. The output driving pulse results from mixing the IQ signals with a intrinsic LO (Fig.~\ref{fig:iqmixer}). Although the Rabi frequency $\Omega$ is proportional to the actual driving amplitude $V_{\text{d}}$, the relationship between $\Omega$ and the input amplitude of IQ signals $V_{\text{IQ}}$ is not always linear due to the semiconductor nature of the IQ mixer (GaAs and similar semiconductor materials). When $V_{\text{IQ}}$ is relatively small, IQ mixer is in the linear work region and $V_{\text{d}}\propto V_{\text{IQ}}$ satisfies. However, the strong amplitude leads to a nonlinear relationship between $V_{\text{d}}$ and $V_{\text{IQ}}$, so that $\Omega\propto V_{\text{IQ}}$ is not valid in the saturation region. This may be caused by the velocity saturation of carriers in the IQ mixer. In order to analytically describe $\Omega$ versus $V_{\text{IQ}}$, we impose the following smooth piecewise function and its inverse:
\begin{eqnarray}
	\Omega&=&
	\begin{cases}
		&\!\!\!\!\eta V_{\text{IQ}},\qquad\qquad\qquad\qquad\qquad\qquad\qquad (V_{\text{IQ}} \leq V^{\text{sat}}_{\text{IQ}})\\
		&\!\!\!\!\Omega_{\text{max}} - \left(\Omega_{\text{max}} - \eta V^{\text{sat}}_{\text{IQ}}\right)e^{-\frac{\eta (V_{\text{IQ}} - V^{\text{sat}}_{\text{IQ}})}{\Omega_{\text{max}} - \eta V^{\text{sat}}_{\text{IQ}}}},\!\!\!\!\qquad(V_{\text{IQ}} > V^{\text{sat}}_{\text{IQ}})\\
	\end{cases}\label{eq:xyamp2rabifreq_fit}\\
	V_{\text{IQ}}&=&
	\begin{cases}
		&\!\!\!\!\frac{1}{\eta}\Omega,\!\qquad\qquad\qquad\qquad\qquad\qquad\qquad\quad 
		(\Omega \leq \eta V^{\text{sat}}_{\text{IQ}})\\
		&\!\!\!\!V^{\text{sat}}_{\text{IQ}}+(\frac{\Omega_{\text{max}}}{\eta}-V^{\text{sat}}_{\text{IQ}})\ln\left(\frac{\Omega_{\text{max}} - \eta V^{\text{sat}}_{\text{IQ}}}{\Omega_{\text{max}} - \Omega}\right),\!\!\!\qquad
		(\Omega > \eta V^{\text{sat}}_{\text{IQ}})\\
	\end{cases}
\end{eqnarray}
where $\eta$, $V^{\text{sat}}_{\text{IQ}}$ and $\Omega_{\text{max}}$ are the parameters to be fitted. Here $\eta$ is the slope in linear region that represents the Rabi frequency corresponding to the unit amplitude of XY driving (IQ signals), $V^{\text{sat}}_{\text{IQ}}$ denotes the critical amplitude before entering the saturation region of IQ mixer, and $\Omega_{\text{max}}$ is the maximum Rabi frequency when $V_{\text{IQ}}\rightarrow\infty$.

\begin{figure}[t]
	\includegraphics[width=0.6\textwidth]{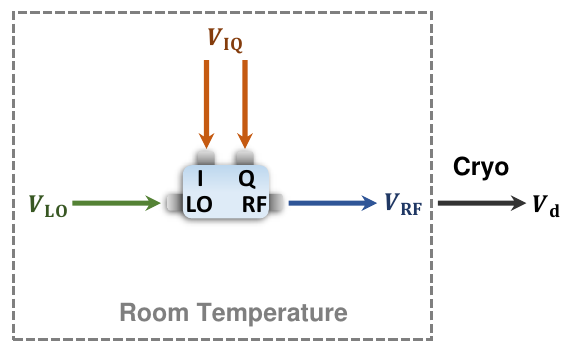}
	\caption{\textbf{Generation of XY drive via frequency mixing.} The intrinsic local oscillation (LO) is generated from a microwave signal source, while the input IQ signals are generated from two channels of the arbitrary waveform generator. The whole circuit is mixed at room temperature and then goes into cryoelectronics (dilution refrigerator). If the amplitude of LO is fixed, the output pulse amplitude will be proportional to the amplitude of IQ signals in small amplitude cases where the IQ mixer is in a linear work region.}\label{fig:iqmixer}
\end{figure}

\subsection{{Origin of multi-qubit crosstalk}}
Now we consider two driven qubits $Q_i$ and $Q_j$ in the circuit (see Fig.~\ref{fig:xydrive2}). The total Lagrangian can be expressed as
\begin{equation}\label{eq:LagrangianTwoQubits}
	\mathcal{L}^{(i,j)}_{\text{driven}}=\sum_{q=i,j}\left(\frac{1}{2}C_q\dot{\Phi}_q^2-\frac{\Phi_q^2}{2L_q}\right)+\frac{1}{2}C_{\text{d},i}\left(V_{\text{d},i}(t)-\dot{\Phi}_i\right)^2+\frac{1}{2}C_{\text{d},j}\left(V_{\text{d},j}(t)-\dot{\Phi}_j\right)^2+\frac{1}{2}C_{ij}\left(\dot{\Phi}_j-\dot{\Phi}_i\right)^2,
\end{equation}
where $C_{ij}$ is the coupling capacitance. The corresponding canonical momentums are
\begin{equation}
\begin{bmatrix}
\tilde{Q}_i \\
\tilde{Q}_j \\
\end{bmatrix}
=
\begin{bmatrix}
\frac{\partial \mathcal{L}^{(i,j)}_{\text{driven}}}{\partial \dot{\Phi}_i} \\
\frac{\partial \mathcal{L}^{(i,j)}_{\text{driven}}}{\partial \dot{\Phi}_j}
\end{bmatrix}
=
\begin{bmatrix}
	{C}_{\Sigma_i}+C_{ij} & -C_{ij}\\
	-C_{ij} & {C}_{\Sigma_j}+C_{ij}
\end{bmatrix}
\begin{bmatrix}
\dot{\Phi}_i \\
\dot{\Phi}_j
\end{bmatrix}
-
\begin{bmatrix}
C_{\text{d},i}V_{\text{d},i} \\
C_{\text{d},j}V_{\text{d},j}
\end{bmatrix},
\end{equation}
where $C_{\Sigma_i}=C_i+C_{\text{d},i}$ and $C_{\Sigma_j}=C_j+C_{\text{d},j}$, and thus
\begin{equation}\label{eq:Phi_ij}
\begin{bmatrix}
\dot{\Phi}_i \\
\dot{\Phi}_j
\end{bmatrix}
=
\frac{1}{\|\mathbf{C}\|}
\begin{bmatrix}
{C_{\Sigma_j}+C_{ij}} & {C_{ij}}\\
{C_{ij}} & {C_{\Sigma_i}+C_{ij}}
\end{bmatrix}
\begin{bmatrix}
\tilde{Q}_i+C_{\text{d},i}V_{\text{d},i} \\
\tilde{Q}_j+C_{\text{d},i}V_{\text{d},i}
\end{bmatrix},
\end{equation}
where ${\|\mathbf{C}\|}={C_{\Sigma_i}C_{\Sigma_j}+C_{\Sigma_i}C_{ij}+C_{\Sigma_j}C_{ij}}$ is the determinant of the capacitance matrix $\mathbf{C}=\begin{bmatrix}
	{C}_{\Sigma_i}+C_{ij} & -C_{ij}\\
	-C_{ij} & {C}_{\Sigma_j}+C_{ij}
\end{bmatrix}$. Substituting Eq.~\eqref{eq:Phi_ij} into Eq.~\eqref{eq:LagrangianTwoQubits}, we obtain
\begin{equation}
	\mathcal{L}^{(i,j)}_{\text{driven}}=\frac{\tilde{Q}_i^2}{2\tilde{C}_{\Sigma_i}}+\frac{\tilde{Q}_j^2}{2\tilde{C}_{\Sigma_j}}+\frac{\tilde{Q}_i\tilde{Q}_j}{\tilde{C}_{ij}},
\end{equation}
with the effective capacitance parameters
\begin{eqnarray}
	\tilde{C}_{\Sigma_i}&=&C_{\Sigma_i}+(C_{\Sigma_j}\|C_{ij})=C_{\Sigma_i}+\frac{C_{\Sigma_j}C_{ij}}{C_{\Sigma_j}+C_{ij}},\\
	\tilde{C}_{\Sigma_j}&=&C_{\Sigma_j}+(C_{\Sigma_i}\|C_{ij})=C_{\Sigma_j}+\frac{C_{\Sigma_i}C_{ij}}{C_{\Sigma_i}+C_{ij}},\\
	\tilde{C}_{ij}&=&\frac{C_{\Sigma_i}C_{ij}+C_{\Sigma_j}C_{ij}+C_{\Sigma_i}C_{\Sigma_j}}{C_{ij}}.
\end{eqnarray}
Then the total Hamiltonian is given by the Legendre transformation:
\begin{eqnarray}
H^{(i,j)}_{\text{driven}}&=&\tilde{Q}_i\dot{\Phi}_i+\tilde{Q}_j\dot{\Phi}_j-\mathcal{L}^{(i,j)}_{\text{driven}}\nonumber\\
&=&\sum_{q=i,j}\!\left(\frac{\tilde{Q}_q^2}{2\tilde{C}_{\Sigma_q}}+\frac{\Phi_q^2}{2L_q}\right)\!+\!\frac{\tilde{Q}_i\tilde{Q}_j}{\tilde{C}_{ij}}\!+\!\left(\frac{C_{\text{d},i}}{\tilde{C}_{\Sigma_i}}V_{\text{d},i}(t)\!+\!\frac{C_{\text{d},j}}{\tilde{C}_{ij}}V_{\text{d},j}(t)\right)\!\tilde{Q}_i\!+\!\left(\frac{C_{\text{d},j}}{\tilde{C}_{\Sigma_j}}V_{\text{d},j}(t)\!+\!\frac{C_{\text{d},i}}{\tilde{C}_{ij}}V_{\text{d},i}(t)\right)\!\tilde{Q}_j.
\end{eqnarray}
Using canonical quantization, we introduce
\begin{eqnarray}
\begin{cases}
&\!\!\!\!\hat{\tilde{Q}}_q=\mathrm{i}\tilde{Q}_{\text{zpf},q}(\hat{a}^{\dagger}_q-\hat{a}_q) \cr
&\!\!\!\!\hat{\Phi}_q=\Phi_{\text{zpf},q}(\hat{a}^{\dagger}_q+\hat{a}_q)
\end{cases}
\end{eqnarray}
with $q\in{\{i,j\}}$, $\tilde{Q}_{\text{zpf},q}=\sqrt{\hbar(\tilde{C}_{\Sigma_q}/L_{\text{c},q})^{\frac{1}{2}}/2}$ and $\Phi_{\text{zpf},q}=\sqrt{\hbar(L_{\text{c},q}/\tilde{C}_{\Sigma_q})^{\frac{1}{2}}/2}$. The quantized Hamiltonian thus is
\begin{eqnarray}
\hat{H}^{(i,j)}_{\text{driven}}&=&\hat{H}^{(i)}_{\text{driven}}+\hat{H}^{(j)}_{\text{driven}}+\hat{H}^{(i,j)}_{\text{int}},
\\
\hat{H}^{(q)}_{\text{driven}}&=&\hbar\omega_q \hat{a}^{\dagger}_q\hat{a}_q-\frac{E_{\mathrm{C}_q}}{2}\hat{a}^{\dagger}_q\hat{a}_q^{\dagger}\hat{a}_q\hat{a}_q+\mathrm{i}\hbar\tilde{\Omega}_q(t)(\hat{a}^{\dagger}_q-\hat{a}_q),\quad q\in{\{i,j\}},\label{eq:local_driving_ham}
\\
\hat{H}^{(i,j)}_{\text{int}}&=&\hbar{J_{i,j}}(\hat{a}^{\dagger}_i-\hat{a}_i)(\hat{a}_j-\hat{a}^{\dagger}_j),
\end{eqnarray}
where the parameters are
\begin{eqnarray}
\hbar\omega_q&=&\sqrt{8E_{\mathrm{C}_q}E_{\mathrm{J}_q}}-E_{\mathrm{C}_q},\quad E_{\mathrm{C}_q}=\frac{\mathrm{e}^2}{2\tilde{C}_q},\quad E_{\mathrm{J}_q}=\frac{\Phi_0^2}{4\pi^2L_{\text{c},q}},
\\
J_{i,j}&=&\frac{\tilde{Q}_{\text{zpf},i}\tilde{Q}_{\text{zpf},j}}{\hbar\tilde{C}_{ij}}=\frac{\sqrt{\tilde{C}_{\Sigma_i}\tilde{C}_{\Sigma_j}}}{2\tilde{C}_{ij}}\sqrt{\left(\omega_i+{\frac{E_{\mathrm{C}_i}}{\hbar}}\right)\left(\omega_j+{\frac{E_{\mathrm{C}_j}}{\hbar}}\right)}\approx\frac{C_{ij}\sqrt{\omega_i\omega_j}}{2\sqrt{(C_{\Sigma_i}+C_{ij})(C_{\Sigma_j}+C_{ij})}},
\\
\tilde{\Omega}_i(t)&=&\epsilon_{ii}\left(V_{\text{d},i}(t)+\frac{\epsilon_{ij}}{\epsilon_{ii}}V_{\text{d},j}(t)\right),\quad\tilde{\Omega}_j(t)=\epsilon_{jj}\left(V_{\text{d},j}(t)+\frac{\epsilon_{ji}}{\epsilon_{jj}}V_{\text{d},i}(t)\right),\label{eq:omega_tilde}
\\
\epsilon_{ii}&=&\frac{\tilde{Q}_{\text{zpf},i}C_{\text{d},i}}{\hbar\tilde{C}_{\Sigma_i}},\quad
\epsilon_{ij}=\frac{\tilde{Q}_{\text{zpf},i}C_{\text{d},j}}{\hbar\tilde{C}_{ij}},\quad
\epsilon_{jj}=\frac{\tilde{Q}_{\text{zpf},j}C_{\text{d},j}}{\hbar\tilde{C}_{\Sigma_j}},\quad
\epsilon_{ji}=\frac{\tilde{Q}_{\text{zpf},j}C_{\text{d},i}}{\hbar\tilde{C}_{ij}}.\label{eq:crosstalk_cij}
\end{eqnarray}
Focusing on Eqs.~\eqref{eq:local_driving_ham}, \eqref{eq:omega_tilde}~and~\eqref{eq:crosstalk_cij}, one can notice that the local driving Hamiltonian of each qubit depends on both external drive $V_{\text{d},i}(t)$ and $V_{\text{d},j}(t)$ due to the presence of coupling capacitance. However, this crosstalk is usually very small. As an example, we take the typical values $C_{\text{d},i}=C_{\text{d},j}=30~\!\mathrm{aF}$, $C_{i}=C_{j}=85~\!\mathrm{fF}$ and $C_{ij}=0.25~\!\mathrm{fF}$. Then we have $\epsilon_{ij}/\epsilon_{ii}=\epsilon_{ji}/\epsilon_{jj}\approx0.3\%$, suggesting a low level of this crosstalk. 
Given the above equations, we note that the local driving Hamiltonian of each qubit is subject to both external drive $V_{\text{d},i}(t)$ and $V_{\text{d},j}(t)$ due to the presence of coupling capacitance. However, this crosstalk is usually very small. In fact, most of the crosstalk comes from the classical microwave crosstalk. The total crosstalk is the sum of the classical microwave crosstalk and the crosstalk due to the coupling capacitance. In the following, we will establish a model to describe the total crosstalk and introduce an efficient method for measuring the crosstalk matrix.

\begin{figure}[t]
	\includegraphics[width=0.9\textwidth]{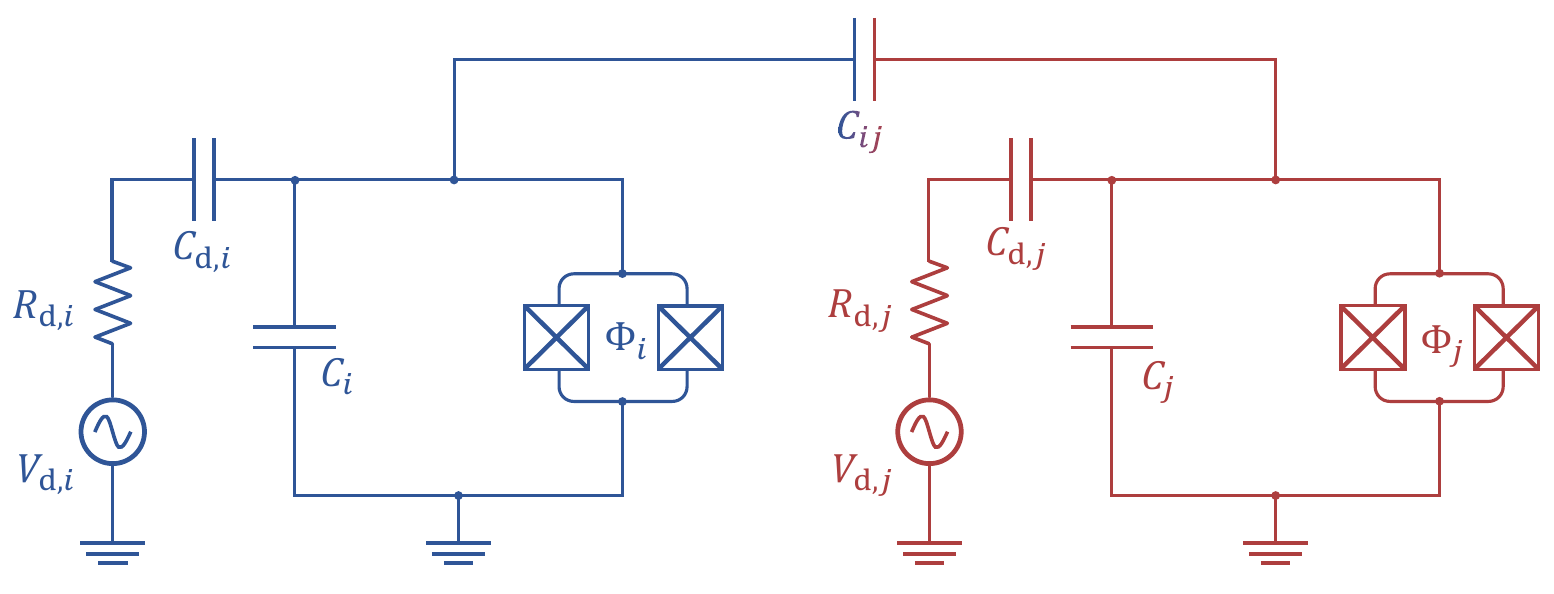}
	\caption{\textbf{Circuit diagram of two driven transmon qubits.} Two qubits are labeled as $Q_i$ and $Q_j$, which are coupled to their respective time-dependent driving voltages $V_{\mathrm{d},i}(t)$ and $V_{\mathrm{d},j}(t)$. The coupling capacitance between the two qubits is represented as $C_{ij}$, and $\Phi$, $C$ and $C_\text{d}$ are the dominant mode flux, the capacitance of the qubit and the capacitance of the drive, respectively.}\label{fig:xydrive2}
\end{figure}

When the microwave signal travels through the medium on the chip, it can be described by the following plane wave form (the medium is assumed to be homogeneous):
\begin{equation}
	V_{\text{d}}(\mathbf{r},t)=V_{\text{d}}(t)e^{\mathrm{i}\mathbf{k}\cdot\mathbf{r}}.
\end{equation}
Here the wave vector $\mathbf{k}$ is generally complex, namely $\mathbf{k}=\mathbf{b}+\mathrm{i}\mathbf{a}$, thus we have
\begin{equation}
	\mathrm{i}\mathbf{k}\cdot\mathbf{r}=-\mathbf{a}\cdot\mathbf{r}+\mathrm{i}\mathbf{b}\cdot\mathbf{r},
\end{equation}
where the first term is the amplitude attenuation induced by the imaginary part of $\mathbf{k}$ and the second term is the phase retardation caused by the real part. Here we define $\xi=\mathbf{a}\cdot\mathbf{r}$ is the amplitude attenuation factor and $\varphi=\mathbf{b}\cdot\mathbf{r}$ is the phase retardation.

\begin{figure}[b]
	\includegraphics[width=0.65\textwidth]{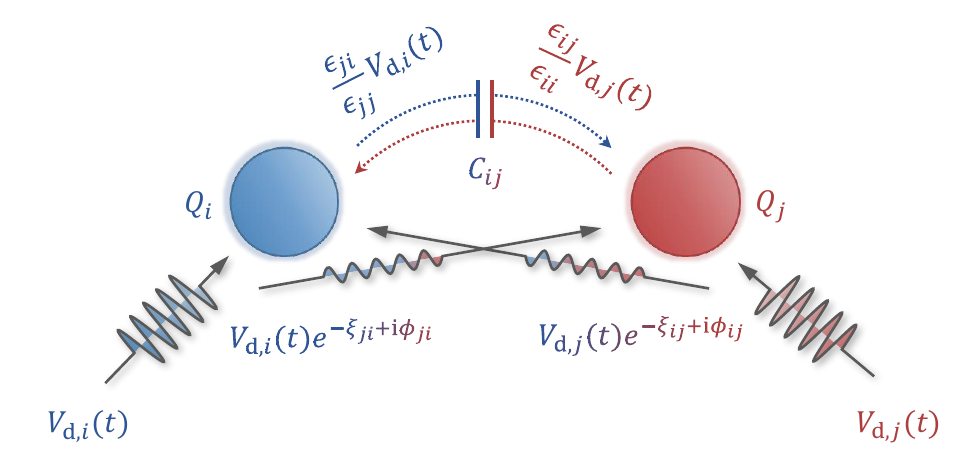}
	\caption{\textbf{Schematic of microwave signal crosstalk.} Here, we take two qubits $Q_i$ and $Q_j$ as an example. Their individual driving voltages $V_{\mathrm{d},i}(t)$ and $V_{\mathrm{d},j}(t)$ induce two types of crosstalk. One type of crosstalk is due to the presence of coupling capacitance $C_{ij}$, which causes the crosstalk only in amplitude. The parameters $\epsilon_{ij}$ and $\epsilon_{ji}$ are explained in Eq.~\eqref{eq:crosstalk_cij}, which depends on the coupling capacitance $C_{ij}$ between the two qubits. The other type of crosstalk is caused by the propagation of microwave signals through the medium on the chip. According to electrodynamics, it will lead to the crosstalk both in amplitude and phase. The parameters $\xi$ and $\phi$ are the amplitude attenuation factor and phase retardation of microwave propagation, respectively.}\label{fig:xytalkdiagram}
\end{figure}

As shown in Fig.~\ref{fig:xytalkdiagram}, the signal $V_{\text{d},i}(t)$ propagates from $Q_i$ to $Q_j$ with a factor $e^{-\xi_{ji}+\mathrm{i}\phi_{ji}}$ attached, which implies the classical microwave crosstalk of $Q_i$ to $Q_j$. Similarly, the classical microwave crosstalk of $Q_j$ to $Q_i$ can be express as $V_{\text{d},j}(t)e^{-\xi_{ij}+\mathrm{i}\phi_{ij}}$. Here we also consider the crosstalk caused by the coupling capacitance as Eq.~\eqref{eq:omega_tilde}. Therefore, the total signals perceived by $Q_i$ and $Q_j$ are
\begin{eqnarray}
	\tilde{V}_{\text{d},i}(t)&=&V_{\text{d},i}(t)+\frac{\epsilon_{ij}}{\epsilon_{ii}}V_{\text{d},j}(t)+V_{\text{d},j}(t)e^{-\xi_{ij}+\mathrm{i}\phi_{ij}},
	\\
	\tilde{V}_{\text{d},j}(t)&=&V_{\text{d},j}(t)+\frac{\epsilon_{ji}}{\epsilon_{jj}}V_{\text{d},i}(t)+V_{\text{d},i}(t)e^{-\xi_{ji}+\mathrm{i}\phi_{ji}},
\end{eqnarray}
or written in matrix form
\begin{equation}
	\begin{bmatrix}
		\tilde{V}_{\text{d},i}(t) \\
		\tilde{V}_{\text{d},j}(t)
	\end{bmatrix}
	=
	\begin{bmatrix}
		1 & {v_{ij}}e^{\mathrm{i}\varphi_{ij}}\\
		{v_{ji}}e^{\mathrm{i}\varphi_{ji}} & 1
	\end{bmatrix}
	\begin{bmatrix}
		V_{\text{d},i}(t) \\
		V_{\text{d},j}(t)
	\end{bmatrix}
\end{equation}
with the definitions of ${v_{ij}}e^{\mathrm{i}\varphi_{ij}}={\epsilon_{ij}}/{\epsilon_{ii}}+e^{-\xi_{ij}+\mathrm{i}\phi_{ij}}$ and ${v_{ji}}e^{\mathrm{i}\varphi_{ji}}={\epsilon_{ji}}/{\epsilon_{jj}}+e^{-\xi_{ji}+\mathrm{i}\phi_{ji}}$. To generalize the above formula to the case of each qubit with crosstalks from all other qubits, we define the vectors $\mathbf{\tilde{V}_{\!d}}(t)=[\tilde{V}_{\text{d},1}(t),\tilde{V}_{\text{d},2}(t),\ldots,~\!\tilde{V}_{\text{d},N}(t)]^{T}$ and  $\mathbf{V_{\!d}}(t)=[V_{\text{d},1}(t),V_{\text{d},2}(t),\dots,~\!V_{\text{d},N}(t)]^{T}$, then
\begin{equation}
	\mathbf{\tilde{V}_{\!d}}(t)=\mathbf{M_{V}}\mathbf{V_{\!d}}(t),
\end{equation}
in which $\mathbf{M_{V}}$ is the signal crosstalk matrix
\begin{equation}
	\mathbf{M_{V}}=\begin{bmatrix}
		1 & {v_{12}}e^{\mathrm{i}\varphi_{12}} & \cdots & {v_{1N}}e^{\mathrm{i}\varphi_{1N}}\\
		{v_{21}}e^{\mathrm{i}\varphi_{21}} & 1 & \cdots & {v_{2N}}e^{\mathrm{i}\varphi_{2N}}\\
		\vdots & \vdots & \ddots & \vdots\\
		{v_{N1}}e^{\mathrm{i}\varphi_{N1}} & {v_{N2}}e^{\mathrm{i}\varphi_{N2}} & \cdots & 1
	\end{bmatrix}.
\end{equation}

\subsection{{Measurement and correction of crosstalk}}
To compensation the crosstalk, we need to measure the total signal crosstalk matrix and perform
\begin{equation}
	\mathbf{V_{\!d}}(t)=\mathbf{M^{-1}_{V}}~\!\mathbf{\tilde{V}_{\!d}}(t),
\end{equation}
where $\mathbf{M^{-1}_{V}}$ is the inverse matrix. However, in practice we cannot obtain $\mathbf{M_{V}}$ directly, we need to characterize the crosstalk matrix of Rabi frequencies $\mathbf{M_{\Omega}}$ and calculate $\mathbf{M_{V}}$ by using
\begin{equation}
	\mathbf{M_{V}}=\bm{\epsilon}~\!\mathbf{M_{\Omega}}~\!\bm{\epsilon^{-1}},
\end{equation}
where $\bm{\epsilon}=\mathrm{diag}\{\epsilon_{11},\epsilon_{22},\ldots,~\!\epsilon_{NN}\}$ and the crosstalk matrix of Rabi frequencies is defined as
\begin{equation}
	\mathbf{M_{\Omega}}=\begin{bmatrix}
		1 & {c_{12}}e^{\mathrm{i}\varphi_{12}} & \cdots & {c_{1N}}e^{\mathrm{i}\varphi_{1N}}\\
		{c_{21}}e^{\mathrm{i}\varphi_{21}} & 1 & \cdots & {c_{2N}}e^{\mathrm{i}\varphi_{2N}}\\
		\vdots & \vdots & \ddots & \vdots\\
		{c_{N1}}e^{\mathrm{i}\varphi_{N1}} & {c_{N2}}e^{\mathrm{i}\varphi_{N2}} & \cdots & 1
	\end{bmatrix}.
\end{equation}
where $c_{ij}$ and $\varphi_{ij}$ are the amplitude and phase crosstalk coefficients to be measured.

\begin{figure}[t]
	\includegraphics[width=0.95\textwidth]{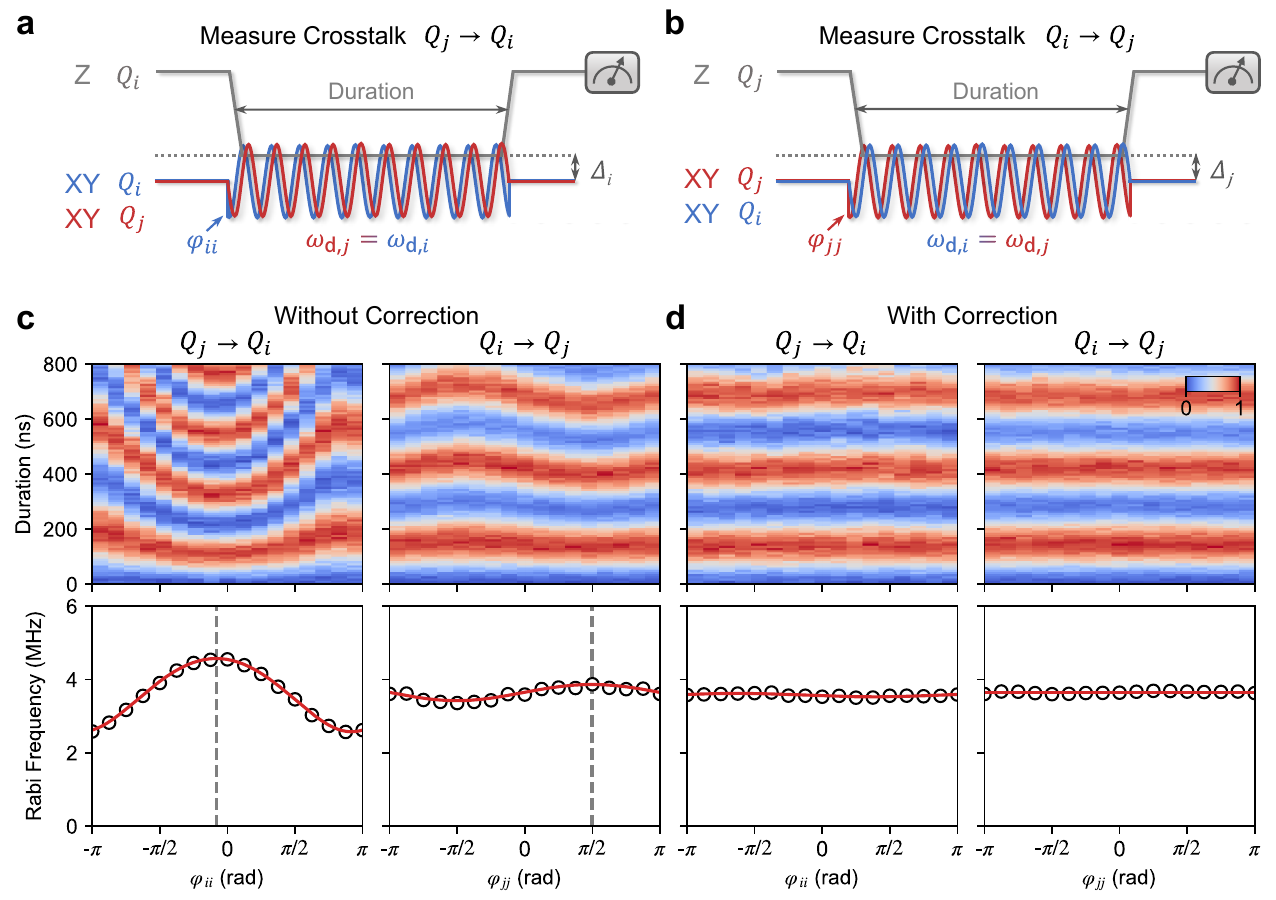}
	\caption{\textbf{Measurement of the microwave crosstalk.} \textbf{a}, Experimental pulse sequence for measuring the crosstalk from $Q_j$ to $Q_i$. \textbf{b}, Experimental pulse sequence for measuring the crosstalk from $Q_i$ to $Q_j$. The parameters $\varphi_{ii}$ and $\varphi_{jj}$ denote the additional phases added into the XY control lines of $Q_i$ and $Q_j$, respectively. The detuning between the qubit frequency and XY drive frequency is defined as $\Delta_q=\omega_q-\omega_{\text{d},q}$, which is usually set to zero. \textbf{c}, Typical experimental data of measuring crosstalk without correction. \textbf{d}, Typical experimental data of measuring crosstalk with correction. The heatmap represents the probabilities of qubit in $\ket{1}$. The black hollow circle denotes the effective Rabi frequency obtained by fitting the Rabi oscillation. The red solid line is the result of fitting the effective Rabi frequency by using Eq.~\eqref{eq:xytalk_fit}. The grey dashed line implies the fitted crosstalk phase.}\label{fig:xytalk}
\end{figure}

In the linear region of IQ mixer, we actually use Eq.~\eqref{eq:xyamp2rabifreq_fit} to describe the relationship between Rabi frequency and the input IQ signal, and thus
\begin{equation}
	\mathbf{M_{V_{IQ}}}=\bm{\eta}~\!\mathbf{M_{\Omega}}~\!\bm{\eta^{-1}},
\end{equation}
where $\bm{\eta}$ is given by $\bm{\eta}=\mathrm{diag}\{\eta_{1},\eta_{2},\ldots,~\!\eta_{N}\}$ with $\eta_i$ being the Rabi frequency of $Q_i$ corresponding to the unit amplitude of IQ signals.

Now, we introduce an efficient method for characterizing $c_{ij}$ and $\varphi_{ij}$ in the crosstalk matrix $\mathbf{M_{\Omega}}$.
. Let us take an example of $Q_i$. As shown in Fig.~\ref{fig:xytalk}\textbf{a}, two resonant microwave signals $\omega_{\text{d},i}=\omega_{\text{d},j}=\omega_{\text{d}}$ are simultaneously input from the XY control lines of $Q_i$ and $Q_j$. Meanwhile, $Q_i$ is biased near the resonant frequency with the detuning $\Delta_i=\omega_i-\omega_{\text{d},i}$. Due to the crosstalk, the effective Hamiltonian of $Q_i$ under the rotation frame becomes $\hat{H}_{d}^{(i)}=\Delta_i\hat{\sigma}^{+}_i\hat{\sigma}^{-}_i+\left(\tilde{\Omega}_i\hat{\sigma}^{+}_i+\mathrm{H.c.}\right)/{2}$ with $\tilde{\Omega}_i=\Omega_ie^{-\mathrm{i}\phi_i}+c_{ij}\Omega_je^{\mathrm{i}(\varphi_{ij}-\phi_j)}$, and the corresponding effective Rabi frequency is
\begin{equation}\label{eq:xytalk_fit}
	\Omega_{\text{R}}^{(i)}=\sqrt{\Delta_i^2+\Omega_i^2+\Omega_{ij}^2+2\Omega_i\Omega_{ij}\cos{(\varphi_{ij}-\varphi_{ii})}},
\end{equation}
where $\Omega_{ij}=c_{ij}\Omega_j$ denotes the crosstalk Rabi frequency from $Q_j$ to $Q_i$, and $\varphi_{ii}=\phi_j - \phi_i$ represents the additional XY phase added in $Q_i$ relative to $Q_j$. By scanning $\varphi_{ii}$ and measure the probabilities of $Q_i$ in $\ket{1}$ as a function of the duration of XY drive, we can obtain $\Omega_{\text{R}}^{(i)}$. Using Eq.~\eqref{eq:xytalk_fit} to fit the results of $\Omega_{R}^{(i)}$, we can determine the crosstalk coefficients $c_{ij}$ and $\varphi_{ij}$. The procedure for determining $c_{ji}$ and $\varphi_{ji}$ is similar as long as we treat $Q_j$ as $Q_i$. Here we show the partial crosstalk matrix between the 24 qubits used in experiments in Fig.~\ref{fig:xytalkmat}.


\begin{figure}[t]
	\includegraphics[width=0.94\textwidth]{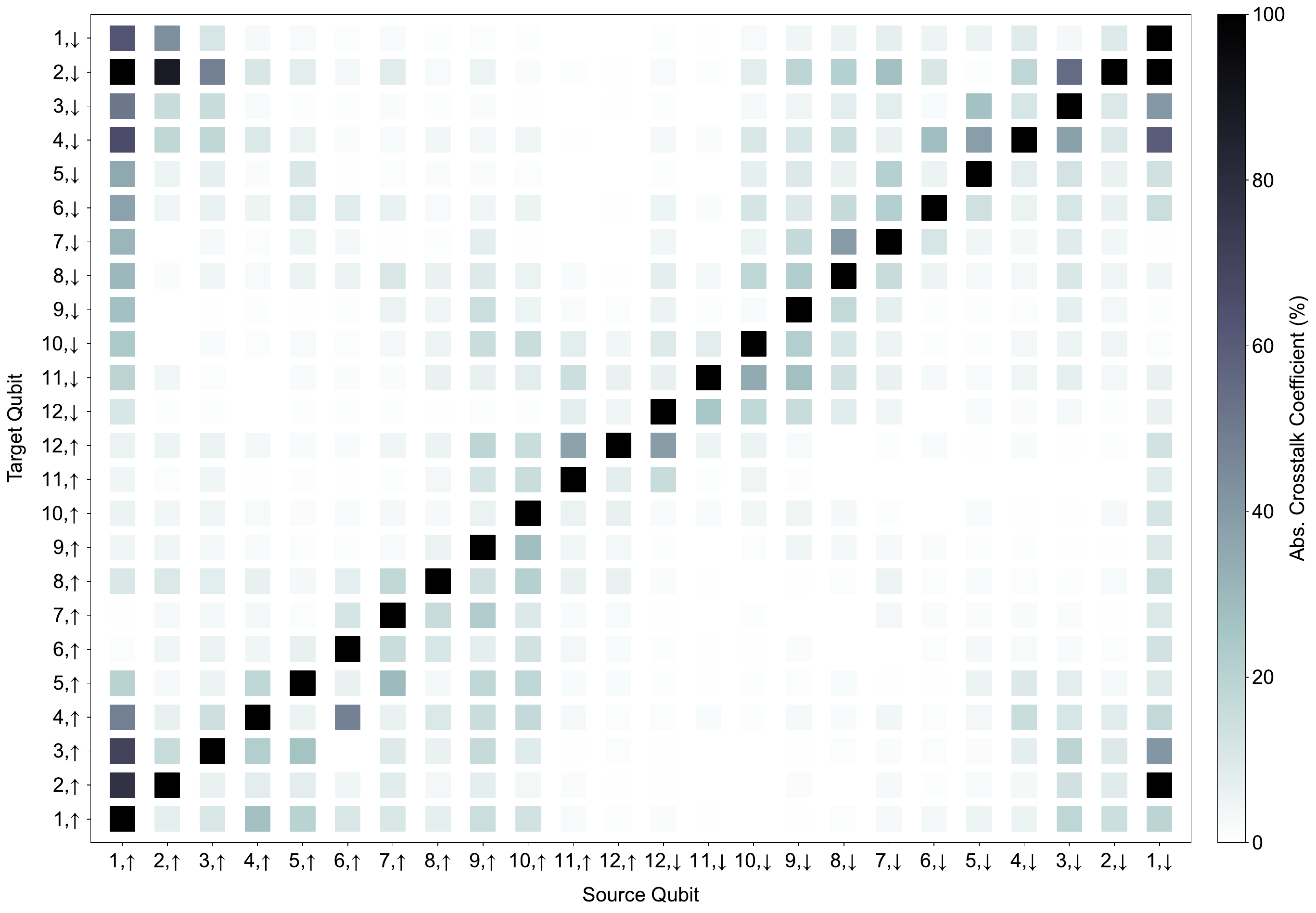}
	\caption{\textbf{Partial crosstalk matrix of XY drive.} The heatmap represents the modulus of the crosstalk coefficient, namely $\left|c_{ij}\right|$. Here, we show the crosstalks between 24 qubits in the ladder.}\label{fig:xytalkmat}
\end{figure}
}

\color{black}
\section{\label{sec:decoherence} The effect of decoherence}

In this section, we discuss the effect of decoherence. Since the conservation of the particle number is essentially important for the observation of spin hydrodynamics, we pay attention to the energy relaxation effect, characterized by the coherence time $T_{1}$. To quantify the impact of decoherence on the particle number, we numerically simulate the dynamics of $n(t)$ by solving the Lindblad master equation
\begin{equation}\label{lindblad}
	\frac{\text{d}\hat{\rho}(t)}{\text{d}t} = i[\hat{H},\hat{\rho}(t)] + \sum_{j=1}^{N}(\hat{L}_{j}\hat{\rho}(t)\hat{L}_{j}^{\dagger} - \frac{1}{2}\{\hat{L}_{j}^{\dagger}\hat{L}_{j},\hat{\rho}(t)\}), 
\end{equation}
where $\hat{\rho}(t) = |\psi(t)\rangle\langle\psi(t)|$ is the density matrix, $\hat{H}$ is the Hamiltonian Eq.~(1) in the main text, and $\hat{L}_{j} = \hat{\sigma}_{j}^{-}/\sqrt{T_{1}}$ represents the Lindblad operators for the energy relaxation, with $T_{1}$ being the energy lifetime.

\begin{figure*}[h!]
	\centering
	\includegraphics[width=0.92\linewidth]{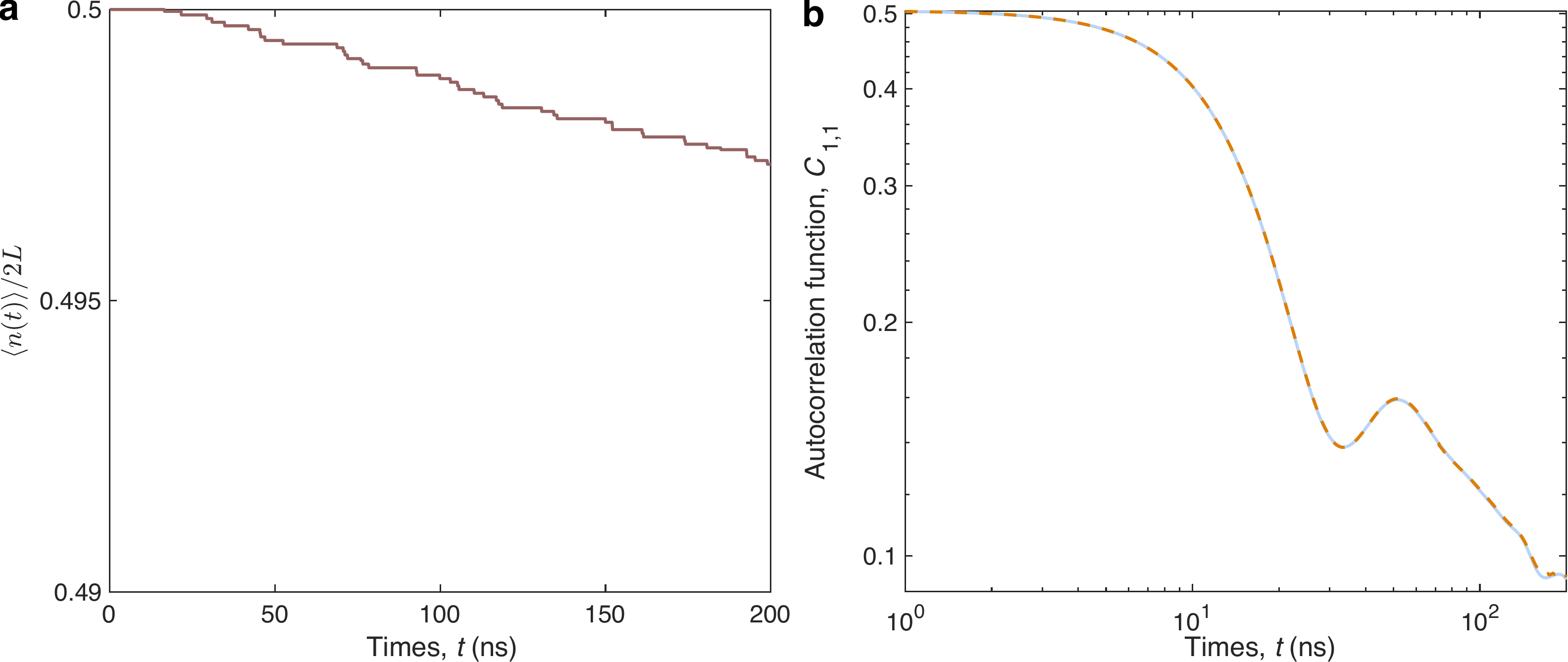}\\
	\caption{\textbf{The effect of decoherence.} \textbf{a}, For the qubit ladder with a length $L=8$ (the number of qubits $N=16$), the dynamics of particle number $\langle n(t) \rangle $ with decoherence, i.e., energy relaxation, quantified by $T_{1} = $ 32  $\mu$s.  \textbf{b},  The dynamics of autocorrelation function $C_{1,1}(t)$ with decoherence (dashed curve), in comparison with the unitary dynamics (solid curve).  }\label{fig_decoherence}
\end{figure*}

For the numerical simulation, we adopt $T_{1} = 32.1$ $\mu$s based on the device information shown in Table.~\ref{tab:paras}. Here, we consider a ladder with the number of qubits $N=16$, and the same initial state shown in the inset of Fig.~\ref{fig_high_p}\textbf{a}. We employ the stochastic Schr\"odinger equation to efficiently solve the Lindblad master equation (\ref{lindblad}). First, we study the dynamics of the particle number $\langle n(t) \rangle $ under the decoherence, and the results are plotted in Fig.~\ref{fig_decoherence}\textbf{a}. With the evolved time $t=200$ ns, the value of $\langle n(t)/2L\rangle $ is around $0.497$, suggesting that decoherence does not significantly influence the conservation of the particle number. We then numerically demonstrate that decoherence does not strongly affect the dynamics of autocorrelation function $C_{1,1}(t)$ with the evolved time up to 200 ns, and the dynamics of $C_{1,1}(t)$ simulated by solving the Lindblad master equation (\ref{lindblad}) is more or less the same to the unitary dynamics (see Fig.~\ref{fig_decoherence}\textbf{b}).

\color{black}
\section{\label{sec:ptd}XY drive approach to generate Haar-random states}


\begin{figure}[b]
	\includegraphics[width=0.97\textwidth]{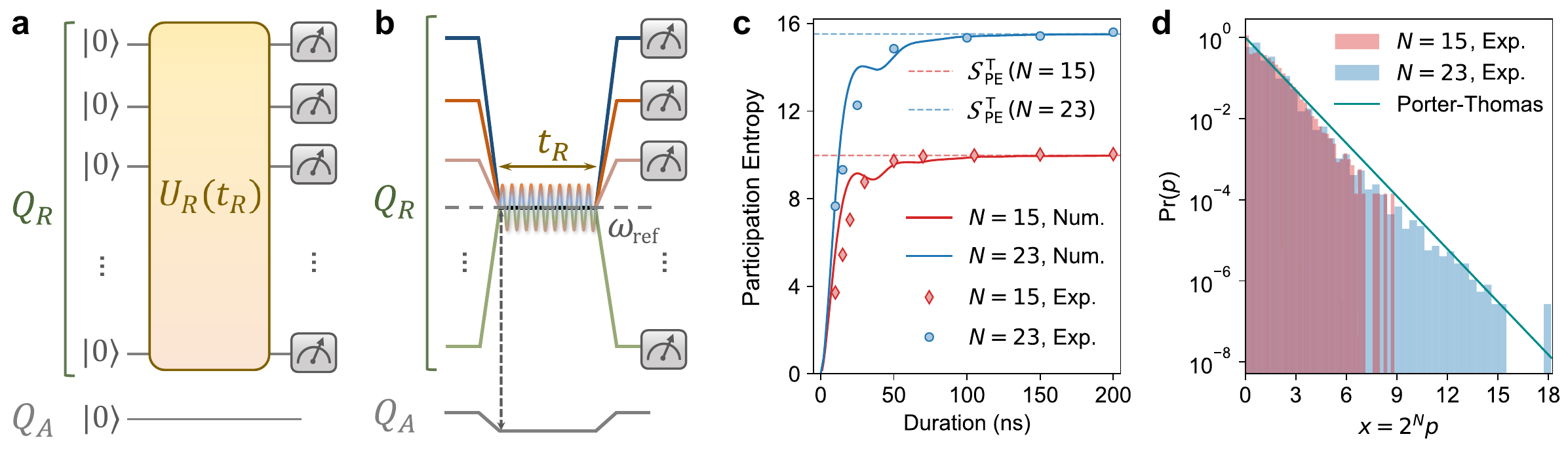}
	\caption{\textbf{Generation and characterization of the XY drive approach to prepare the Haar-random states.} \textbf{a}, The schematic diagram of the quantum circuit. \textbf{b}, The corresponding experimental pulse sequence. We bias the auxiliary qubit $Q_A$ away from the resonance frequency and apply the XY drive pulses on all the remainder qubits $Q_R$ participating in the resonance at frequency $\omega_{\text{ref}}\approx4.534~\!\mathrm{GHz}$, with a duration $t_{R}$. \textbf{c}, The evolution of {participation} entropy $S_{\text{PE}}$ vs. the duration of XY drive. The dashed line represents the {participation} entropy of $N-$qubit Haar-random state. Here, we fix $Q_{1,\uparrow}$ as $Q_A$, and $N$ is the total number of $Q_R$. \textbf{d}, The bitstring histogram of the measured $D=2^{N}$ joint probabilities. The solid line shows the ideal results of Poter-Thomas distribution. For $N=15$ and $N=23$, we perform $N_{\text{s}}=5\times10^5$ and $N_{\text{s}}=3\times10^7$ single-shot measurements, respectively}\label{fig:ptd}
\end{figure}

For the Haar-random state $\ket{\psi^R}$, we can define the probability with respect to the computational basis $|k\rangle$ as $p_{k} = |\langle k| \psi^R\rangle|^{2}$. It has been shown that the distribution of the probabilities $\{p=p_{k}\}$ will approximate the so-called Porter-Thomas distribution~[35-37]
\begin{equation}
\mathrm{Pr}(p)=De^{-Dp},
\end{equation}
where $D=2^N$ is the total dimension of the Hilbert space. To generate the Haar-random states via the evolution $\hat{U}_R$ in this experiment (seen in the main text or Fig.~\ref{fig:ptd}\textbf{a}), we bias the auxiliary qubit $Q_A$ away from the resonance frequency and apply the XY drive pulses on all the remainder qubits $Q_R$ participating in the resonance. The experimental pulse diagram is shown in Fig.~\ref{fig:ptd}\textbf{b}. After a time $t_R$, we perform joint readout of $Q_R$ with $N_{\text{s}}$ single-shot measurements to obtain the joint probabilities, and then calculate the {participation} entropy
\begin{equation}
S_{\text{PE}}(t_R)=-\sum_{k=1}^{D}p_k(t_R)\ln{p_k}(t_R),
\end{equation}
where $p_n$ is the joint probabilities of all $D=2^N$ bitstrings.
As shown in Fig.~\ref{fig:ptd}\textbf{c}, the {participation} entropy increases rapidly and then tends to a stable value. This value matches the {participation} entropy of the Haar-random state, namely
\begin{equation}
S_{\text{PE},\ket{\psi^R}}=-D\int_{0}^{\infty}\!\mathrm{d}p\;\mathrm{Pr}(p)~\!p\ln{p}=\ln{D}-1+\gamma,
\end{equation}
where $\gamma\approx0.577$ is the Euler constant. The final state after a long-time evolution is therefore closer to a Haar-random state, which shows the Poter-Thomas distribution of the bitstring joint probabilities in the statistical histogram, see Fig.~\ref{fig:ptd}\textbf{d}. In the experiment, we select $t_R=200~\!\mathrm{ns}$ to generate the Haar-random state and use this state as the initial state for subsequent interactions.

We note that the von Neumann entanglement entropy (EE) can also characterize the Haar-random states by achieving the Page value $S_{\text{Page}}\simeq \log m - m/2n$, where $m$ and $n$ represent the dimension of Hilbert space of the subsystem and the remainder, respectively. However, experimental measurement of EE requires additional single-qubit rotations, which can influence the accuracy of the results, especially for large system sizes. Here, we adopt the participation entropy, which can be directly measured by single-shot readout in $z$-direction, without rotations of qubits.

\begin{figure}[t]
	\includegraphics[width=0.97\textwidth]{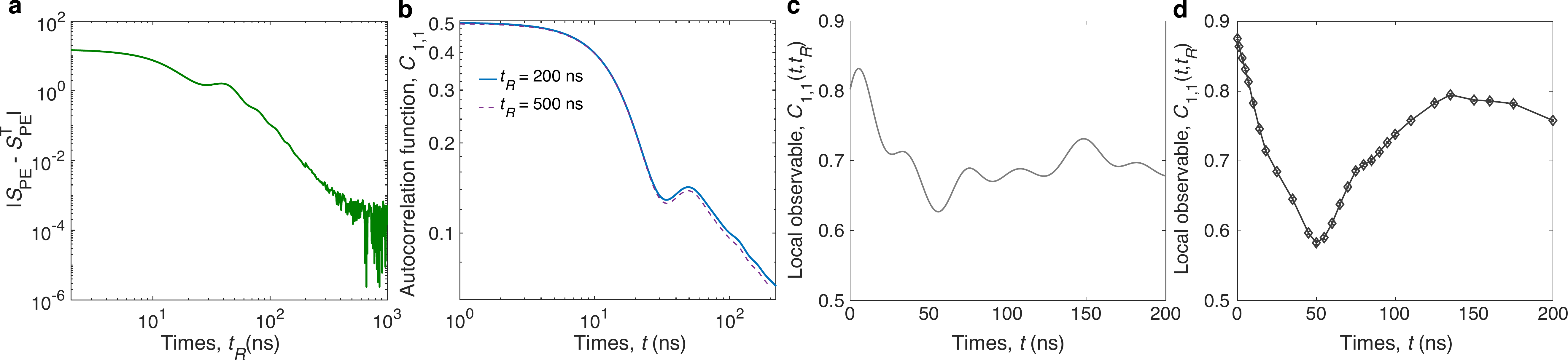}
	\caption{\textbf{Impact of different $t_{R}$ for generating Haar-random states.} \textbf{a}, The difference between the participation entropy at an evolved time $t_{R}$ and that corresponding to Haar-random states $S_{\text{PE}}^{T}$, i.e., $|S_{\text{PE}}(t_{R})-S_{\text{PE}}^{T}|$. \textbf{b}, The numerical results of autocorrelation function $C_{1,1}$ for the qubit ladder with $L=12$, and different states generated from $\hat{U}_{R}(t_{R})$ with $t_{R}=200~\mathrm{ns}$ and $500~\mathrm{ns}$. \textbf{c}, The numerical simulation of the dynamics of the local observable $C_{1,1}(t_{R},t)$ with a fixed $t_{R}=15~\mathrm{ns}$. \textbf{d}, The experimental data for the dynamics of the local observable $C_{1,1}(t_{R},t)$ with a fixed $t_{R}=15~\mathrm{ns}$.}\label{fig:suppadd1}
\end{figure}

{We now discuss the impact of different evolved time $t_{R}$ for generating Haar-random states on the measurement of infinite-temperature autocorrelation function $C_{1,1}$. In Fig.~\ref{fig:suppadd1}\textbf{a}, we plot the numerical results of the difference between the participation entropy of the quenched state at $t=t_R$ and the participation entropy corresponding to the Haar-random state, i.e., $|S_{\text{PE}}(t_{R})-S_{\text{PE}}^{T}|$ with the evolved time $t_{R}$ up to 1 $\mu$s. It can be seen that with $t_{R}\simeq 200~\mathrm{ns}$, the difference reaches $|S_{\text{PE}}(t)-S_{\text{PE}}^{T}|\sim 10^{-1}$, and a lower difference can be achieved for longer evolved time $t$. However, as shown in Fig.~~\ref{fig:suppadd1}\textbf{b}, the dynamical behaviors of autocorrelation function $C_{1,1}$, with the states generated by different evolved time of $\hat{U}_{R}(t_{R})$ with $t_{R}\geq 200~\mathrm{ns}$, do not have a significant change, which indicates that the evolved time $t_{R}\simeq 200~\mathrm{ns}$ is sufficient to generate a faithful Haar-random state for measuring the infinite-temperature spin transport. }

{We then extensively study the dynamics with short $t_{R}$. In this case, the state  $|\psi^{R}\rangle$ is far away from Haar-random states, and the local observable $\langle \psi_{\beta}^{R}| \hat{\sigma}_{\alpha}^{z}(t) |\psi_{\beta}^{R} \rangle $, with $ |\psi_{\beta}^{R} \rangle = \hat{U}_{R}(t_{R})\otimes_{i\in Q_{R}}|0\rangle_{i}$, can no longer be approximate with the infinite-temperature correlation function $\text{Tr}[\hat{\sigma}_{\alpha}^{z}(t)\hat{\sigma}_{\beta}^{z}]/D$. Consequently, we denote the quantity as local observable $C_{1,1}(t_{R},t)$, with $t_{R}$ and $t$ being the evolved time for $\hat{U}_{R}(t_{R})$ and $\hat{U}_{H}(t)$ shown in the quantum circuit shown in Fig.~1\textbf{c} in the main text, respectively. In Fig.~\ref{fig:suppadd1}\textbf{c} and \textbf{d}, we plot the numerical and experimental data for the dynamics of the local observable $C_{1,1}(t_{R},t)$ with a fixed short time $t_{R}=15~\mathrm{ns}$, respectively. For a small $t_{R}=15~\mathrm{ns}$, after an initial drop, the local observable has an oscillation around a value larger than 0.5. This can be explained by the fact that   the state $|\psi^{R}\rangle$ is close to the initial state $|00...0\rangle$ with small $t_{R}$, and when $t_{R}=0$ and $t=0$, actually, based on Eq.~(3) of the main text, $c_{1,\uparrow;1,\uparrow} = c_{1,\uparrow;1,\downarrow} = c_{1,\downarrow;1,\downarrow} = c_{1,\downarrow;1,\uparrow} = 1$, which leads to the local observable $C_{1,1}(0,0)=1$. }

\section{\label{sec:finite1}Finite-size effect for the spin transport in the clean superconducting qubit ladder}

\begin{figure}[t]
	\includegraphics[width=0.46\textwidth]{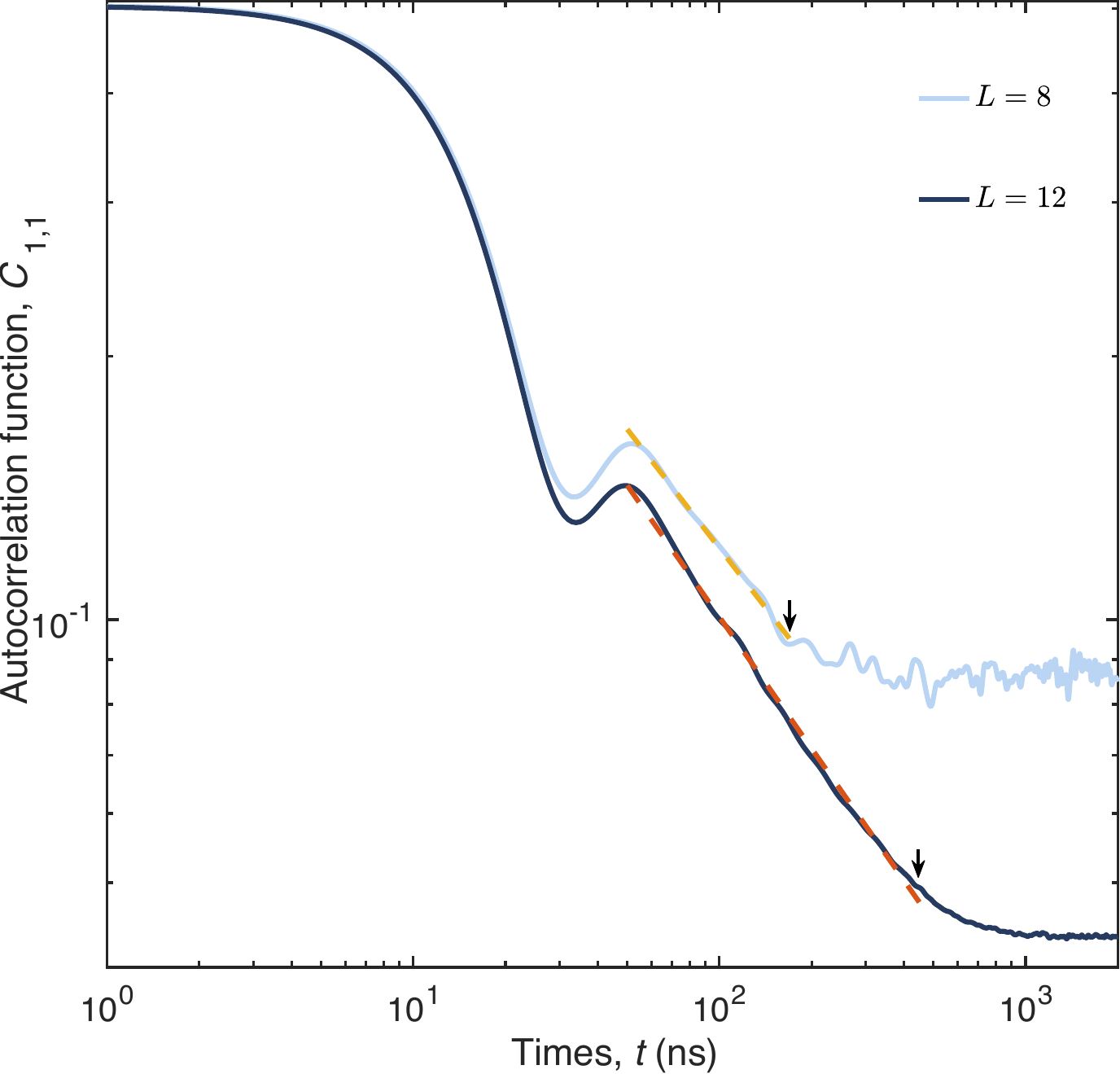}
	\caption{\textbf{Finite-size effect.} Numerical simulation of the autocorrelation function $C_{1,1}(t)$ for the qubit ladder with different system sizes. For $L=12$, the system consists of 24 qubits, i.e., $Q_{1,\uparrow},...,Q_{12,\uparrow}$ and $Q_{1,\downarrow},...,Q_{12,\downarrow}$.  For $L=8$, the system consists of 16 qubits, i.e., $Q_{1,\uparrow},...,Q_{8,\uparrow}$ and $Q_{1,\downarrow},...,Q_{8,\downarrow}$. The dashed lines show the power-law fitting of the numerical results in the time interval $t\in[50 ~\mathrm{ns}, 170~\mathrm{ns}]$ for $L=8$ and $t\in[50~\mathrm{ns}, 450~\mathrm{ns}]$ for $L=12$. }\label{fig:suppadd2}
\end{figure}

{In this section, we discuss the finite-size effect of the spin transport. We consider the clean superconducting qubit ladder without disorder or linear potential as an example, where the diffusive transport is expected to occur. We numerically simulate a long time evolution with the final time $t=2000~\mathrm{ns}$ ($t\overline{J^{\parallel}}\simeq 91.2$). As shown in Fig.~\ref{fig:suppadd2}, due to the finite-size effect, the $C_{1,1}(t)$ will saturate to a stable value for long time. The time interval with the power-law decay $C_{1,1}\propto t^{-z}$ becomes longer for larger $L$. For $L=8$ and $12$, the estimated time intervals with the power-law decay are $t\in[50 ~\mathrm{ns}, 170 \text{ns}]$ and $t\in[50 ~\mathrm{ns}, 450 ~\mathrm{ns}]$ (highlighted by the arrows in Fig.~\ref{fig:suppadd2}), respectively. By fitting the numerical data in the time interval for $L=8$ in $t\in[50 ~\mathrm{ns}, 140 ~\mathrm{ns}]$ and $L=12$ in $t\in[50 ~\mathrm{ns}, 450~\mathrm{ns}]$, we obtain the exponent $z\simeq 0.45$ for $L=8$ and $z\simeq 0.5$ for $L=12$. In short, the signature of diffusive transport becomes more clear for larger system size. }

\section{\label{sec:finite2}Finite-time effect for the spin transport in disordered systems}

\begin{figure}[t]
	\includegraphics[width=0.8\textwidth]{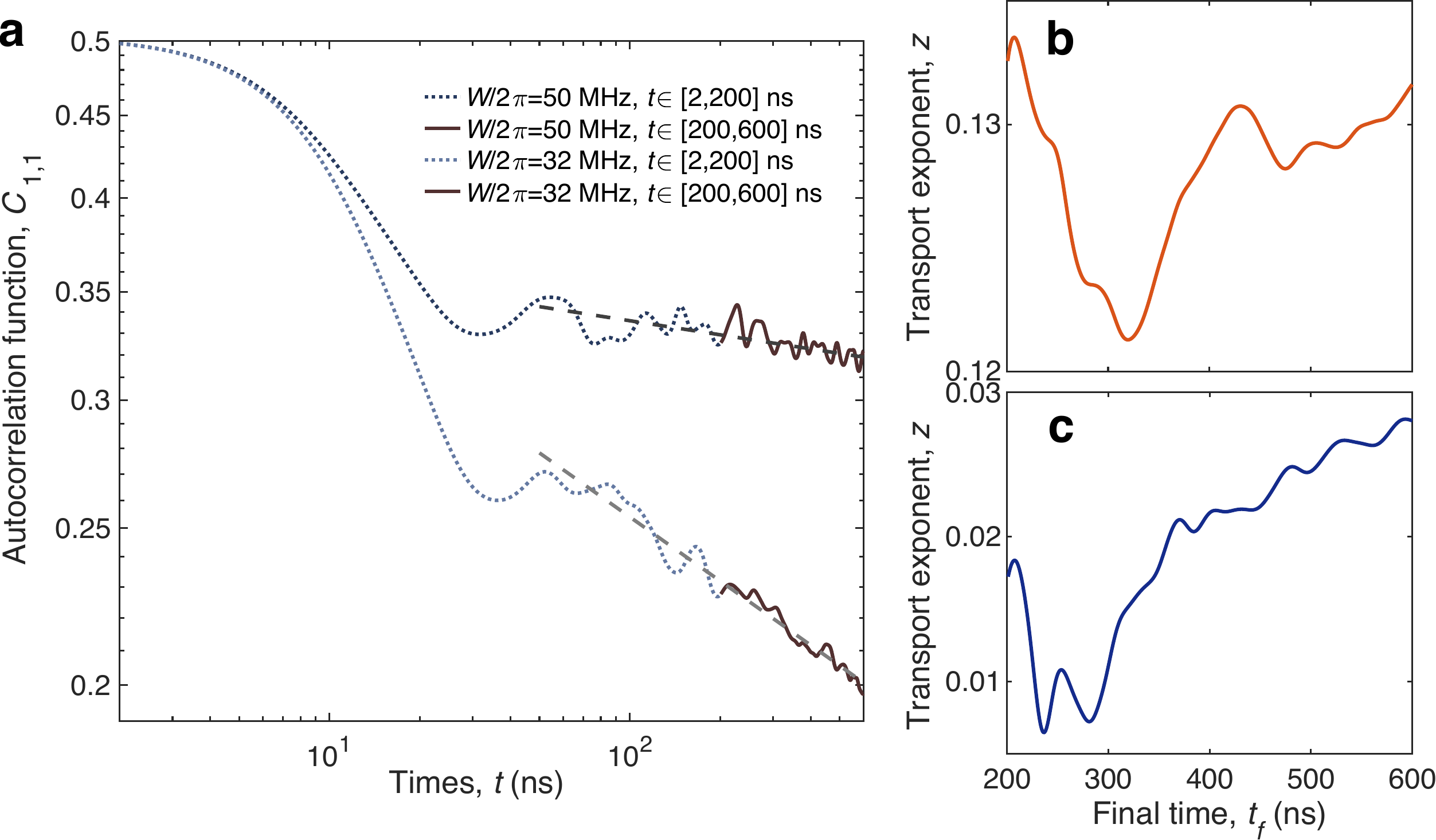}
	\caption{\textbf{Impact of the finite-time effect.} \textbf{a}, Numerical results for the time evolution of autocorrelation function $C_{1,1}(t)$ for the qubit ladder with $L=12$, and two values of disorder strengths $W/2\pi = 32~\mathrm{MHz}$ and $50~\mathrm{MHz}$. The evolved time is up to a longer time $t=600~\mathrm{ns}$. The dashed lines show the power-law fitting $C_{1,1} \propto t^{-z}$. \textbf{b}, For the disordered system with $W/2\pi=32~\mathrm{MHz}$, the transport exponent $z$ obtained from the power-law fitting for the numerical results with the time interval $t\in [t_{i},t_{f}]$, $t_{i}=50~\mathrm{ns}$, and different $t_{f}$. \textbf{c} is similar to \textbf{b}, but for the disordered system with $W/2\pi=50~\mathrm{MHz}$.  }\label{fig:suppadd3}
\end{figure}

{Here, we consider a longer evolved final time  $t=600~\mathrm{ns}$, and study the impact of longer final time on the transport exponent $z$ obtained by the power-law fitting $C_{1,1} \propto t^{-z}$. We focus on the disordered systems with $W/2\pi=32~\mathrm{MHz}$ and $50~\mathrm{MHz}$.  With the time window $t\in[50,200]$ ns, as shown in the Fig.~3\textbf{b} of the main text, $z\simeq 0.02$ and $z\simeq 0.13$ for $W/2\pi = 50~\mathrm{MHz}$ and $32~\mathrm{MHz}$, respectively. With the time window $t\in[50,600]$ ns, the fittings are shown in Fig.~\ref{fig:suppadd3}\textbf{a} with $z\simeq 0.03$ and $z\simeq 0.13$ for $W/2\pi = 50~\mathrm{MHz}$ and $32~\mathrm{MHz}$, respectively. It is seen that with the time window $t\in[50,600]$ ns, the transport exponents $z$ are slightly larger than those for the time window $t\in[50,200]$ ns. }

{We also plot the transport exponent $z$ obtained from the power-law fitting in the time interval $t\in[t_{i},t_{f}]$, with a fixed initial time $t_{i}=20$ ns, and different $t_{f}$ in Fig.~\ref{fig:suppadd3}\textbf{b} and \textbf{c} for $W/2\pi=32~\mathrm{MHz}$ and $50~\mathrm{MHz}$, respectively. It is shown that with longer final time $t_{f}$, the transport exponent $z$ exhibits a propensity to increase.  }

\section{\label{sec:an}Additional numerics and discussions}

In this section, we numerically study another type of autocorrelation functions which are defined by the average over a product state $|\psi_{0}\rangle$. In the main text, we focus on the infinite-temperature autocorrelation function $C_{\textbf{r},\textbf{r}} = \text{Tr}[\hat{\rho}_{\textbf{r}}(t)\hat{\rho}_{\textbf{r}}]/D$ with $D$ being the dimension of the Hilbert space. Alternatively, one can also consider the autocorrelation function average over a product state $|\psi_{0}\rangle$, i.e., 
\begin{equation}\label{c_psi0}
C_{\textbf{r},\textbf{r}} (|\psi_{0}\rangle)=\langle \psi_{0}| \hat{\rho}_{\textbf{r}}(t)\hat{\rho}_{\textbf{r}} |\psi_{0}\rangle. 
\end{equation}
Here, we reveal that the autocorrelation function $C_{\textbf{r},\textbf{r}} (|\psi_{0}\rangle)$ cannot show generic properties of spin transport, and the dynamics of $C_{\textbf{r},\textbf{r}} (|\psi_{0}\rangle)$ is highly dependent on the choice of $|\psi_{0}\rangle$. 

\begin{figure}[t]
	\includegraphics[width=0.97\textwidth]{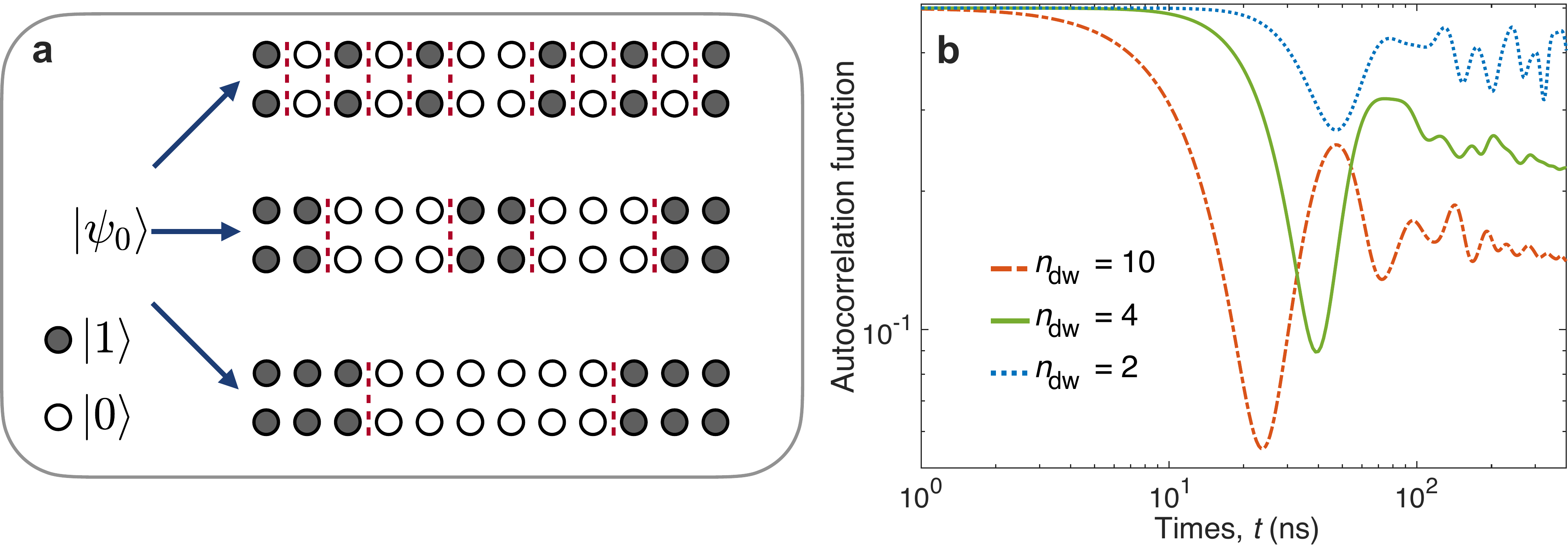}
	\caption{\textbf{Additional numerical results for the spin transport on the titled superconducting qubit ladder.} \textbf{a}, Schematic diagram of three different product states $
		|\psi_{0}\rangle$ for the definition of the autocorrelation function $C_{1,1}=\langle \psi_{0}|\hat{\rho}_{1}(t)\hat{\rho}_{1}|\psi_{0}\rangle$. From the top to bottom, the domain wall number of product states $|\psi_{0}\rangle$ is $n_{\text{dw}} = 10$, $4$, and $2$, respectively. \textbf{b}, Time evolution of the autocorrelation function $C_{1,1}=\langle \psi_{0}|\hat{\rho}_{1}(t)\hat{\rho}_{1}|\psi_{0}\rangle$ with the product states shown in \textbf{a} for the titled superconducting qubit ladder with $W_{S}/2\pi = 60$ MHz. }\label{fig:stark_add}
\end{figure}

We consider the titled superconducting qubit ladder consisting of 24 qubits with $W_{S}/2\pi = 60$ MHz, and the slope of the linear potential $\gamma/2\pi\simeq 11$ MHz. Three chosen product states $|\psi_{0}\rangle$ for the autocorrelation function (\ref{c_psi0}) are shown in Fig.~\ref{fig:stark_add}\textbf{a}. The product states with the domain wall number $n_{\text{dw}}=10$, $4$, and $2$ are labeled as $|\psi_{0}^{(10)}\rangle$, $|\psi_{0}^{(4)}\rangle$, and $|\psi_{0}^{(2)}\rangle$, respectively. It can be directly calculated that the $\langle  \psi_{0}^{(10)} | \hat{H} |\psi_{0}^{(10)}\rangle = \langle  \psi_{0}^{(4)} | \hat{H} |\psi_{0}^{(4)}\rangle = \langle  \psi_{0}^{(2)} | \hat{H} |\psi_{0}^{(2)}\rangle$. The results of the time evolution of $C_{\textbf{r},\textbf{r}} (|\psi_{0}\rangle)$ with $\textbf{r}=1$ are presented in Fig.~\ref{fig:stark_add}\textbf{b}. It is seen that for the product state with $n_{\text{dw}}=2$, the decay of $C_{\textbf{r},\textbf{r}} (|\psi_{0}\rangle)$ can be neglected, while the decay becomes stronger when we consider $C_{\textbf{r},\textbf{r}} (|\psi_{0}\rangle)$ with $n_{\text{dw}}=4$ and $10$. 

Actually, in ref.~[15], it has been shown that the infinite-temperature autocorrelation function can be expanded as 
\begin{equation}\label{expand}
C_{\textbf{r},\textbf{r}} = \frac{1}{D}\text{Tr}[\hat{\rho}_{\textbf{r}}(t)\hat{\rho}_{\textbf{r}}]  = \frac{1}{D} \sum_{k=1}^{D} \langle k |\hat{\rho}_{\textbf{r}}(t)\hat{\rho}_{\textbf{r}}| k \rangle, 
\end{equation}
where $|k\rangle = |\sigma_{1,\uparrow}\sigma_{2,\uparrow}...\sigma_{12,\uparrow};\sigma_{1,\downarrow}\sigma_{2,\downarrow}...\sigma_{12,\downarrow}\rangle$ is the product states in the $\sigma^{z}$ basis. As shown in Fig.~\ref{fig:stark_add}\textbf{b}, a single term $\langle k |\hat{\rho}_{\textbf{r}}(t)\hat{\rho}_{\textbf{r}}| k \rangle$ in (\ref{expand}) cannot capture the properties of infinite-temperature spin transport. In our work, we employ the quantum circuit shown in Fig.~1\textbf{c} to directly measure the infinite-temperature autocorrelation function, without the need of sampling different product states. 

